\documentclass[12pt,preprint]{article}
\pdfoutput=1           
\usepackage{graphicx,times}
\usepackage{natbib}
\usepackage{amssymb,amsmath}
\usepackage[dvips]{epsfig}
\usepackage{pstricks}
\usepackage{pst-plot}
\usepackage{pstricks-add}
\bibpunct{(}{)}{;}{a}{}{,}

\usepackage[letterpaper=true,pagebackref=true]{hyperref}
\hypersetup{pdftitle = The title of my PDF, pdfauthor = My name, pdfsubject= The subject, pdfkeywords = keyword1 keyword2 keyword3} 
\hypersetup{colorlinks = true, linkcolor = green, anchorcolor = red, citecolor = blue, filecolor = red, pagecolor = red, urlcolor = red}
\title{A thick-disk galaxy model and simulations of equal-mass galaxy pair collisions}

\author{Guillermo Arreaga-Garc\'{\i}a\\
        Departamento de Investigaci\'on en F\'{\i}sica \\
        Universidad de Sonora \\
        Apdo. Postal 14740, C.P. 83000, Hermosillo, Sonora, Mexico.\\
        {\it guillermo.arreaga@unison.mx} }

\begin{document}

\maketitle

\begin{abstract}
We implement a numerical model reported in the literature to simulate the evolution of a galaxy composed of four 
matter components, such as: a dark-matter halo; a rotating disk of stars; a spherical bulge of stars and 
a ring of molecular gas. We show that the evolution of this galaxy model is stable at least 
for 10 Gyr (Gyr=$10^{9}$ years). We characterize the resulting configuration of this galaxy model 
by figures of the circular velocity and angular momentum distribution; the tangential and radial components of the 
velocity; the peak density evolution and the radial density profile. Additionally, we calculate several 
models of equal-mass galaxy binary collisions, such as: (i) frontal and 
(ii) oblique (with an impact parameter), (iii) two models with initial conditions taken from a 
2-body orbits and (iv) a very close passage. To allow comparison with the galaxy model, we characterize 
the dynamics of the collision models in an analogous way. Finally, we determine the de Vaucouleurs fitting curves 
of the radial density profile, on a radial scale of 0-100 kpc, for all the collision models irrespective 
of the pre-collision trajectory. To study the radial mass density and radial surface density profiles at a smaller radial 
scale, 0-20 kpc, we use a four-parameters fitting curve.
\end{abstract}

{\bf keywords--galaxies: kinematics and dynamics--galaxies: interaction--methods: numerical}

  
%
\newpage
\section{Introduction}           
\label{sect:intro}

The pioneering work of~\citet{Toomre1972} showed that the gravitational interaction between galaxies 
results in a profound morphological transformation of the participating galaxies and even leads to the formation 
of new types of galaxies. In that paper, the authors used a dynamic approach, in which the two 
colliding galaxies were represented as point masses, while the disk of the galaxies was represented with 
particles that had no gravitational interaction between them. In their simulations, these authors managed to 
reproduce systems of galaxies in which long lines or bridges appeared, whose similarity with 
real systems, such as the Antenna galaxy (NGC 4038-39) or the Mice galaxy (NGC 4676), was very encouraging.

Since the 1970s, the simulation of the formation, evolution and interaction of galaxies has been a huge research 
area of computational astrophysics and continues to be of interest even today, see~\citet{atha}.
Although many simulations of galaxy collisions have been carried out over the last few decades, in the first 
years only gas-free models were considered. Recently, gas began to be included to study its 
effects. In particular, numerical simulations aimed to follow the gas dynamics 
in a collision between a pair of comparable-mass galaxies, have a 
long history. Pioneering works were done 
by \citet{negroponte83}, in which the gas was represented by spherical 
particles of variable radius; by \citet{noguchi}, in which the influence 
of the tidal force of a perturbing galaxy on the gas dynamics of 
the companion galaxy was studied. A new generation of papers, in which 
the SPH technique was already used, were done by \citet{barnes1991}, 
\citet{barnes1996} and \citet{MihosHernquist96}, among others. 

The papers \citet{barnes1991} and  
\citet{barnes1996} showed that a strong concentration of gas 
takes place in the central region of the remnants and for this 
reason, they argued about the possible occurrence of a 
starburst in the central region. Noteworthy, \citet{barnes1996} demonstrated that 
the gas and the stars showed a different behavior whether the 
galaxy model evolved as an isolated system or during a galactic 
collision. \citet{barnes1996} also noted that the morphology of 
a merger remnant can be strongly affected by the dynamics of the gas.

Even more recently, ~\citet{naab} presented a large set of simulations of uneven-mass galaxy collision models 
to understand the influence of a gas component on the global structure of mergers remnants. These authors 
found that the presence of a gas component changes the shape of the merger remnants. ~\citet{burkert} also found that 
some physical properties of the merger remnants depends both on the initial mass ratio of the colliding galaxies 
and on the gas fraction that they contain. 

Numerical simulations of the interaction between a pair of galaxies have been considered in statistical terms 
by \citet{matteo}, who studied a total sample of 240 interactions. They first determined the star formation rate in 
their galaxy model to then compare these results with those obtained in their models of galaxy interaction.  
The images they show in their section ''A gallery of galaxy interactions'' are impressive, as they were able 
to compare the time evolution of several matter components of their models, see for instance their Figs. 3,4 and 5.

\citet{Gabassov2006} presented a rotating 
galaxy model in which three matter components were included, namely: a dark-matter halo, a spherical 
star-bulge, and a rotating star-disk. However, they did not include gas in their model. After six times of the 
rotation period 
of their galaxy model, the galactic evolution generates a bar, so that their model successfully 
reproduces the dynamics of a barred spiral galaxy. Subsequently, \citet{Luna2015} (following the work 
of~\citet{Gabassov2006} presented several collision models in which the spiral bar galaxy 
introduced by \citet{Gabassov2006}, was the only element of collision. \citet{Luna2015} also did not 
include gas in their models.

In the present paper, we also follow the galaxy model of~\citet{Gabassov2006} and~\citet{Luna2015} regarding the 
initial dynamics of the three matter components mentioned earlier, but here we also include gas, which 
is initially distributed as the disk. ~\citet{moster} considered a five components galaxy model: dark-matter 
halo, stellar disk, stellar bulge, gaseous disk and gaseous halo. Consequently, our galaxy 
model includes a gaseous disk component. It is also 
important to emphasize that all four matter components considered in this work interact gravitationally and, 
as expected, the computational cost increases significantly with respect to papers in which the gravitational 
interaction is modeled or suppressed entirely.

This galaxy model proves to be 
stable. Therefore, it can be useful to represent the M82 Galaxy, that was 
originally cataloged as an irregular amorphous galaxy, as reported by~\citet{Mayya}, whose galaxy model 
showed the formation of an elongated disk-shaped structure in 
the central region. We characterize the dynamic of this galaxy model by calculating 
the time evolution of the density peak and the radial density profile of all the matter components at 
an advanced evolution stage.
 
It must be noted that the parameter space of hydrodynamical simulations is enormous, even in their most basic 
implementation, so a new paper can almost always find a new possible variation of 
these parameters. In the case of this paper, the width of the disk is greater than what is 
commonly used in many papers. 

It should be emphasized that a galaxy model like this is physically possible 
and interesting, because collisions between galaxies of very different masses can thicken the disk of the most 
massive galaxy, see~\citet{quinn}. \citet{villa} presented SPH simulations to explore the 
problem of thick disc formation by means of minor collisions 
between a satellite galaxy hitting on a host galaxy with a pre-existing thin disc. The scaleheight of the initial thin disk 
is 0.35 kpc while the resulting vertical structure of the thick disk indicates a scaleheight within 1-2 kpc, see their 
Fig.10.

In this paper we also consider a small sample of equal-mass thick-disk galaxy collision models, like the ones 
obtained by \citet{villa}, so that 
we repeat the characterization analysis on the merger remnants, to assess 
the effects of the collision process on the dynamics of the matter components, particularly on the 
gas component. We find that the gas forms rapidly rotating structures with a peak density in the central 
region.   

We then calculate the radial density profile of the merger remnants and report the 
values of the parameters $b_e$, $\log(\rho_e)$ and $R_e$ that best fit it by using a de Vaucouleurs 
function, so that this fitting curve apparently does not depend on the particular geometry of the collision 
process on a radial scale of 0-100 kpc. It should be noted that ~\citet{aguilar} found that the de Vaucouleurs 
surface brightness profile does not change significantly after a couple of galaxies have undergone a tidal 
encounter, both of which started their evolution with the de Vaucouleurs density profile with other parameters. 

There are many empirical formula available in the literature 
to obtain fitting curves in addition to the de Vaucouleurs, such as the S\'ersic function, core-S\'ersic, S\'ersic-type 
transition model, Nuker model, for a review see~\citet{ferrarese}. In addition, 
~\citet{kormendy2009} found that the S\'ersic 
functions fit very well the surface brightness profiles of elliptical and spheroidal galaxies in 
the Virgo cluster. They then tried to distinguish between elliptical and spheroidal galaxies by noting the 
differences in these fits for small radii, so that these differences can be interpreted as signatures of 
the galaxy formation mechanism. 

These formula are improvements to the de Vaucouleurs function. In spite of this and in addition to the 
fact that there is no astrophysical reason known to highlight the de Vaucouleurs function over the other formula, we will show in 
this paper a similar result to that found by ~\citet{aguilar} that is, the merger remnants manage to 
take a radial density profile with form of the de Vaucouleurs function, irrespective of the collision model. It 
must be emphasized that the reconfirmation of this result is now obtained by using a more complete galaxy model, 
because ~\citet{aguilar} used 3000 particles per galaxy model. We complement these results with the de Vaucouleurs 
function, whose details are shown in Appendix A, by testing with another formula in the radial 
scale of 0-20 kpc, which has given good results as a fitting model for the 
radial profiles of the HI surface density for 42 galaxies, as \citet{wang} demonstrated 
recently. These four-parameters fitting function is described in Appendix B.           

The rest of this paper is structured as follows. In Sections~\ref{subsec:inimass} to~\ref{subsec:inivel}, we  
explain the generation of the initial conditions of the galaxy model. In 
Section~\ref{subs:code}, we present the evolution code. In 
Section~\ref{sec:results} we show the results obtained: first, for the evolution of 
the galaxy model in Section~\ref{subs:evol} and second, for the collision models in 
Section~\ref{subs:galcol}. A dynamic characterization of the matter components between different 
collision models is presented in Section~\ref{subs:dyncaracol}. Some results 
of this paper are discussed in Section~\ref{sec:disc}.  
In Section \ref{sec:comp} we will try to establish the consistency of the simulations 
presented in this paper by comparing our main results with other simulations, with 
observations and with virtual observations as well. Finally, the main conclusions of this paper are 
summarized in Section~\ref{sec:conclu}.  

\section{The galaxy model}
\label{sec:galm}

In this paper, we use the SPH (smooth particle hydrodynamics) technique, in 
which a fluid is represented by a finite set of particles (see \citet{liu} and 
references there in), so that the galaxy model has four types of 
particles, one for each type of matter component: halo, bulge, disk and gas. It is important 
to note that there is a difference between these particle types from the computational point 
of view, as will be mentioned in Section~\ref{subs:code}.  

As usual, each particle must have a mass, a position and a velocity at time $t = 0$. We show in 
Sections~\ref{subsec:inimass},\ref{subsec:inipos} and \ref{subsec:inivel}, how the mass, the positions 
and finally, the velocities are assigned, respectively.

\subsection{Initial mass of particles}
\label{subsec:inimass}

Because the total mass of each component of the galaxy model is very different, the number 
of particles will also be very different. This is mainly because we want to use only a single 
magnitude of elementary mass for all the matter components. It is shown elsewhere that the 
simulations of this kind produce better computational results than those with particles 
having very different elementary masses.

In Table~\ref{tab:param}, we list the matter 
component and its properties, as follows. To achieve the total mass per matter component 
given in column 3, the mass of the elementary particle, which is of 412,500 $M_{\odot}$, must be multiplied by 
the number of particles given in column 2. The fractions that this matter component 
represent in the entire galaxy model are shown in column 4. 

The total number of particles in the galaxy model is 676237, such that the total mass 
is 2.79 $\times 10^{11} \, M_{\odot}$ that extends over a sphere of radius of 240 kpc. The average density 
of the system is $3.02 \times 10^{-28}$ g/cm$^3$.

The masses reported in Table~\ref{tab:param} were suggested 
in the papers by \citet{Gabassov2006} and~\citet{Luna2015}. The meaning of the other columns of 
Table~\ref{tab:param} are explained below.    

\begin{table}[ph]
\caption{Parameters of the galaxy model.}
{\begin{tabular}{|c|c|c|c|c|c|c|} \hline
matter    & number of & total mass      & mass fraction & initial radial  & final radial    & a [kpc] \\
component & particles &  [$M_{\odot}$]  &               & extension [kpc] & extension [kpc] &         \\
\hline
\hline
gas   &  10000  &  4.12$\times 10^{9}$   &   0.0147  &  16-20 &  180  & --\\
\hline
disk  &  99950  &  4.12$\times 10^{10}$  &   0.147   &  0-16  &   60  & 3.3\\
\hline
bulge &  33205  &  1.4$\times 10^{10}$   &   0.049   &  0-60  &  160  & 1.66\\
\hline
halo  &  533082 &  2.2$\times 10^{11}$   &   0.788   &  0-240 &  $\ge$ 200 &  4\\
\hline
\hline
\end{tabular} }
\label{tab:param}
\end{table}
\subsection{Initial position of particles}
\label{subsec:inipos}

The Monte Carlo method is used, so that 
the particles are located randomly in the space available of the galaxy model. In column 5 of 
Table~\ref{tab:param}, we show the initial radial extension achieved by the initial distribution of 
particles. The number of particles in a ring of radial width $R$ and $R + \delta R$ is determined 
to satisfy the radial density profile that has been reported in the literature. For example, for the 
dark-matter halo, we use the density profile reported by \citet{Dehnen}, which includes the length 
parameter $a_h$. For the bulge we use the profile reported by \citet{Hernquist1990}, which also includes 
a parameter of length $a_b$; for the disk we use the formula reported by \citet{Freeman}, with a 
length parameter $a_d$. The length parameter $a$ determines the radius in which the density curve falls 
with respect to the radius of the galaxy. The values of these parameters are reported 
in column 7 of Table~\ref{tab:param}. It should be emphasized again that these formulas have been taken 
from the papers of~\citet{Gabassov2006} (see equations 1, 2 and 3) and~\citet{Luna2015} 
(see equations 1, 2 and 3) and therefore we do not reproduce them again here. 

The gas particles were initially located between an inner and outer radii, so that the 
gas particle radius is always greater than the initial radial extent of the star disk 
reported in column 5 of Table~\ref{tab:param} and smaller than the radial limit of 20 kpc. In other words, 
the gas was uniformly distributed in a ring in the range 16--20 kpc. In this 
case, there is no parameter $a_g$, as reported in Table \ref{tab:param}. The width of the star disk must 
also be specified, which was set in this work at the value of $z_0$ = 1 kpc.

The gas component can be located initially forming a ring, as is usually observed to be in spiral galaxies, 
see \citet{extra}. The typical inner and outer radii of this molecular gas ring for spiral galaxies 
are 3 kpc $<$ R $<$ 8 kpc with an scale-height of 0.09 kpc. Atomic hydrogen gas can be observed up to a radius 
of $R<25$ kpc with a scale-height of 0.2 kpc. However, as we mentioned in Section \ref{sect:intro}, in this paper 
we consider the case of a thick disk of gas, which can be the result of several collisions between a 
large disk and small companions, as was modeled by \citet{quinn}, who demonstrated that the 
original disk is not destroyed (as usually happens in the case of major mergers) but is slowly 
disturbed, so that the resulting disk spreads in radius and inflates vertically until it eventually 
settles into a new equilibrium configuration. For this reason, the ring of gas in this paper has initially been 
located as explained above.
\subsection{Initial velocity of particles}
\label{subsec:inivel}

We determine the initial velocities of the particles by means of a distribution function. For example, assuming 
that the halo and bulge are isotropic, they then follow a Maxwell distribution function, with only a radial 
velocity dispersion, denoted by $\sigma_r$, which determines the opening of the distribution function 
curve (a Gaussian curve). In general, the radial dispersion of the velocity for the halo and bulge depends on 
the radial coordinate of the galaxy. For anisotropic cases, velocity dispersions in all coordinate directions must 
also be included, for example, in spherical coordinates with $\theta$ and $\phi$ the 
azimuthal and polar angles, then $\sigma_{\theta}$ and $\sigma_{\phi}$ are the velocity 
dispersions needed. In this work, as a first approximation, we only consider isotropic velocity 
distributions for the halo and bulge (see equations 4 and 6 of the paper by 
\citet{Luna2015}).

Likewise, it should be emphasized that we use the escape velocity, defined by 
$V_{\rm esc}=\sqrt{2 \, G \, M(R)/R} $, of each matter component as an upper 
limit of velocity magnitude in the velocity distribution function. Here, $M(R)$ is the total mass of each 
matter component contained up to the radius $R$ and $G$ is Newton's gravitational constant.

This means that all velocities greater 
than the escape velocity were re-defined with the value of the escape velocity. Consequently, we consider 
that the resulting velocity distribution may be characterized by comparing the average 
velocity of each matter component with its corresponding escape velocity, see Table~\ref{tab:vel}.

\begin{table}[ph]
\caption{The average velocity of the galaxy model obtained from a distribution function.}
{\begin{tabular}{|c|c|c|} \hline
matter    & escape velocity & average velocity\\
component &     [km/s]      &     [km/s]      \\
\hline
\hline
gas   &  42   &  13 \\
\hline
disk  &  148  & 106\\
\hline
bulge &  44   & 30 \\
\hline
halo  &  88   &  87\\
\hline
\hline
\end{tabular} }
\label{tab:vel}
\end{table}

The particles of the disk have an assigned angular velocity, the value of which depends on the 
radial coordinate of the particle and the value of its gravitational potential in that radial 
coordinate. We emphasize that the average value of the 
angular velocity of the disk particle distribution is $1.3 \times 10^{-16}$ radians per second. 

The velocity distribution functions for the disk are characterized by three dispersion functions, in cylindrical 
coordinates $(r, \theta, z)$ these are $\sigma_r$, $\sigma_{\theta}$ and $\sigma_z$. The velocity dispersion in the radial 
coordinate depends on the value obtained from the angular velocity and the radius of the disk, so that 
the mathematical formula was taken from the article by \citet{Hohl} (see his equation 8).

Following the papers of~\citet{Gabassov2006} and~\citet{Luna2015}, the velocity dispersion in the $z$ coordinate, 
$\sigma_z$, is given in terms of the radial dispersion, as follows $\sigma_z = \sigma_r / 2$. Following the paper 
by \citet{Hohl}, the dispersion of the velocity in the tangential direction becomes equal in magnitude to the radial 
dispersion, thus $\sigma_{\theta} = \sigma_r$. These choices are somewhat arbitrary and can be changed according to 
the galactic dynamics desired. In this case, the galaxy model rotates differentially 
(not as a rigid body), with the Z-axis as the axis of rotation, so that the rotation period of 
the galaxy is about 1.5 Gyr.

The gas component described in Section \ref{subsec:inipos}, was endowed with a radial velocity 
dispersion similar to the disk, except that we now use the average angular velocity of the disk for all 
of the gas particles, rather than the angular velocity at each radial position of the particle. With this 
procedure, we try that the ring of gas rotates as a rigid body with constant angular velocity. The 
tangential component of the velocity of each particle increases linearly with its radius and is proportional to  
a constant, which is called the epicyclic frequency of the system, see for example equations 
6--63 on page 371 of ~\citet{BinneyTremaine}.

\begin{figure}
\begin{center}
\begin{tabular}{cc}
\includegraphics[width=2.5 in]{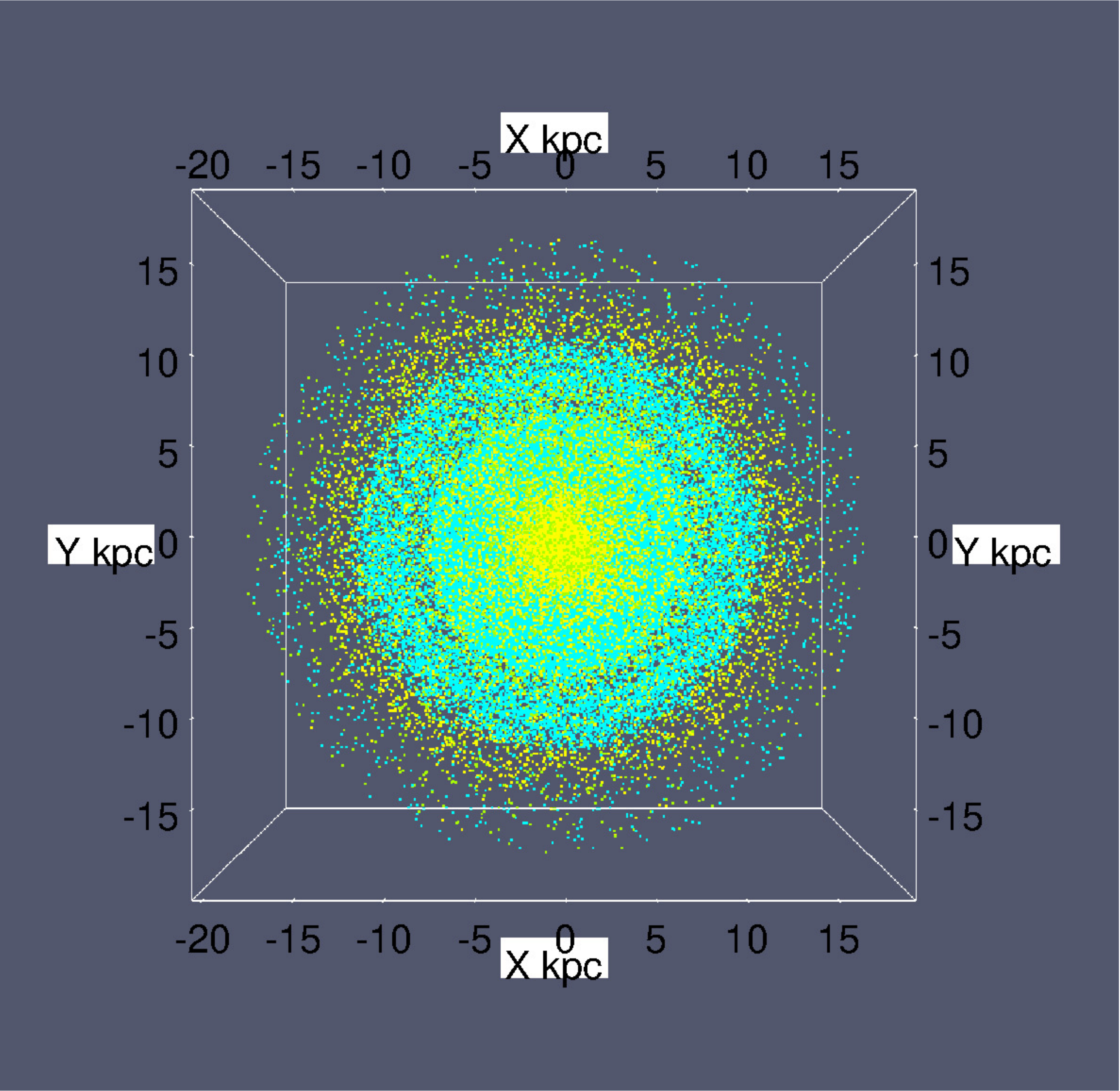}&\includegraphics[width=2.5 in]{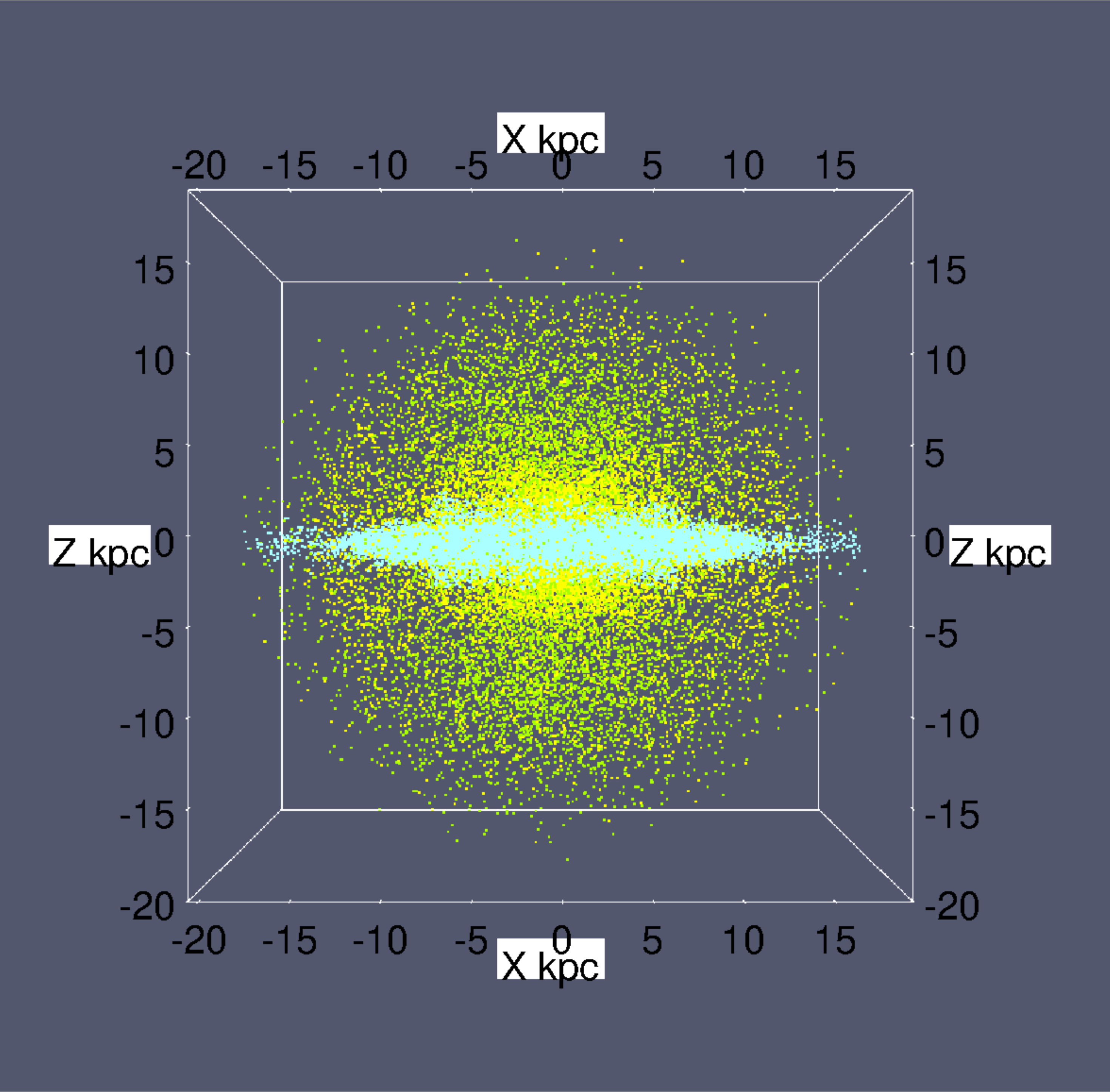}\\ 
\end{tabular}
\caption{\label{initialcond} Initial configuration for all the particles at time t = 0 of the three 
matter components: a view of the XY plane is shown in the left panel, while a view of the ZX plane 
is shown in the right panel. The region shown is within the interval (-20,20) kpc for the 
X-axis, the Y-axis and the Z-axis. The colors indicate the matter components according to bulge-yellow, 
gas-green and disk-blue.}
\end{center}
\end{figure}

In Fig.~\ref{initialcond} we show the initial configuration of all the particles for three matter 
components, because the dark-matter halo is not shown. The evolution of 
these initial conditions is carried out using the public code Gadget2, which is presented in 
Section \ref{subs:code}. The initial conditions were evolved up to 10 Gyr or equivalently, for almost 6.6 times the 
rotation period of the galaxy model. To carry out this evolution, 
almost 200 hours of computation were necessary running in parallel in 40 processors of the Cuetlaxcoapan 
Supercomputer of the LNS-BUAP. The results obtained are presented below in Section \ref{sec:results}, by means of 
figures in which the matter components are shown separately by a color set, assigned by the public code see ~\citet{paraview}.
\subsection{The evolution code}
\label{subs:code}

In this paper, we use the particle-based code Gadget2, which is
based on the tree-PM method for computing the gravitational forces
and on the standard SPH method for solving the Euler equations of
hydrodynamics, see~\citet{Springel}. The Gadget2 program has implemented
a Monaghan--Balsara form for the artificial viscosity,
see~\citet{balsara1995}. The strength of the
viscosity is regulated by setting the parameter $\alpha_{\nu} = 0.75$ and
$\beta_{\nu}=\frac{1}{2}.
\times \alpha_v$, see equations 11 and 14
in~\citet{Springel}. We have fixed the Courant factor to be $0.1$.

In Gadget2, the SPH sums are 
evaluated using the spherically symmetric M4 kernel 
and so gravity is spline-softened with this same kernel. 
There is a smoothing length, denoted here by $h$, which establishes the compact support, so 
that only a finite number of neighbors to each 
particle contribute to the SPH sums. In particular, each particle has its own smoothing
length, which evolves with time so that the mass contained in
the kernel volume is a constant for the estimated density. Particles are also 
allowed to have individual gravity softening lengths, denoted by $\epsilon$, which evolve in step 
with the smoothing length $h$, so that the ratio $\epsilon/h$ is of
order unity. The $\epsilon$ determines the smallest possible separation for 
two individual particles, so that the spatial resolution of a simulation is set by the choice of $\epsilon$.
In Gadget2, $\epsilon$ is set equal to the minimum 
smoothing length $h_{\rm min}$, calculated over all particles at the end of each time
step. 

As we mentioned at the beginning of Section~\ref{sec:galm}, there are six types of particles 
defined in the Gadget2, which are 
labeled from 0 to 5. When the gravitational interaction is computed, all the of particles are 
treated in the same way by the Gadget2, irrespective of the particle type. However, it is allowed that 
each particle type can have a different gravitational softening. 

In this paper, the gas particles have been assigned the Gadget2's particle type 0; in this case, there is 
an hydrodynamical force to be calculated in addition to the gravitational force, so that the former includes 
a pressure gradient generated by differences in the spatial distribution of the thermal pressure 
field. Consequently, these particles are considered as collisional particles. 

All of the other particle types of the Gadget2 are considered as collision-less particles. Thus, the dark-matter 
particles have been assigned the Gadget2's particle type 1, which means that they are 
treated as collision-less particles of unknown nature. The disk particles have been assigned a Gadget2's particle 
type 2. The bulge particles have been assigned the Gadget2's particle type 3, so that 
the disk and the bulge are both composed of collision-less stars, but the Gadget2 allows them to have  
different masses. 

However, due to our implementation procedure based on an elementary mass 
particle, described in Section~\ref{subsec:inimass}, in this paper the only difference 
between the disk and bulge components is the number of elementary particles that are used to represent 
their total masses, see~Table~\ref{tab:param}. 

Finally, it should be noted that there are no other differences between these particle types in addition to 
their collision or collision-less nature. Moreover, this 
paper has not considered the Gadget2's particle type 4, labeled in the code 
as ''Stars'', which allows the implementation of a star formation algorithm. Gadget2's particle 
types will be very useful in Section~\ref{sec:results}, where plots 
will be presented in which the particle types are handled separately, so that we will follow the spatial 
distribution of each particle type in both the galaxy model and in the collision models.        
\section{Results}
\label{sec:results}

\subsection{Evolution of the galaxy model and its dynamic characterization}
\label{subs:evol}

\begin{figure}
\begin{center}
\begin{tabular}{cc}
\includegraphics[width=2.5 in]{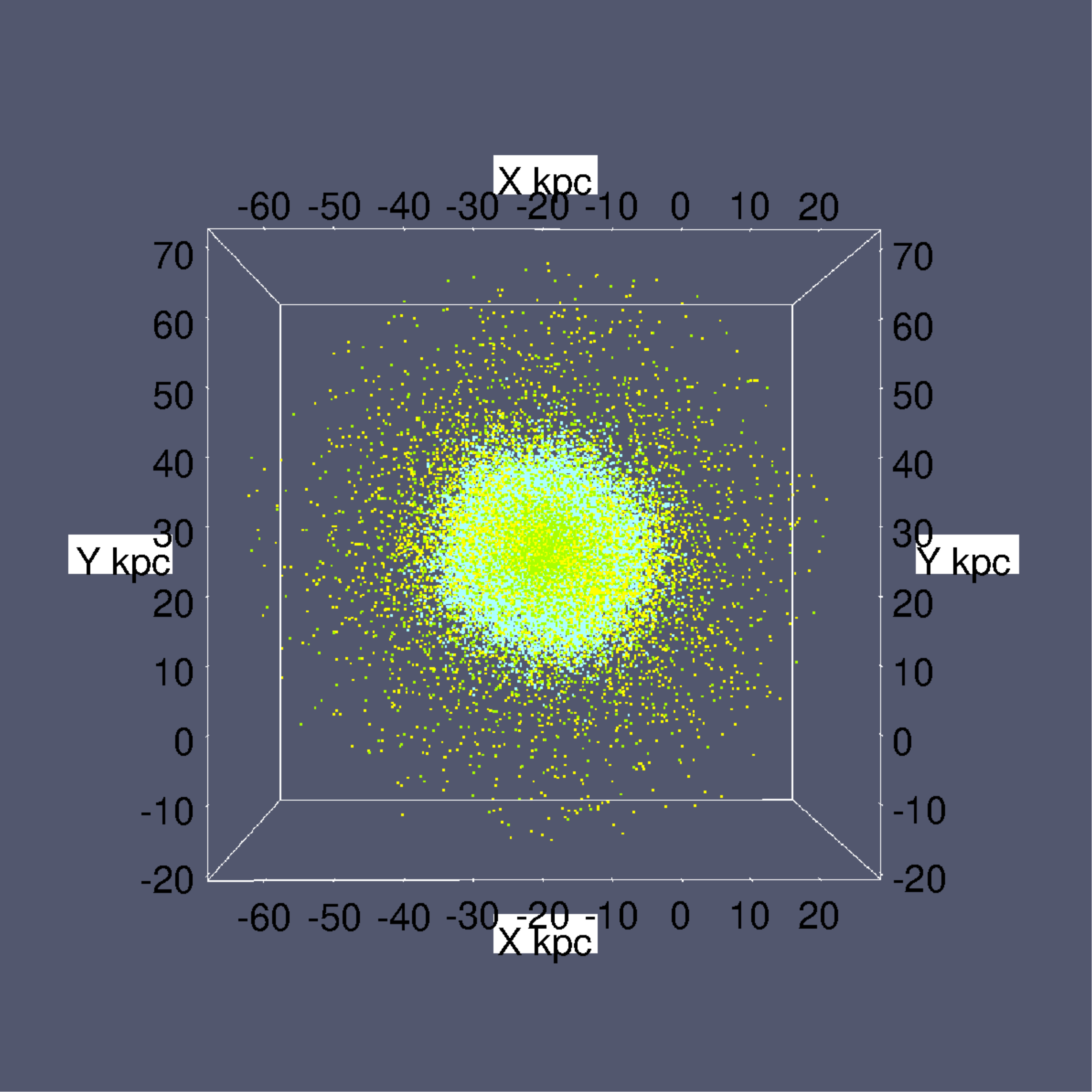}&\includegraphics[width=2.5 in]{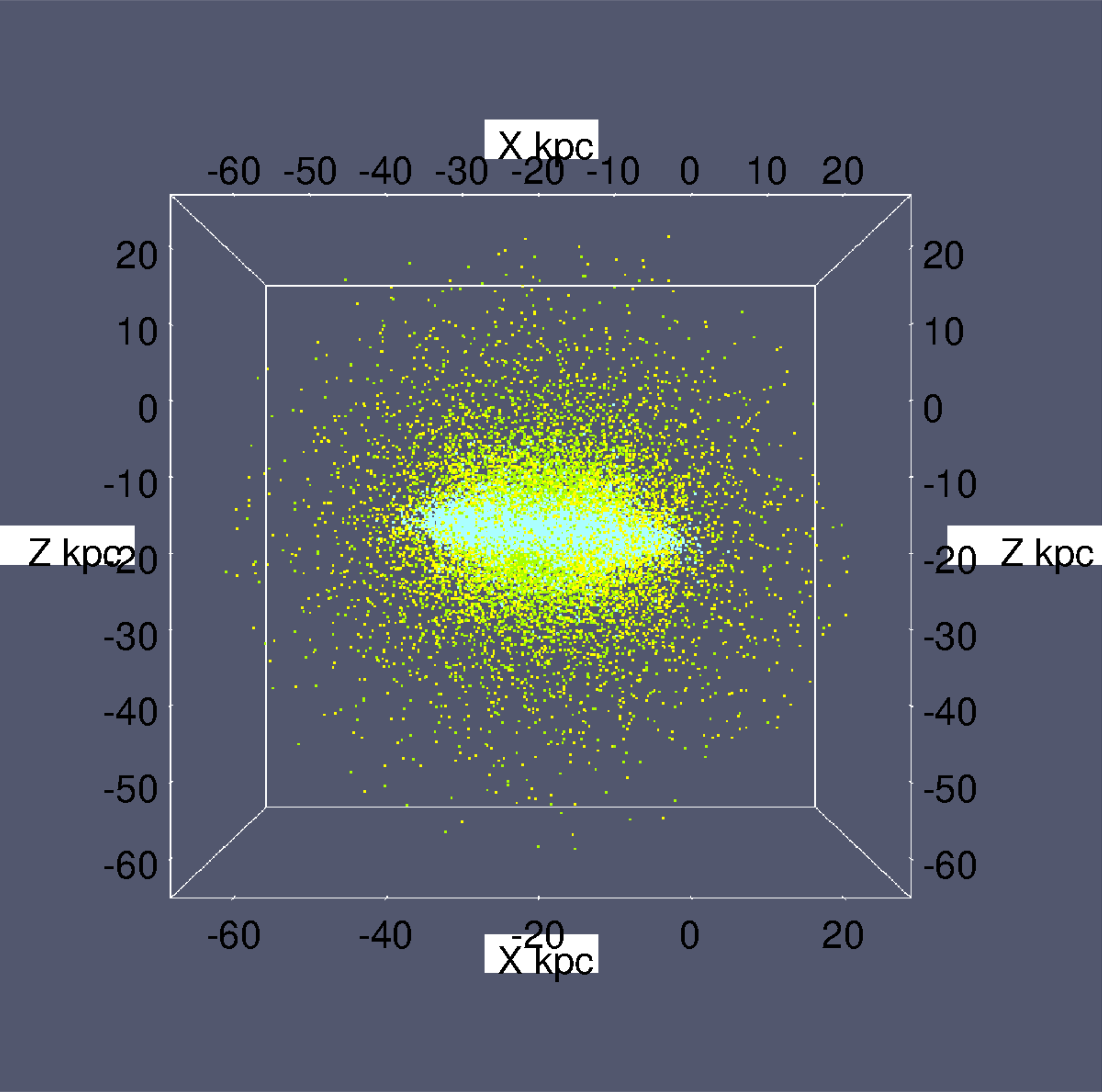}\\
\includegraphics[width=2.25 in]{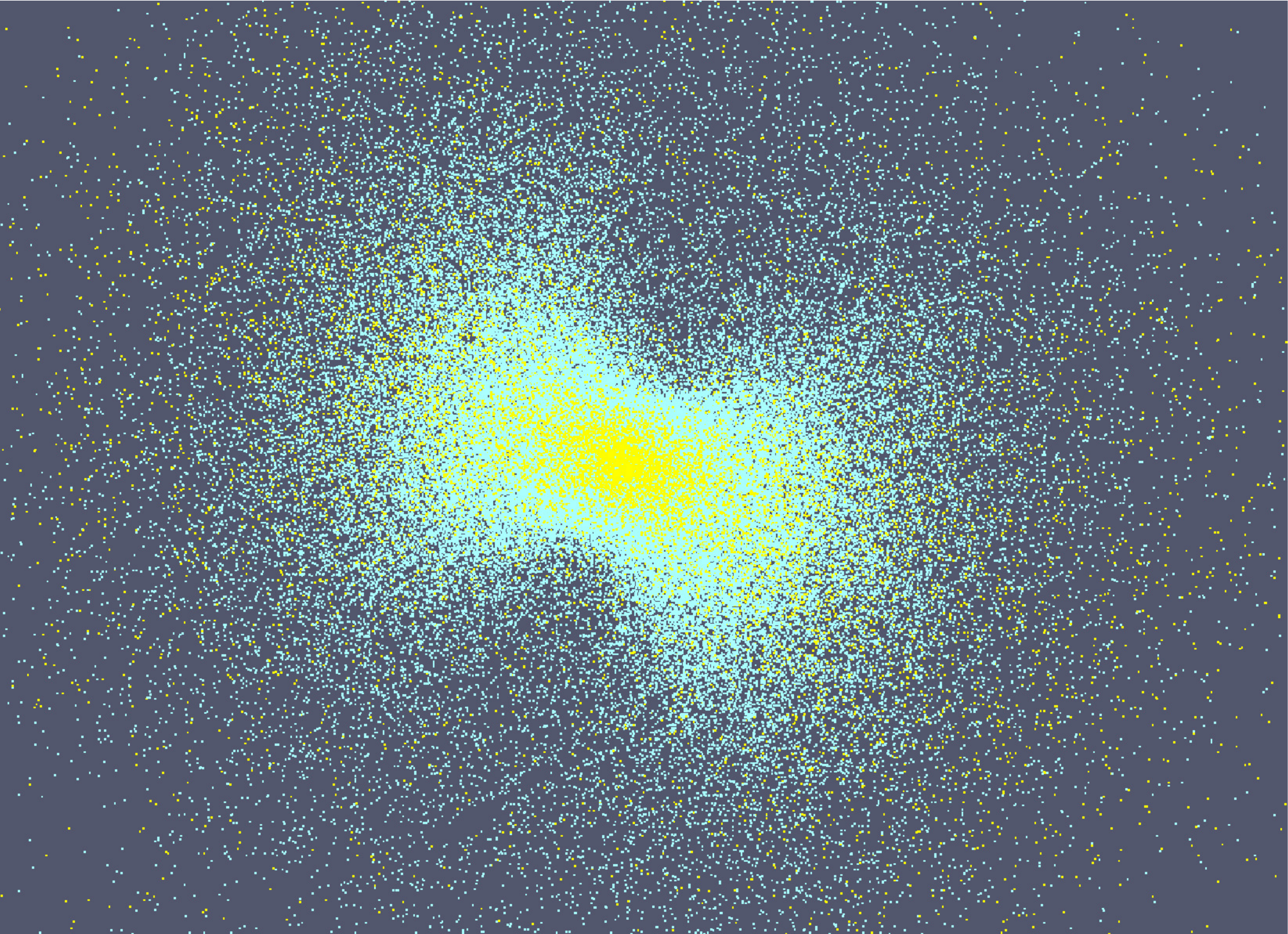} &\includegraphics[width=2.5 in]{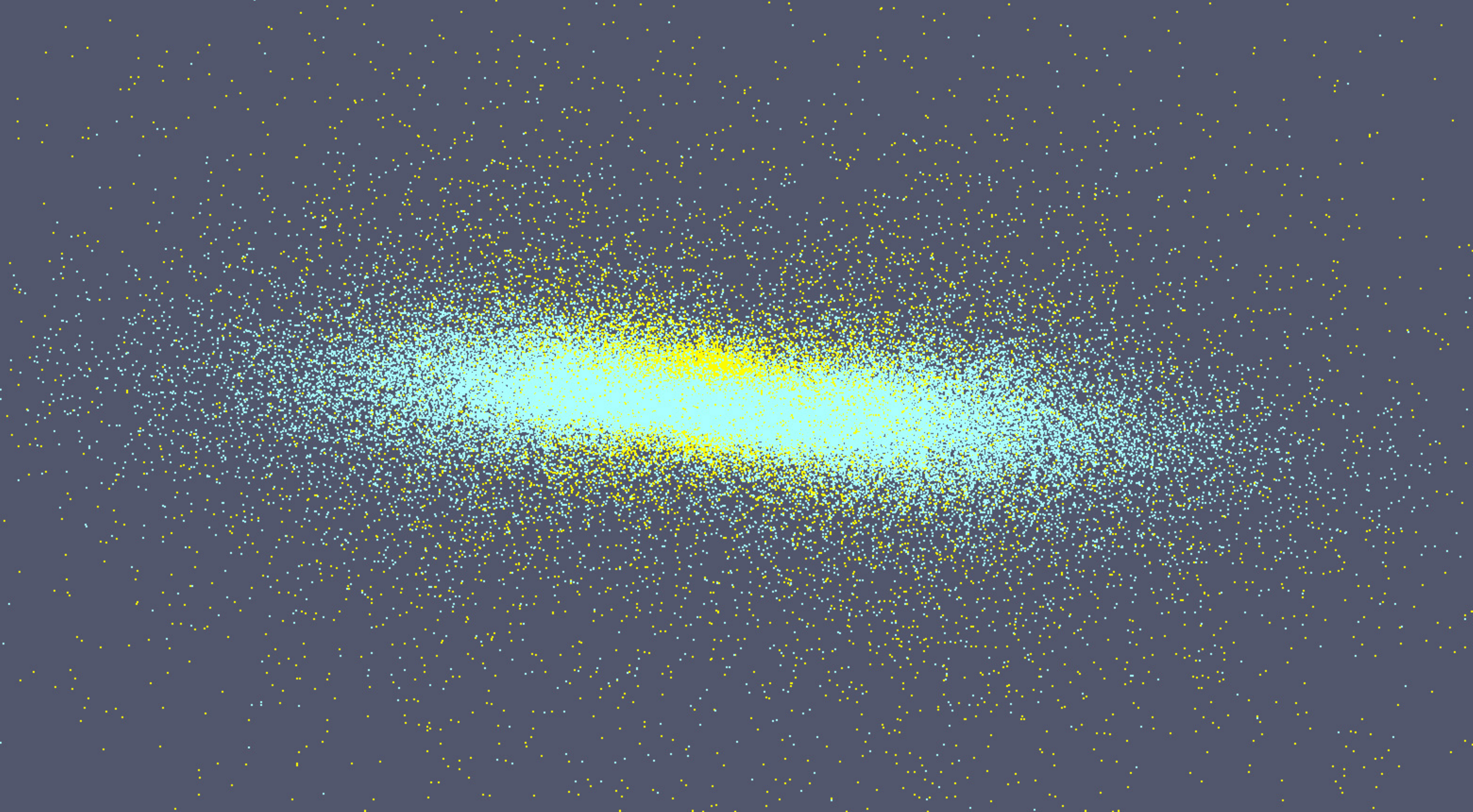}  
\end{tabular}
\caption{\label{finalcond} At time t = 9.8 Gyr, equivalent to 6.5 times the rotation period of the 
galaxy model, we show a XY view at the top left panel and a ZX view at the top right panel. The 
region shown is within the interval (-60,20) kpc in all the axes. Two amplifications of these panels are 
shown in the panels on the line below, respectively, so that the region amplified is now within the interval (-40,0) kpc in the 
X-axis, (0,40) kpc in the Y-axis and (-20,0) kpc in the Z-axis. The colors indicate 
the matter components according to bulge-yellow, gas-green and disk-blue.}
\end{center}
\end{figure}

In Fig.~\ref{finalcond}, we show the evolution obtained from the galaxy model up to 9.8 Gyr time or equivalently, to 
6.5 times the rotation period of the galaxy model. It is seen that the overall system has developed an elongated 
shape in the central region. The disk has also experienced an expansion in its central region. However, at the ends 
there is and additional extension in the form of a bow tie, which can be seen in the left-hand bottom panel of 
Fig.~\ref{finalcond}. The right-hand bottom panel Fig.~\ref{finalcond} indicates that the disk keeps its 
elongation along XY plane, such that in the ZX view, and it still looks like a thick disk. The bulge keeps 
wrapping the central part of the disk. 

\subsubsection{The circular velocity profile and the time evolution of the angular momentum for the galaxy model.}
\label{subs:cirvelgalaxy}

\begin{figure}
\begin{center}
\begin{tabular}{cc}
\includegraphics[width=3.0 in]{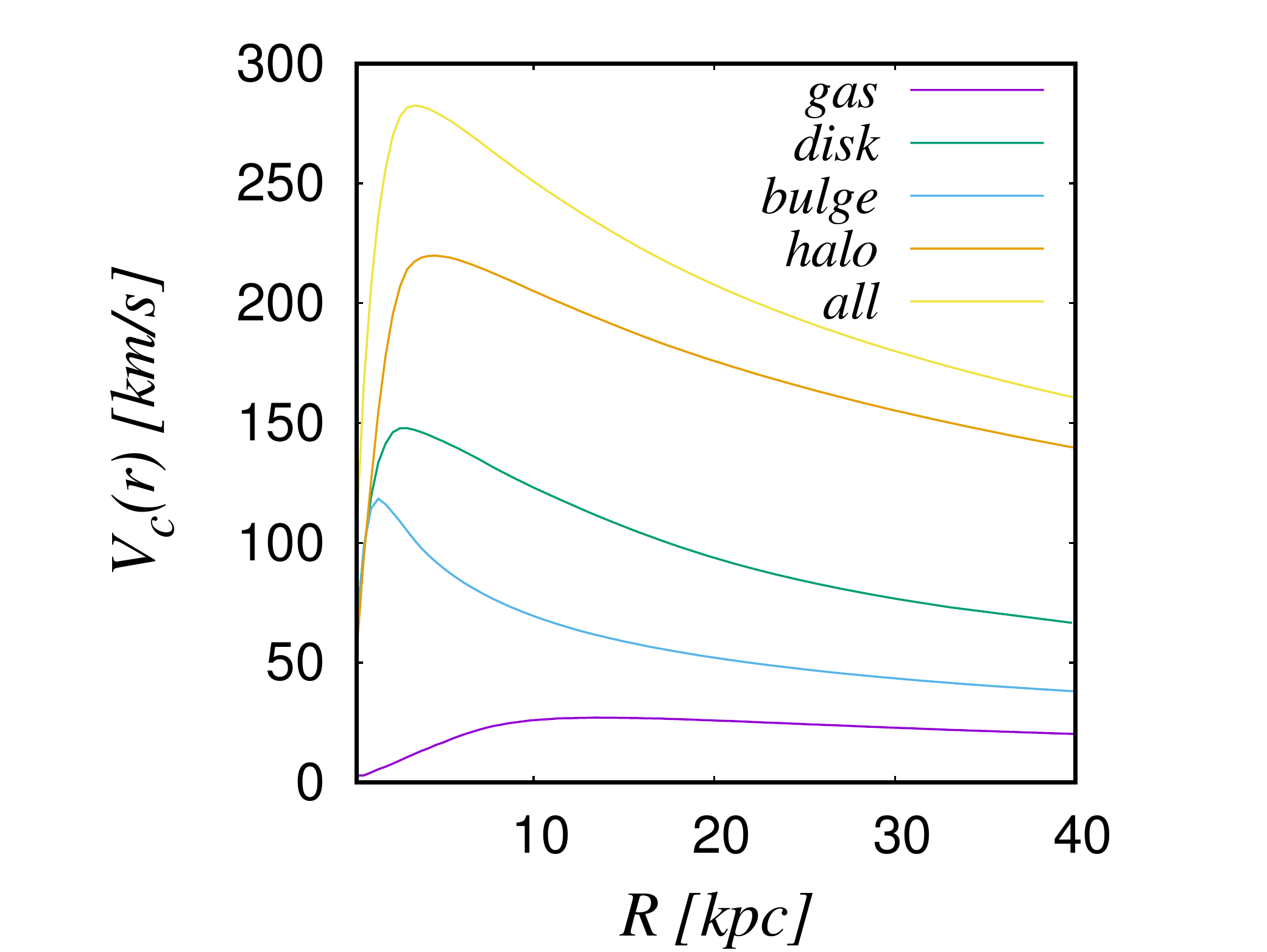}&\includegraphics[width=3.0 in]{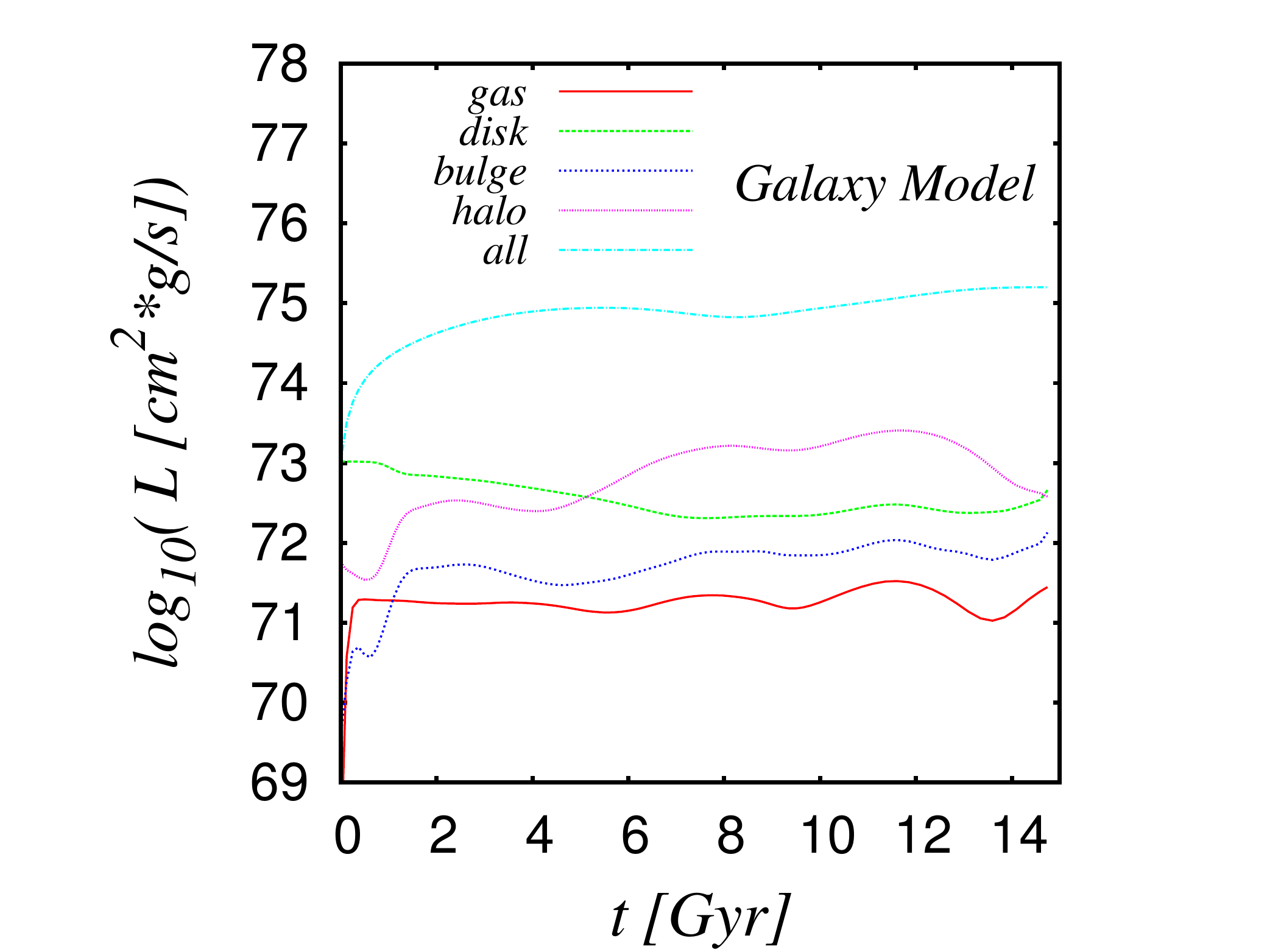} 
\end{tabular}
\caption{\label{circularvel} (left) Circular velocity of the galaxy model at time 
t = 13.7 Gyr. (right) The magnitude of the angular 
momentum L, in cgs units. On the horizontal axis, the evolution time in Giga-years. Each curve of both panels 
is generated by taking into account the center of mass of each matter component separately. }
\end{center}
\end{figure}

To characterize the mass distribution obtained at the evolution end of the galaxy model, in the left-hand panel  
of Fig.~\ref{circularvel} we show the circular velocity curves of the galaxy model, so that the matter 
components are considered separately. It must be clarified that to make this plot, we take a radial 
partition of the galaxy model in $n_{\rm bin}$ bins, starting from the center of mass of each matter component, up 
to a maximum radius of 100 kpc. Next, we accounted for all the particles contained in each radial bin by 
taking into account their matter component type, so that the total mass $M(R)$ of each matter component 
contained up to the 
radius $R$, is calculated, and we then get the circular velocity, which is defined as 
$V_{\rm cir}=\sqrt{ G \, M(R)/R} $, where $G$ is Newton's gravitational constant and $R$ is the radius 
shown on the horizontal axis. The curve labeled ''all'' includes all the particles irrespective of their 
matter component type.      

The velocity curves reach their maximum velocity at a very small radius; for greater radii, the curves 
fall very quickly as the distance to the center of the galaxy increases. Gas is an exception because its circular 
velocity curve remains practically constant for every radius greater than 10 kpc. 
It is must be emphasized that both the bulge and the gas have been extended spatially to a scale 
of 160 and 180 kpc, respectively. The disk remains more concentrated in the central region of the galaxy 
model, but some part of it reaches a length extension of 40 kpc. 

These curves can be compared with 
those calculated by~\citet{Kuijken} (see his Figure 4) who reports rotation curves with very 
pronounced drop for the bulge and the disk. It should be emphasized that the observations of the 
rotation curves of the M82 galaxy, reported by~\citet{Mayya} also show a very pronounced 
drop, just as this galaxy model does in this work.The shape of the circular velocity curves 
obtained in this paper are similar to those obtained by ~\citet{Meza}, in which the formation of an 
elliptical galaxy in a cosmological simulation is calculated.   

In the right-hand panel of Fig.~\ref{circularvel} we show the time evolution 
of the magnitude of the total angular momentum for the galaxy model. We emphasize that the 
magnitude $L$ was calculated using all the particles in the simulation, and in the case of the 
curve labeled ''all'', irrespective of the matter component type. To take into account the 
type of the matter component separately in the calculation 
of the angular momentum for the galaxy model, so that the other four curves of the 
right-hand panel of Fig.~\ref{circularvel} must be now considered.  

For the galaxy model, we observe that all these matter components 
have zero initial angular momentum (extrapolating the behavior of the curve near 
the origin of coordinates) except for the disk, whose angular momentum was given 
initially a non-zero value. However, in less than 2 Gyr of evolution, all these 
matter components very quickly acquire a significant total angular momentum that is comparable in 
magnitude to the angular momentum of the disk.

Although the total mass of the gas is considerably smaller than the total mass of the other 
matter components, the gas follows a circular movement and rapidly gains angular momentum. Its angular 
momentum is clearly smaller in magnitude than of the rest of the components. Nevertheless, its 
magnitude is very significant, because it indicates that its angular velocity should be very large.

\begin{figure}
\begin{center}
\begin{tabular}{cc}
\includegraphics[width=3.0 in]{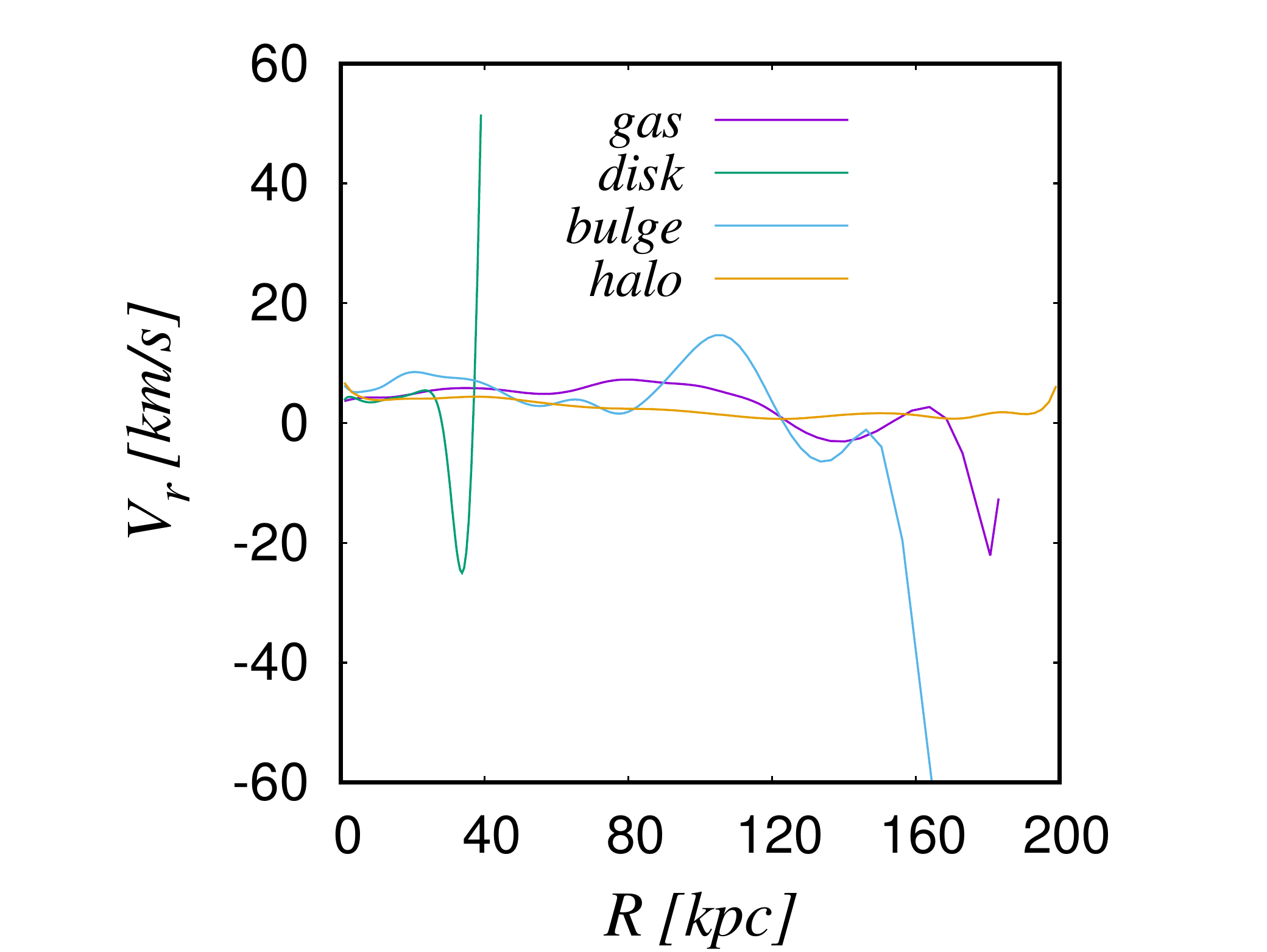}&\includegraphics[width=3.0 in]{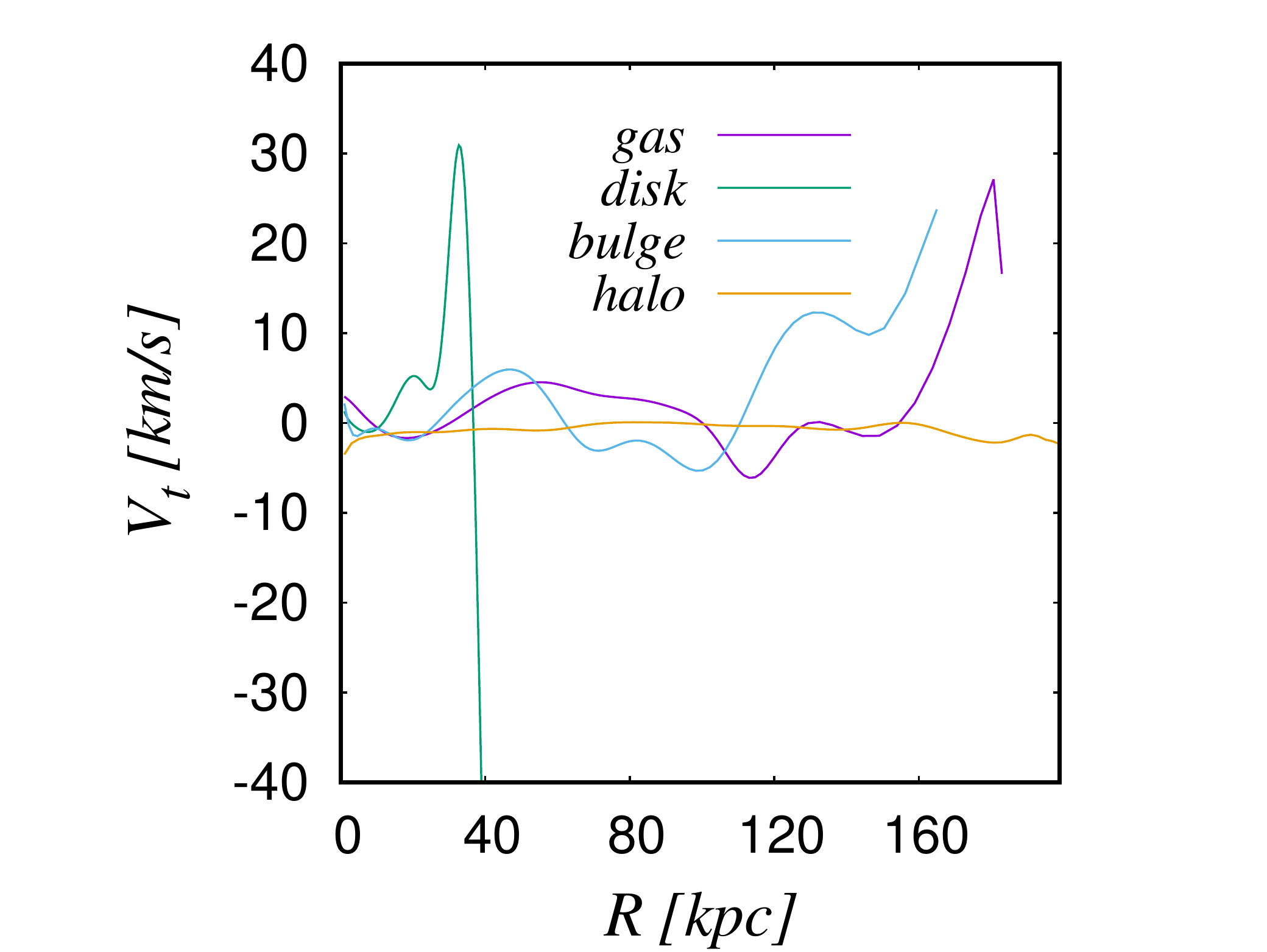} 
\end{tabular}
\caption{\label{componentesvel} At the same time of Fig. \ref{circularvel}, in the left panel we show the 
radial component of the velocity and in the right panel the corresponding tangential component of the velocity.}
\end{center}
\end{figure}

In Fig.~\ref{componentesvel}, we show the radial and tangential components of the 
velocity for each matter component of the galaxy model. It can be seen that the disk maintains its initial 
nature of a rotating rigid body. Meanwhile, the other components, such as the halo and the bulge, do not show 
any appreciable circular movement. It should be noted that the radial length extended as much as was necessary, 
to take into account the radial bins where there were still particles.

\subsubsection{Time evolution of the density peak and radial density profile for the galaxy model}
\label{subs:densitypeakgalaxy}

\begin{figure}
\begin{center}
\begin{tabular}{cc}
\includegraphics[width=3.0 in]{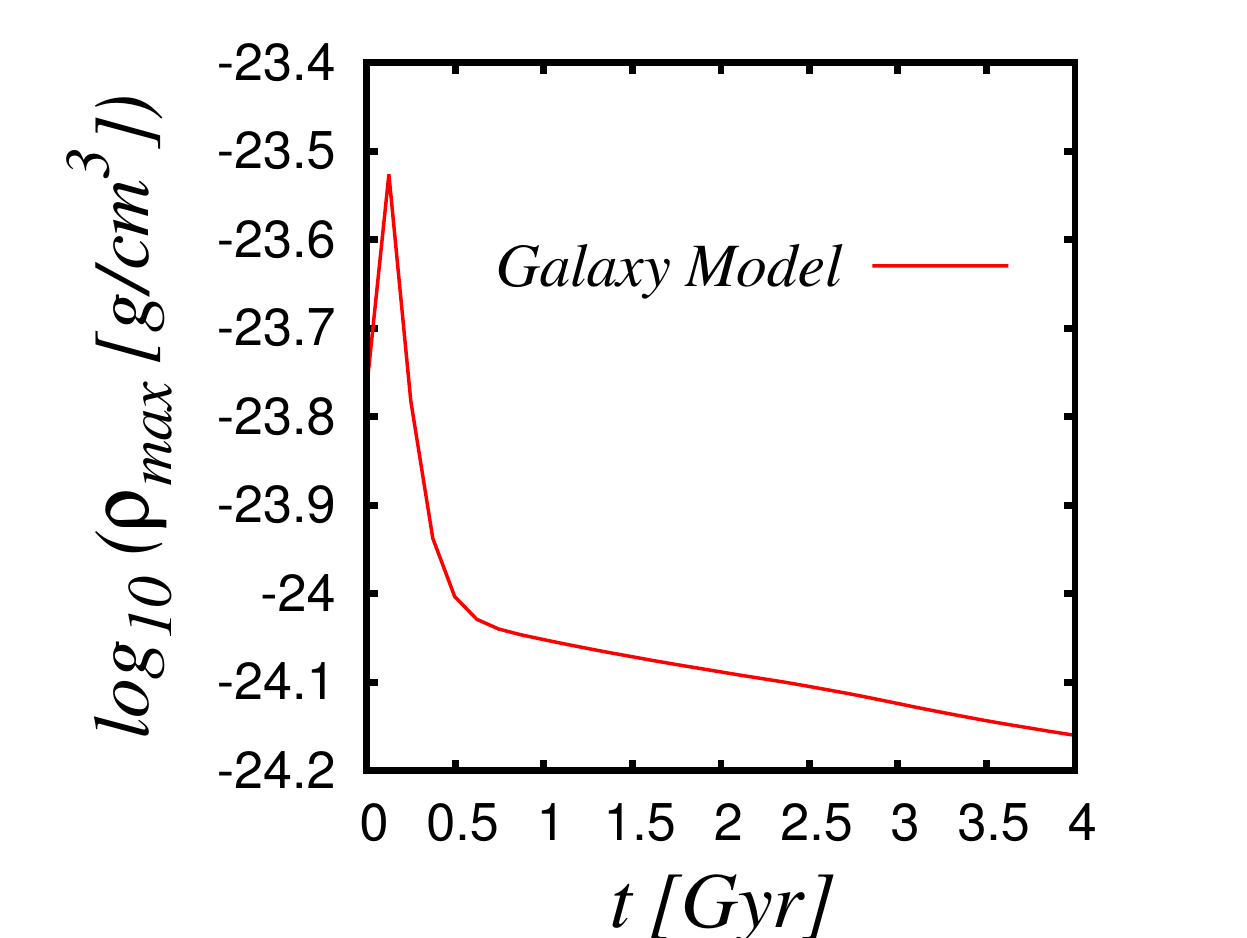} & \hspace{-2 cm} \vspace{-0.5 cm} \includegraphics[width=3.5 in]{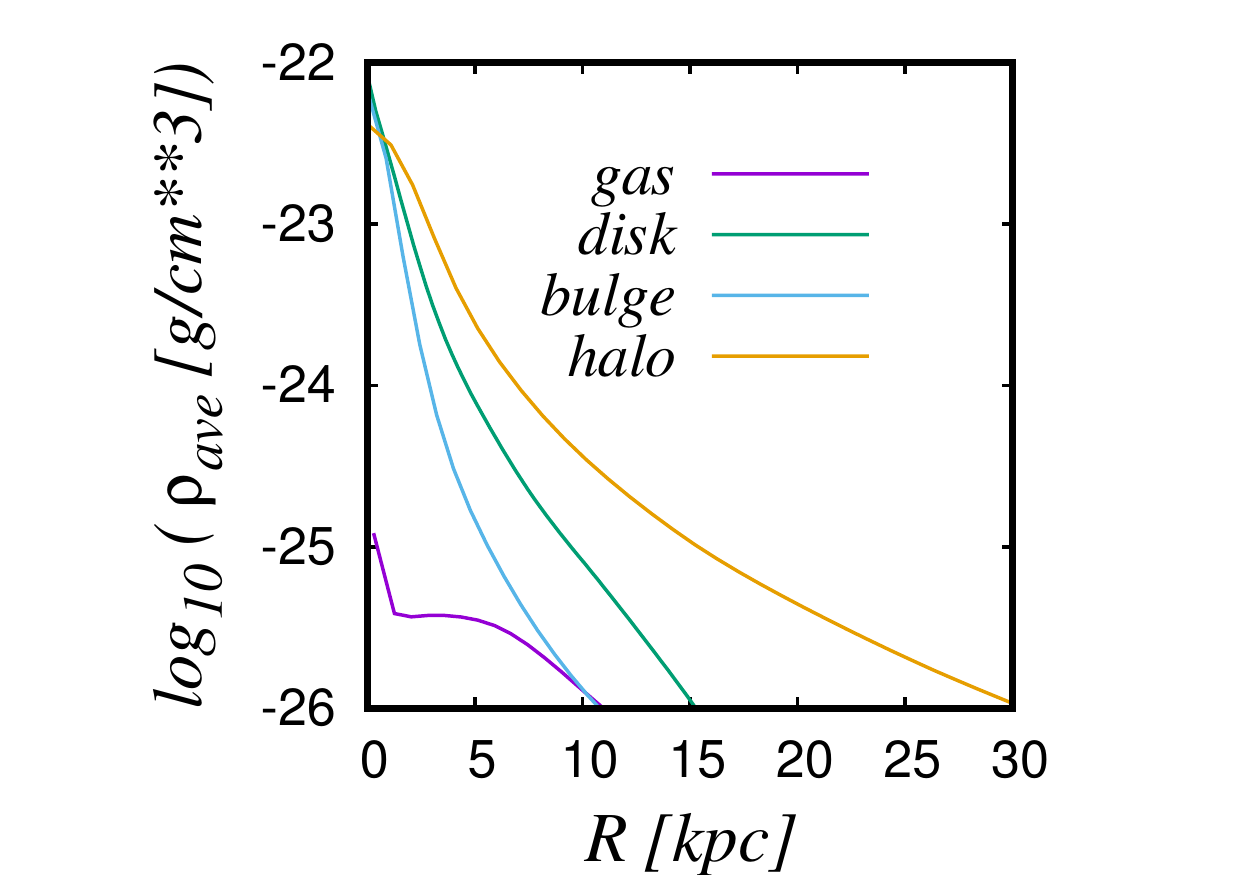}
\end{tabular}
\caption{\label{densityprofile} At time t = 13.7 Gyr, we show (left panel) the time evolution of the peak density 
for the gas component and (right panel) the radial density profile for each matter component 
of the galaxy model.}
\end{center}
\end{figure}

In the left-hand panel of Fig.~\ref{densityprofile}, we show the time evolution of the peak density for the gas 
component up to 4 Gyr of evolution, despite the fact that the final snapshot was taken at a time of 
13.7 Gyr. Therefore, this panel will be useful for comparison with the characterization of the 
collision models to be presented in Section~\ref{subs:galcol} and whose results will be discussed in 
Section~\ref{subs:gasdyn}. 

Recall that in the galaxy model considered in this paper, the 
gas component is initially located in a ring with radii in the range within 16 to 20 kpc. The left-hand panel of 
Fig.~\ref{densityprofile} indicates that the gas get moved rapidly towards the central 
region of the galaxy model and then get expanded radially, so that the final radial extension of the model is 
in the range of 0--180 kpc, see column 6 of Table~\ref{tab:param}

In the right-hand panel of Fig.~\ref{densityprofile}, we show the radial density profile for each matter component of the 
galaxy model up to a radius of 100 kpc, despite the fact that in the final snapshot most of the 
particles are concentrated within a radius of 80 kpc. This is done in this way to allow comparison with 
the characterization of the collision models to be discussed in Section~\ref{subs:gasdyn}. 

As indicated in the right-hand panel of Fig.~\ref{densityprofile}, there is a strong concentration of all types of 
matter in the center of the galaxy and to the extent that we move away from the center-- say in radii 
greater than 40 kpc - the density drops up to 7 orders of magnitude. These curves can be compared 
with those calculated by~\citet{Kuijken}. In addition, it should be remembered that the average density of the system 
is 3.02 $\times 10^{-28}$ g/cm$^3$, so it should be noted that the increase in density in the central region of the 
galaxy model is of 5 orders of magnitude; that is, it reaches up to 3.0 $\times 10^{-23}$ g/cm$^3$.

It should be noted that the curves for the gas and bulge are very similar for large radii, except in the central 
region, so that for a radius smaller than 10 kpc, the curve of the bulge is steeper than the curve of the gas, as 
can be seen in the right-hand panel of Fig.~\ref{densityprofile}. In addition, the curve for the disk falls very 
quickly with the radius while the curve for the halo falls softly.

We have determined the extreme spatial extension of each component by the end of the simulation. We found that the 
gas has reached a huge spatial expansion; on the contrary, the disk component remains more or less bounded to 
the center. It should be noticed that the bulge component has expanded more than the disk component. Consequently, the 
gas density is obviously lower than the density of the other mass components, as the gas spreads at large radii 
from the galaxy center.    

Then, based on the results shown both in the left-hand panel of Fig.~\ref{circularvel} and in the right-hand panel 
of Fig.~\ref{densityprofile}, it can be concluded 
that all of the matter components of the galaxy model are strongly concentrated in the central 
region, so that the mass contained up to the radius $R$ grows very quickly with the radius. 
In fact, for radii a little smaller than 10 kpc from the galaxy center, the total mass contained has 
already reached their asymptotic value, and for this reason, all the curves shown in the left-hand panel 
of Fig.~\ref{circularvel} decrease as $1/\sqrt{R}$ for large $R$.  

It has been observed that the galaxy model is stable after almost 10 rotation periods of evolution (equivalent to 14 Gyr 
of evolution) because it has reached a state of dynamic equilibrium.

\subsection{Models of galaxy collision.}
\label{subs:galcol}

The most important application of the basic galaxy model characterized 
in Section \ref{subs:evol} is the study of collision models between equal-mass galaxies. In this paper we 
only consider a few collision models, which are described below and summarized 
in Table~\ref{tab:collision}. The evolution of each collision model was carried out with the 
public code Gadget2 described in Section \ref{subs:code}, during 100 hours of CPU, running in parallel 
with 20 processors of the Cuetlaxcoapan Supercomputer of the LNS-BUAP.  

\begin{table}[ph]
\caption{Models of galaxy collisions.}
{\begin{tabular}{|c|c|c|c|} \hline
model & impact parameter  & initial positions                   &                      initial velocities \\
      &     [kpc]         & $(x,y,z)_1$ and $(x,y,z)_2$  [kpc]  & $(vx,vy,vz)_1$ and $(vx,vy,vz)_2$  [km/s]\\
\hline
\hline
S02   &  0   &  (-200,0,0);(200,0,0)    & (75,0,0);(-75,0,0) \\
\hline
S02b  & 100  &  (-200,0,-50);(200,0,50) & (75,0,0);(-75,0,0) \\
\hline
Orb   &  0   &  (-197,0,0);(197,0,0)    & (0,-6.19,0);(0,6.19,0)\\
\hline
Tom   &  20  &  (20,-20,0);(0,0,0)      & (0,136,136);(0,0,0)\\
\hline
Rot   &  0  &  (-90,49,0);(90,-49,0)    & (31.3,-6.39,0);(-31.3,6.39,0) \\
\hline
\hline
\end{tabular} }
\label{tab:collision}
\end{table}

In the first two models of Table~\ref{tab:collision}, two equal-mass galaxies 
are initially separated by almost 400 kpc along the X-axis, so that the galaxies move 
with respect to each other with an initial translation velocity of 75 km/s, such that the 
approach velocity before the collision is 150 km/s. 

\begin{figure}
\begin{center}
\begin{tabular}{cc}
\includegraphics[width=2.5 in]{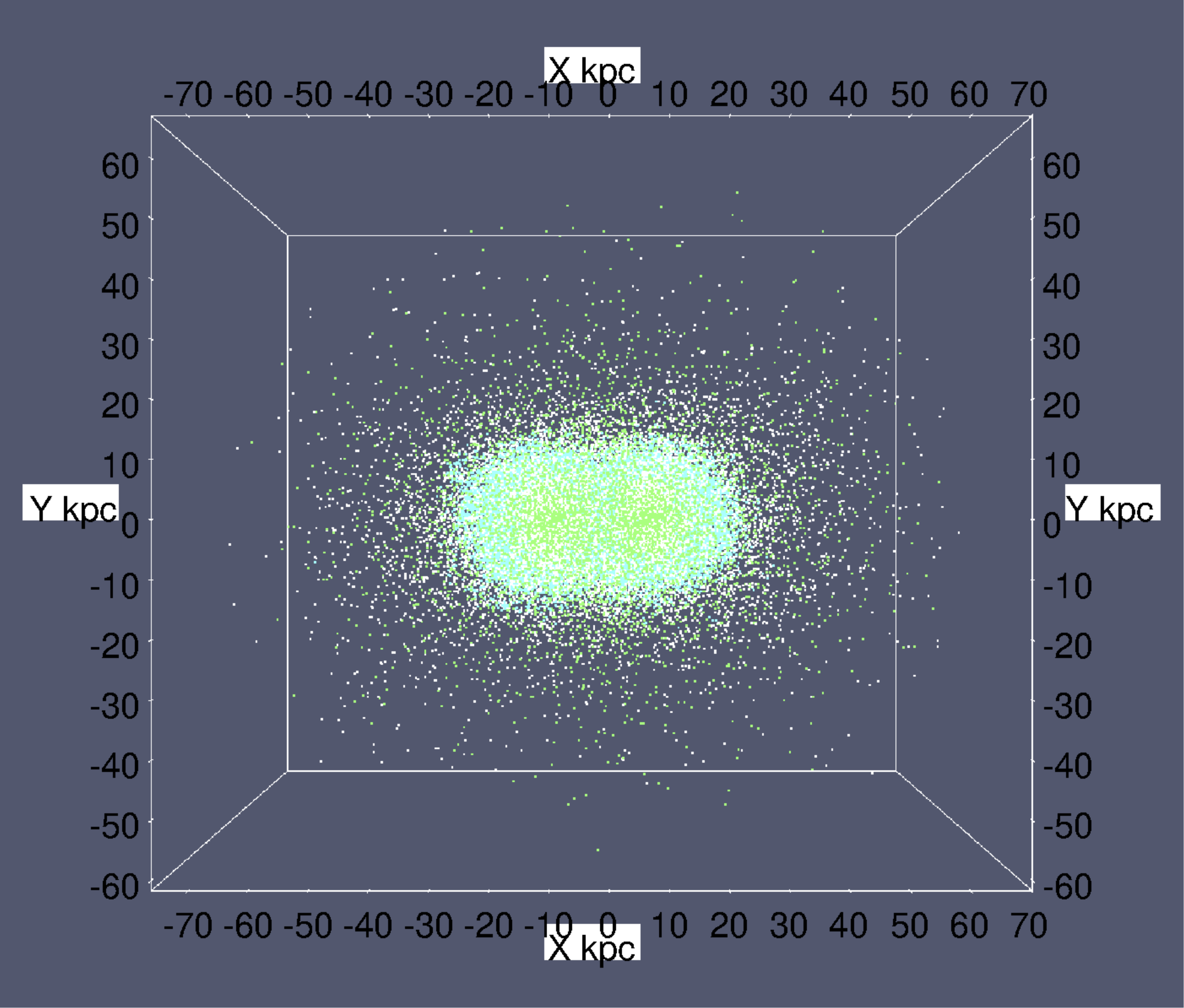}&\includegraphics[width=2.5 in]{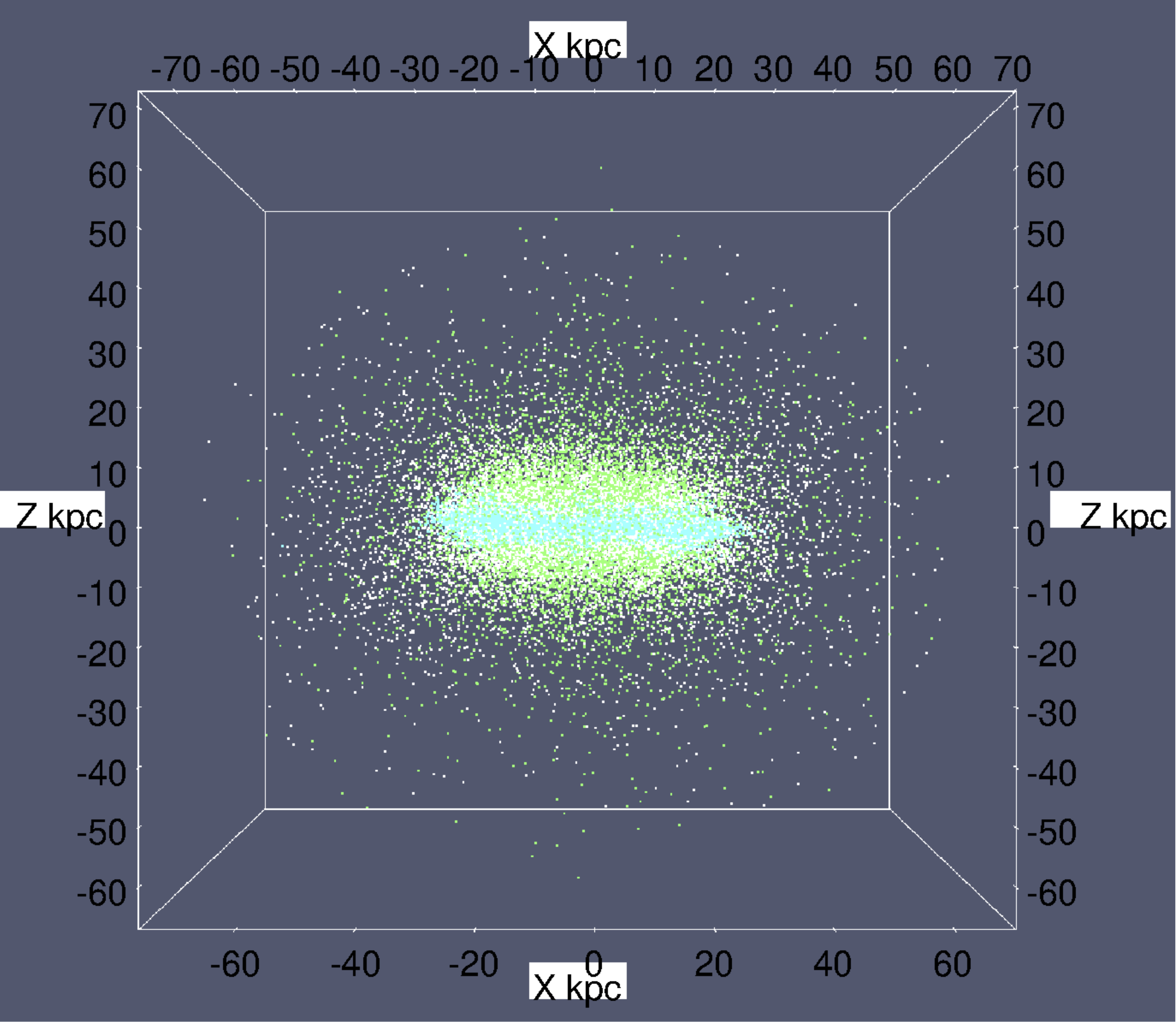}\\
\end{tabular}
\caption{\label{coalescenciaS02} Coalescence of the collision model S02 at the time 1.9 Gyr 
equivalent to 1.26 times the rotation periods of the galaxy model; the XY view is displayed on 
the left panel while the ZX view is shown on the right panel. The region shown is within the 
interval (-70,70) kpc in the X-axis, (-60,60) in the Y-axis and (-70,70) in the Z-axis. The 
colors indicate the matter components according to bulge-yellow, gas-green and disk-blue.}
\end{center}
\end{figure}

In the case of the model S02, both galaxies collide 
directly, so that a merging process starts at a time of 1.76 Gyr of evolution, in which the center of mass 
of each galaxy are very close to each other. The merging process seems to finish at the time of 2.1 Gyr, in which one 
sees only one center of mass oscillating along the X-axis. Fig.~\ref{coalescenciaS02} shows a snapshot of 
the merging process of both galaxies at a time of 1.9 Gyr. It is interesting to note that the gas and 
bulge expand spatially during both the pre-collision period of translation and the merging process. It is also 
interesting to mention that the disk remains elongated in the new galactic 
structure formed after the collision.

\begin{figure}
\begin{center}
\begin{tabular}{c}
\includegraphics[width=2.5 in]{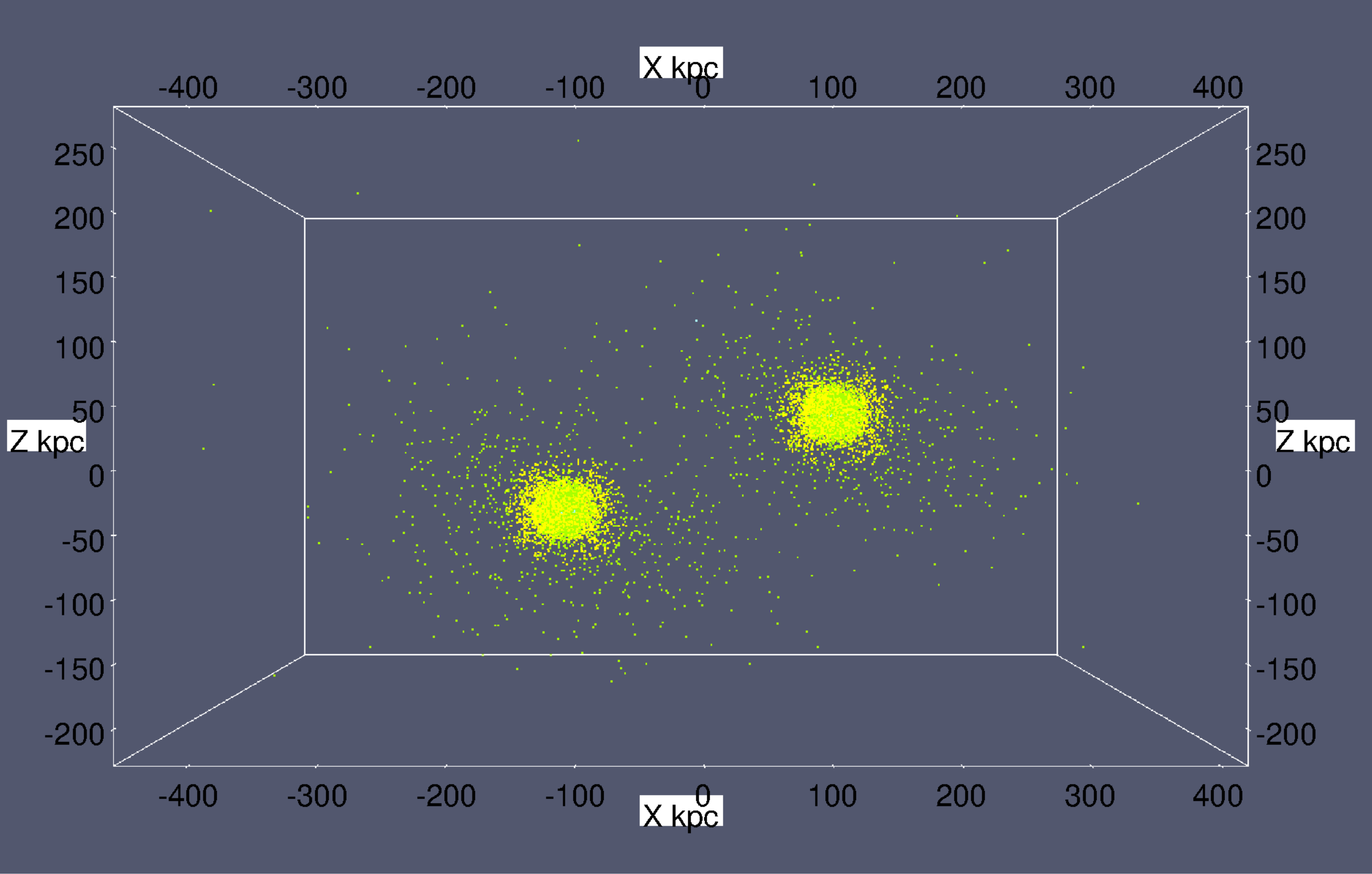}\\
\includegraphics[width=2.5 in]{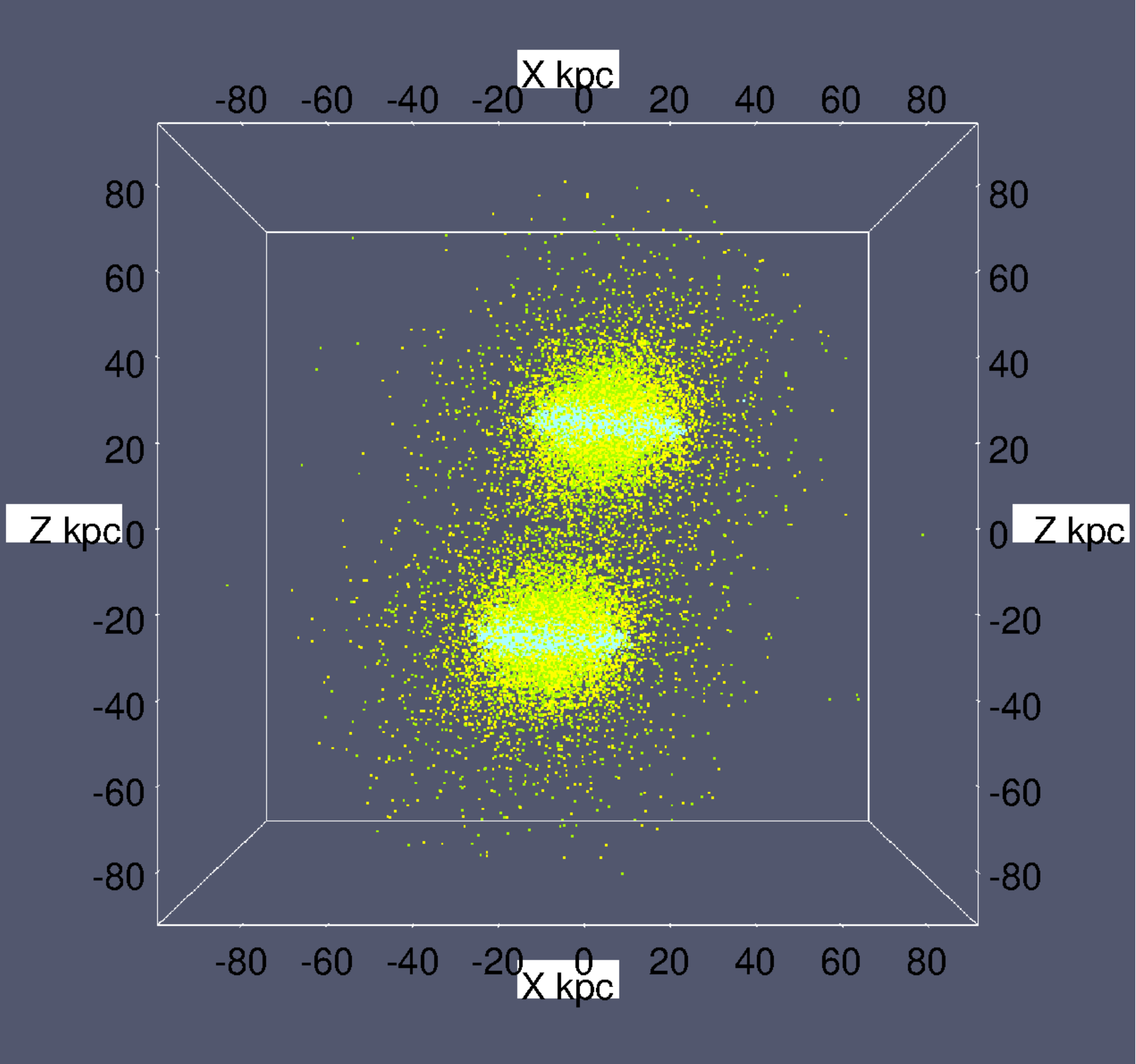}\\
\includegraphics[width=2.5 in]{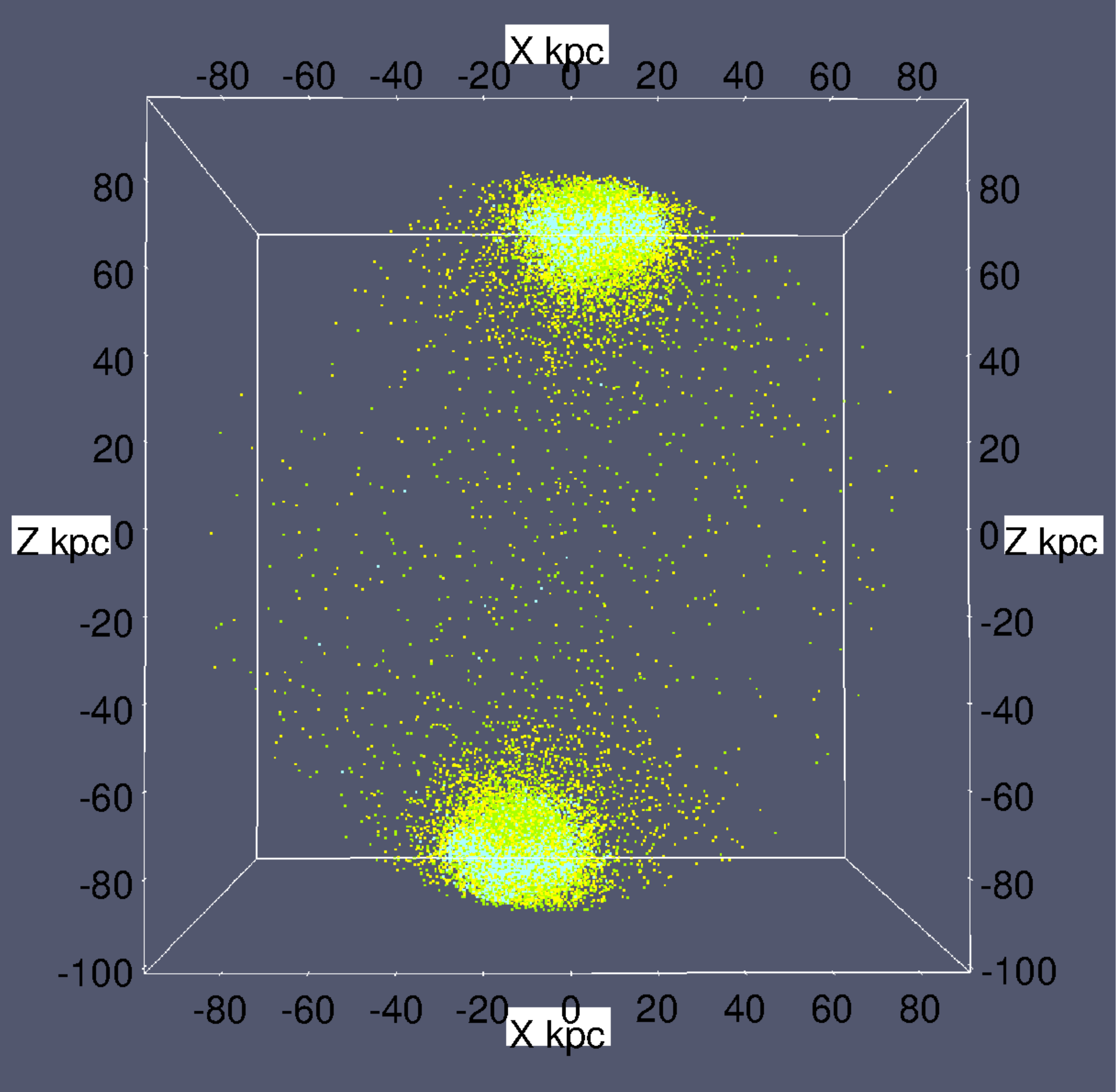} 
\end{tabular}
\caption{\label{colisionoblicuaS02b} The ZX view of the oblique collision model S02b. In the top 
panel the 100 kpc length lever arm is seen 
implemented along the Z-axis; the approach speed is 75 km/s on each side; it corresponds to 
an evolution time of 0.86 Gyr, which is equivalent to 0.57 times 
the rotation period of the galaxy model; the region shown is (-400,400) kpc in the X-axis 
and (-200,200) in the Z-axis. The middle panel corresponds to 2.0 Gyr or 
equivalent to 1.3 times the rotation period of the galaxy model. The bottom panel corresponds to 
3.1 Gyr or 2 times the rotation period of the galaxy model. The region shown in both 
the middle and bottom panels is (-80,80) kpc both in the X-axis and in the Z-axis. The colors 
indicate the matter components according to bulge-yellow, gas-green and disk-blue.}
\end{center}
\end{figure}

In the case of the oblique collision model S02b, both galaxies are 
additionally displaced a distance of 50 kpc along the Z axis. Consequently, this separation acts as an effective 
impact parameter of 100 kpc for the collision model (this is the only difference 
with respect to the frontal collision model S02). The galaxies move along the X-axis, so that the 
point of maximum pre-collision approach is reached at the time of 2 Gyr of evolution. The galaxies enter in 
orbit one with respect to the other, as shown in Fig.~\ref{colisionoblicuaS02b}, in which the evolution time 
increases from the top panel to the bottom panel.

Around an evolution time of 2.5 Gyr, the binary system has made a complete turn over its orbit. By the time of   
2.75 Gyr, the galaxies of the binary system start separating again. We follow the evolution of the model S02b 
up to a time of 6.54 Gyr, equivalent to 4.3 times the rotation period of the galaxy model. We do not observe a 
subsequent approach of the galaxies. Therefore, it is very likely that this system is not sufficiently bound 
to maintain its galaxies in orbit, so that a galaxy will eventually escape from the gravitation field of 
the other. It is again interesting to mention that the disk remains elongated while the bulge has 
spread to connect both disks during their orbital motion.

The initial conditions of the collision model labeled Orb in Table~\ref{tab:collision}, are calculated 
according to the exact solution of the gravitational 2-body problem, so that both galaxies are modeled 
in the exact solution as particles of a total mass equal to the sum of the all the masses 
reported in Table \ref{tab:param}. In this case, the free parameters of the exact solution 
are the total energy $E_{\rm cm}$ and angular momentum $L_{\rm cm}$ of the system 
with respect to the center of mass. The values given to these parameters in this 
collision model are as follows:  $E_{\rm cm}=-1.66 \, \times 10^{58}\,$ erg and 
$L_{\rm cm}=4.21 \, \times 10^{74}\,$ g cm$^2$/s, respectively.

\begin{figure}
\begin{center}
\begin{tabular}{cc}
\includegraphics[width=2.5 in]{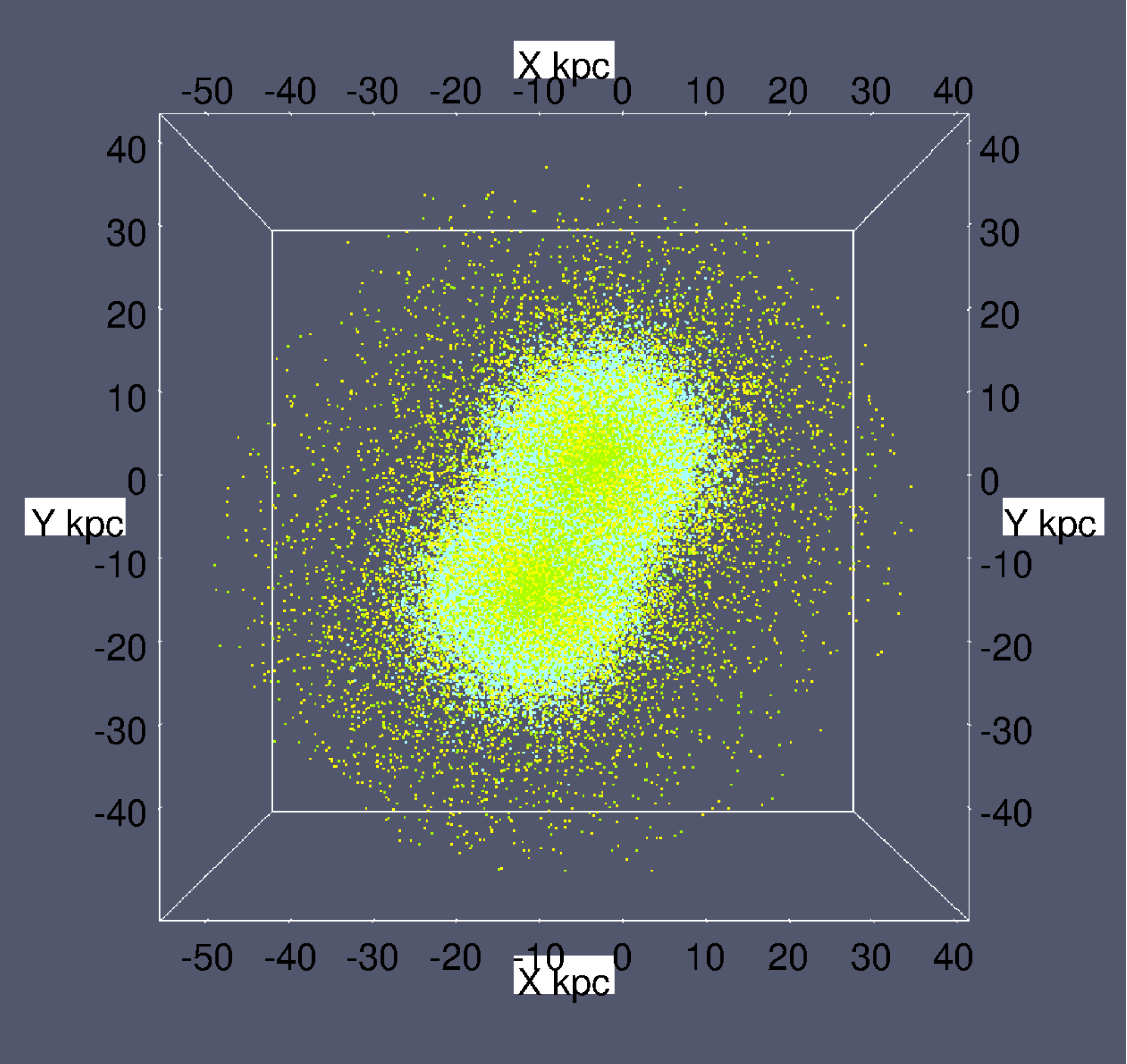}&\includegraphics[width=2.5 in]{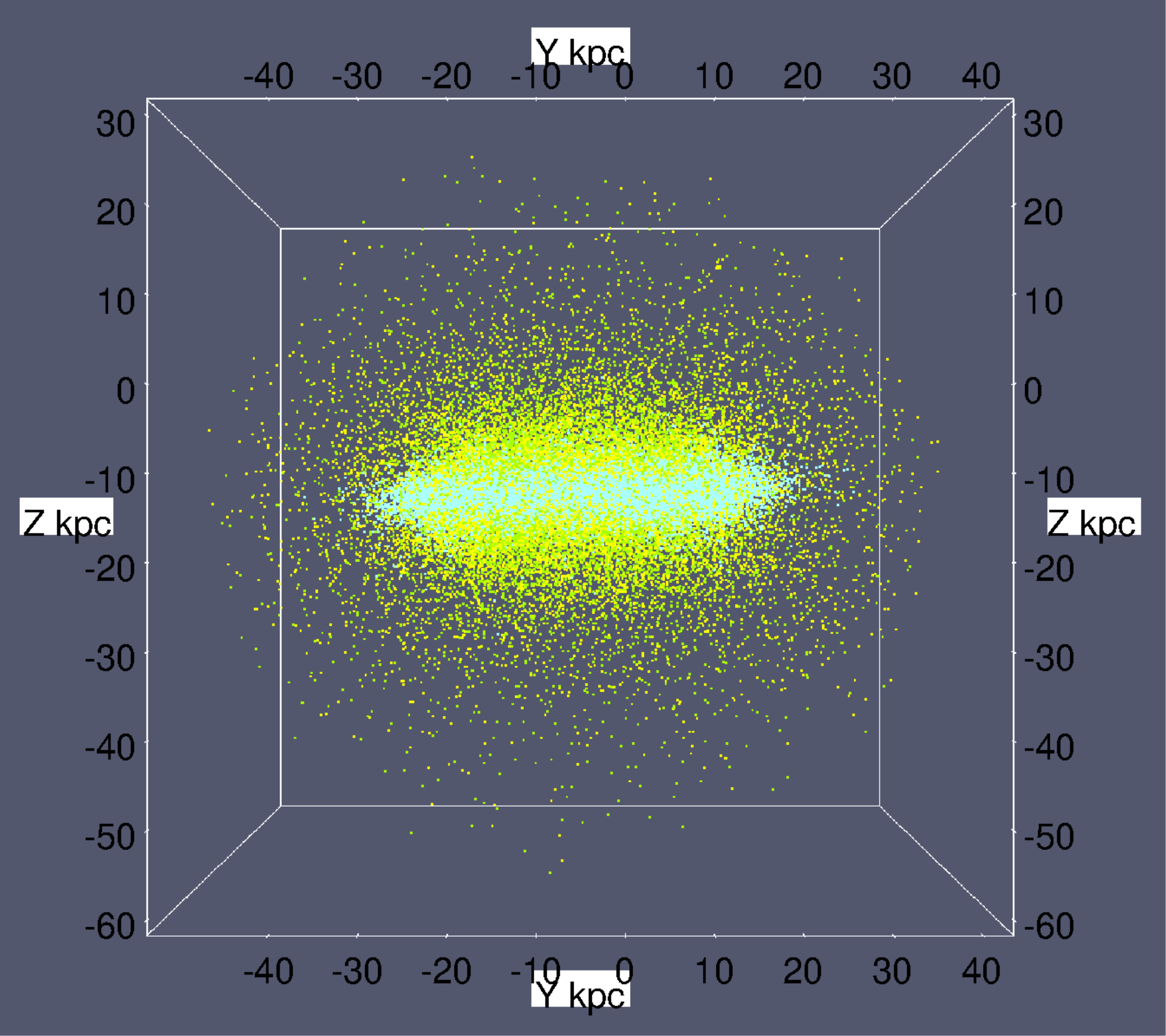}
\end{tabular}
\caption{\label{colisionorb} The collision model Orb, at the time of 5.4 Gyr, equivalent to 3.69 times 
the rotation period of the galaxy model. The XY view is shown on the left panel and the ZY view on 
the right panel. The region shown is (-50,50) kpc in the X-axis, (-50,40) in the Y-axis and (-60,30) in 
the Z-axis. The colors indicate the matter components according to bulge-yellow, gas-green and disk-blue.}
\end{center}
\end{figure}

In this model we observe a soft approach of the galaxies. Therefore, the center of mass of each 
galaxy enter in orbit one with respect to the other, in such a way that many complete turns of 
the orbital motion are observed in the central region during the evolution time within the 
short interval of 5.4-5.6 Gyr. At the evolution time of 6.14 Gyr, the merging process is 
completed. Therefore, one only sees a single rotating dense core, see Fig.~\ref{colisionorb}. This 
dynamic can be captured by following the gas component, as is explained in the last paragraph of 
Section~\ref{subs:gasdyn}.

The initial conditions of the collision model labeled Tom in Table~\ref{tab:collision} are taken 
from one of the collision models described by ~\citet{Toomre1972}. The two galaxies are placed in the 
XY plane very close to each other. The center of mass of one galaxy 
has initial velocity directed outward of the XY plane in the positive direction 
of the Z-axis, while the second galaxy is at rest. Consequently, we say that there 
is an effective impact parameter on the Y-axis, as indicated in 
Table~\ref{tab:collision}.

\begin{figure}
\begin{center}
\begin{tabular}{cc}
\includegraphics[width=2.5 in]{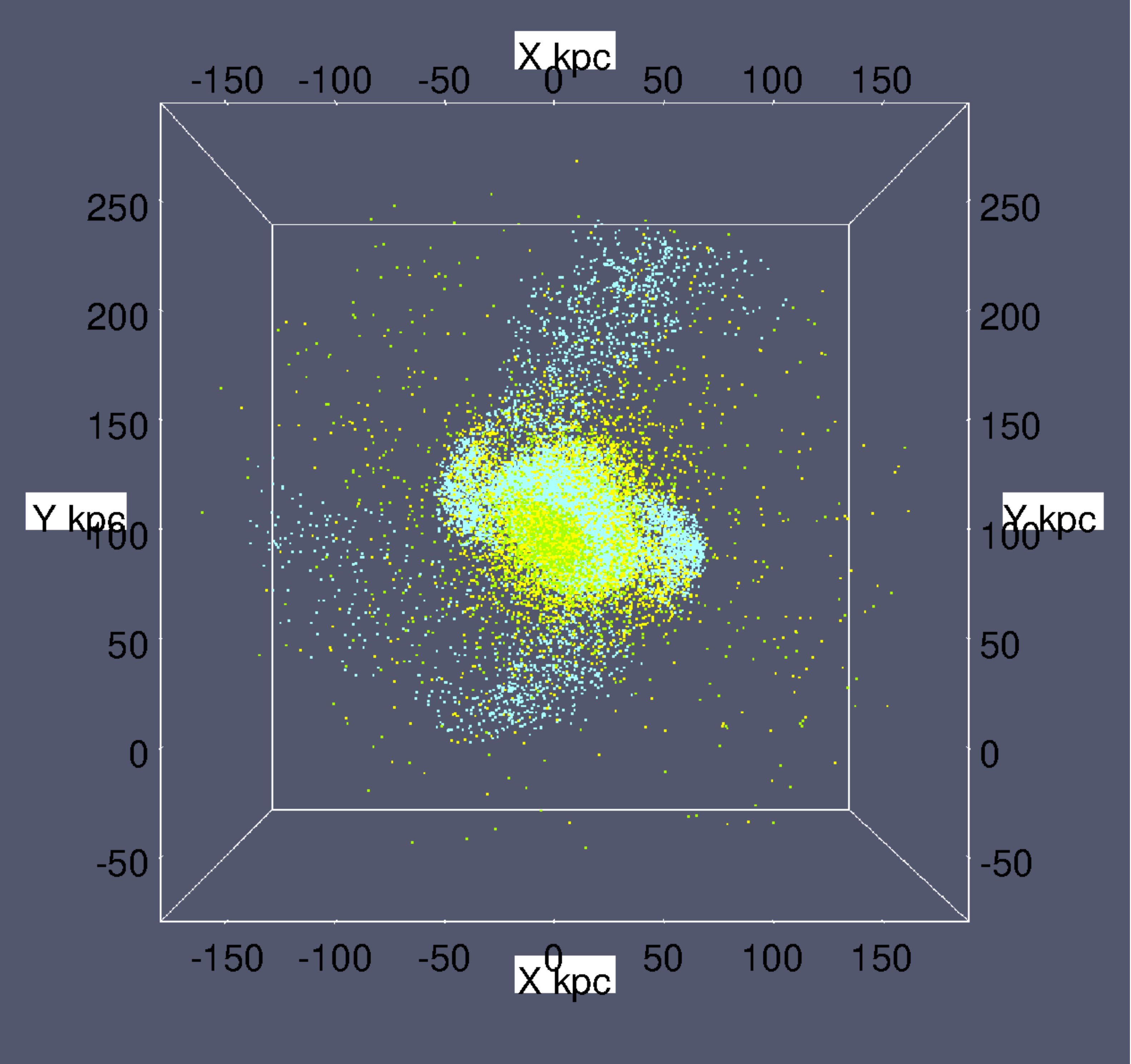}&\includegraphics[width=2.5 in]{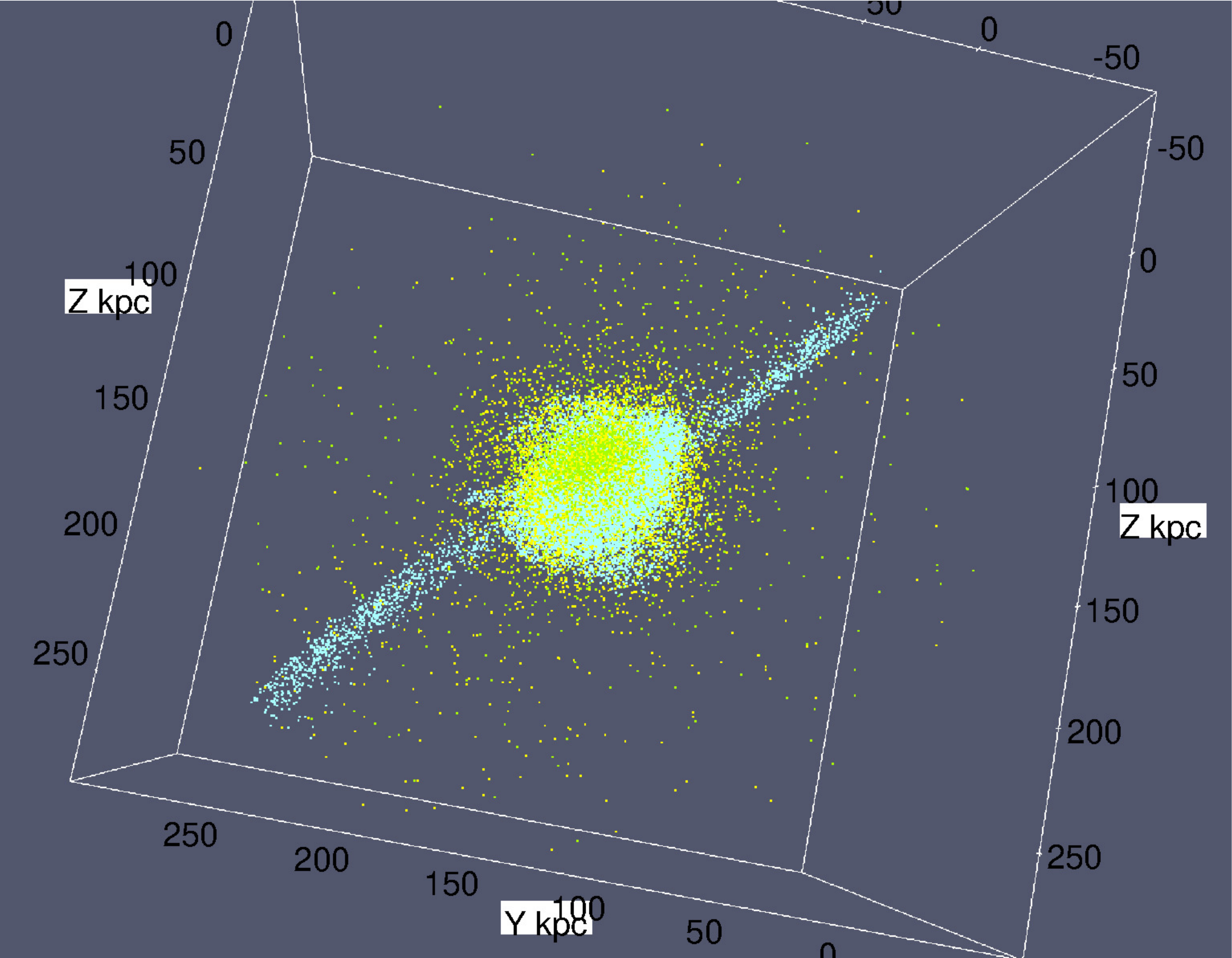}\\ 
\end{tabular}
\caption{\label{colisiontom} The collision model Tom, at the time 1.68 Gyr, equivalent to 
1.12 times the rotation period of the galaxy model. In the left panel we show the XY view and in the 
right panel the ZY view rotated arbitrarily. The region shown is (-150,150) kpc in the 
X-axis, (-50,250) in the Y-axis and (-50,250) in the Z-axis. The colors indicate the matter 
components according to bulge-yellow, gas-green and disk-blue.}
\end{center}
\end{figure}

It is observed that the moving galaxy describes an arc and finally falls onto the motionless 
galaxy, so that the system develops an appreciable orbital movement, which causes 
spiral arms to develop, as can be seen in Fig.~\ref{colisiontom}. It must be emphasized that these 
spiral arms are mainly composed of disk particles. In this model, both disks have lost their initial 
elongation. Most of the bulge surrounds the central part of the disk while an small fraction of the bulge particles 
also follows slightly the spiral arms.      

\begin{figure}
\begin{center}
\begin{tabular}{c}
\includegraphics[width=2.5 in]{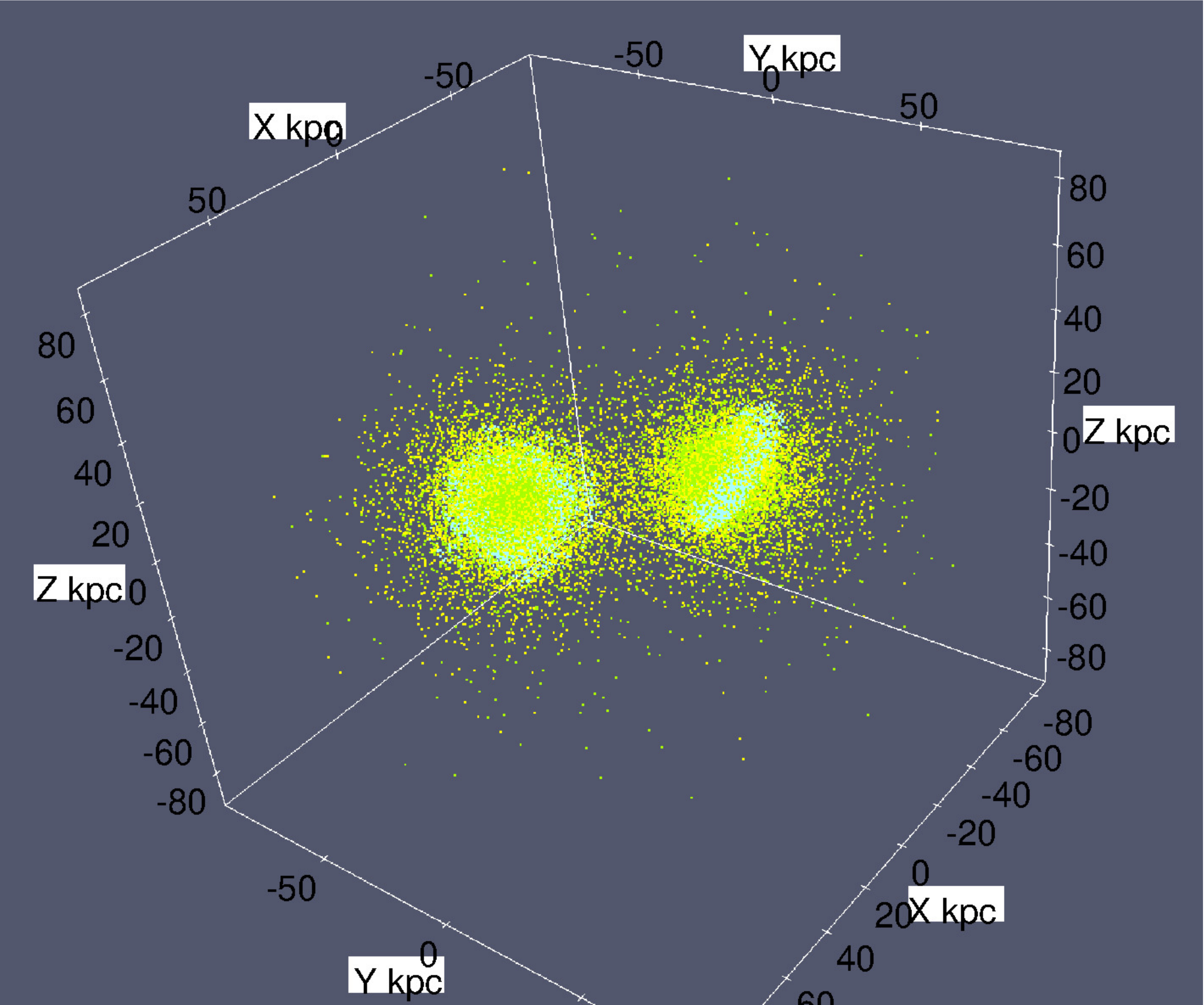}\\
\includegraphics[width=2.5 in]{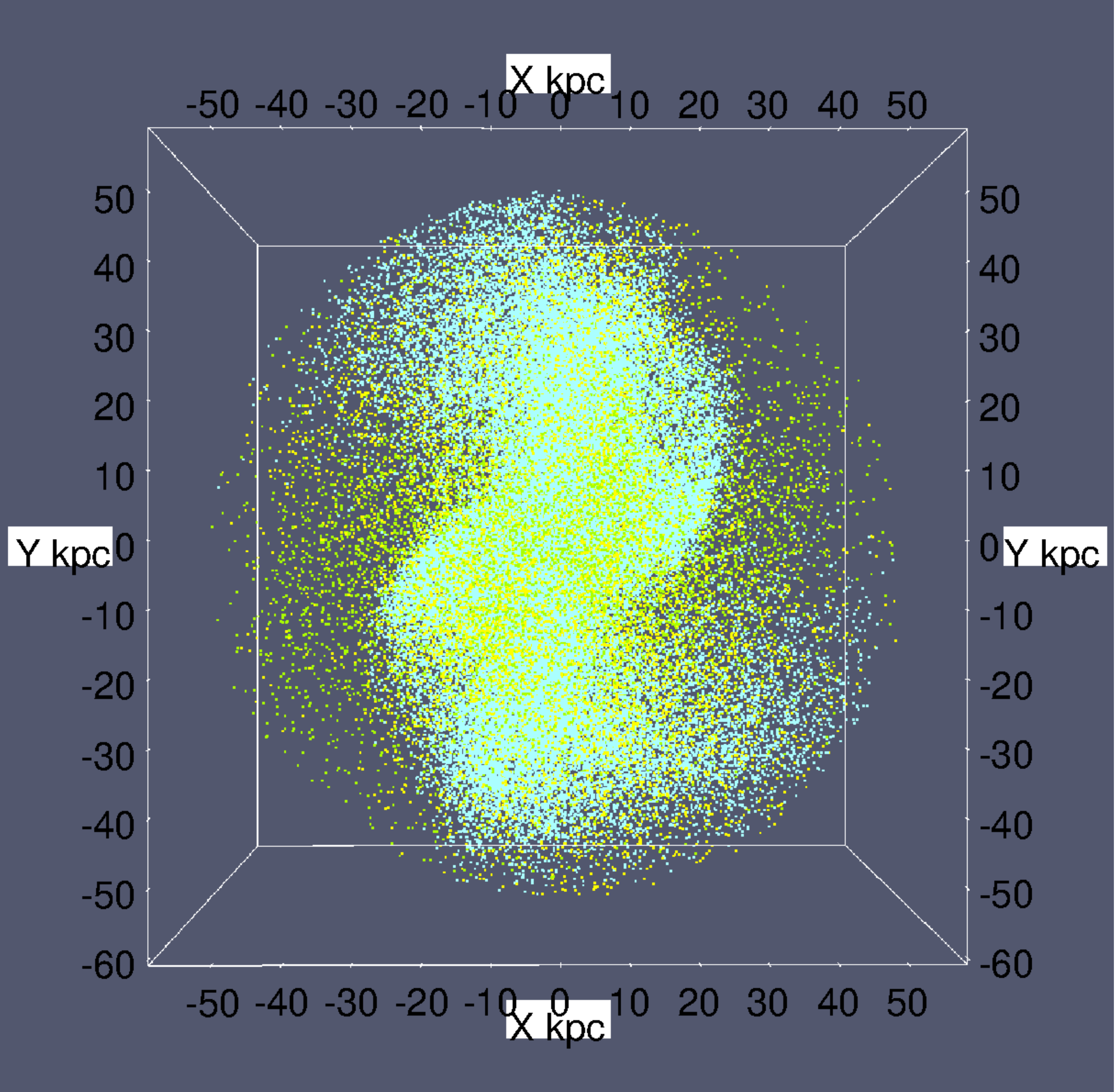}\\
\includegraphics[width=2.5 in]{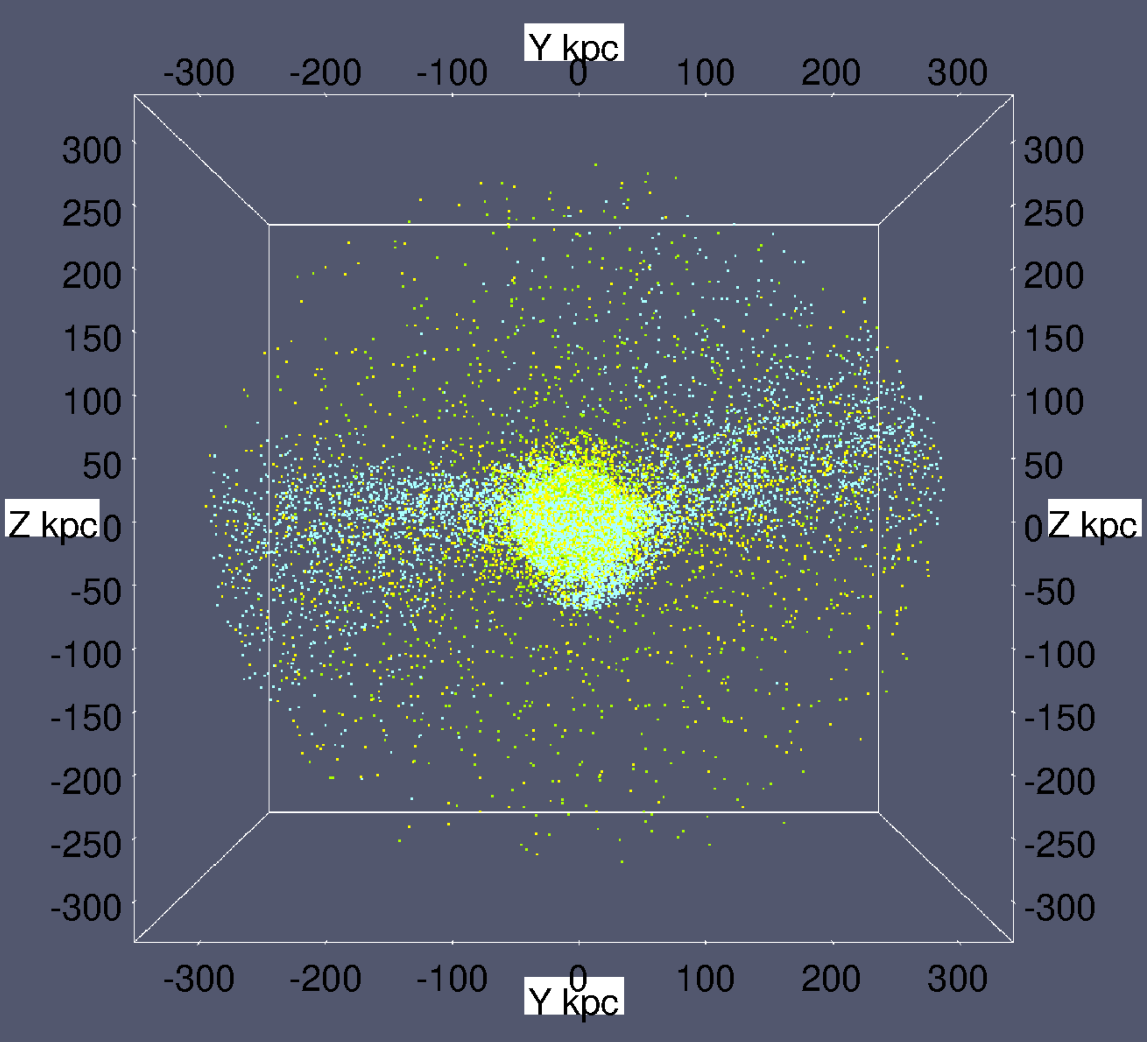}
\end{tabular}
\caption{\label{colisionOrbRot} The collision model Rot, at the top panel we show the 
snapshot taken at the time 1.23 Gyr, equivalent to 0.82 times the rotation period of the galaxy 
model, which is seen from a view rotated arbitrarily; the region shown is (-80,80) kpc 
in all the axes. In the middle panel we show the XY view corresponding to 
1.56 Gyr or 1.04 times the rotation period of the galaxy model. The region shown 
is (-50,50) kpc in the X-axis and (-50,50) in the Y-axis. It should be noticed the elongated 
configuration formed, in which small spiral arms are seen.  Finally, in the bottom 
panel we show the a view rotated arbitrarily at the time 3.33 Gyr or 2.22 times the 
rotation period of the galaxy model. The region shown is (-300,300) 
in the Y-axis and (0,300) in the Z-axis. It must be noted the long spiral arms formed at the 
merged system. The colors indicate the matter components according to bulge-yellow, 
gas-green and disk-blue.}
\end{center}
\end{figure}

The last collision model considered in this paper was labeled Rot in Table~\ref{tab:collision}. This model 
is very similar to model Orb. In fact, in model Rot the disk planes defined in model Orb 
are rotated, as can be seen in the top panel of Fig.~\ref{colisionOrbRot}. The dynamic evolution 
shows the formation of an elongated bar, in which both the disk and the bulge take part in this 
rotating structure, as can be seen the middle panel of Fig.~\ref{colisionOrbRot}. 
Finally, a new structure is formed at the end of the evolution time as a merger remnant, such that 
a mass concentrated in the center is surrounded by a couple of spiral arms, which are mainly composed 
of both disk and the bulge particles, as can be seen in the bottom panel of 
Fig.~\ref{colisionOrbRot}.

It must be emphasized that most of the collision models considered so far (with exception of the model S02b) 
led to the formation of a new galaxy structure, presumably with different physical properties to those observed for 
the original galaxy model, out of which the new structure is formed as a merger remnant. To get more 
details about the physical properties of these new structures, in Section~\ref{subs:dyncaracol} we try 
to characterize them by looking at (i) the dynamic behavior of the peak density of the gas component 
in Section~\ref{subs:gasdyn} and (ii) the evolution of the angular momentum of the collision 
models in Section~\ref{subs:angmomcolmod}.

\subsection{Dynamic characterization of the merger remnants of the collision models.}
\label{subs:dyncaracol}

Using the visualization software pv-wave version 8, we managed to track 
the evolution of the gas and 
show short movies at the web address: 
\url{https:\\drive.google.com/open?id=1VUhCAZhWWnOHsh_fWKkWrhYjh8WD08I5}.
It must be noted that 
the gas shows an interesting dynamic despite the fact that it is always bounded gravitationally during 
the evolution time either to the central region of the galaxy model or to the merged system.       
\subsubsection{Time evolution of the density peak and radial density profile for the collision models}
\label{subs:gasdyn}

As we mentioned in Section~\ref{subs:evol}, the 
gas of the galaxy model expanded very quickly to reach an equilibrium configuration, as characterized by 
an almost flat curve of the peak density. 

\begin{figure}
\begin{center}
\includegraphics[width=3.0 in]{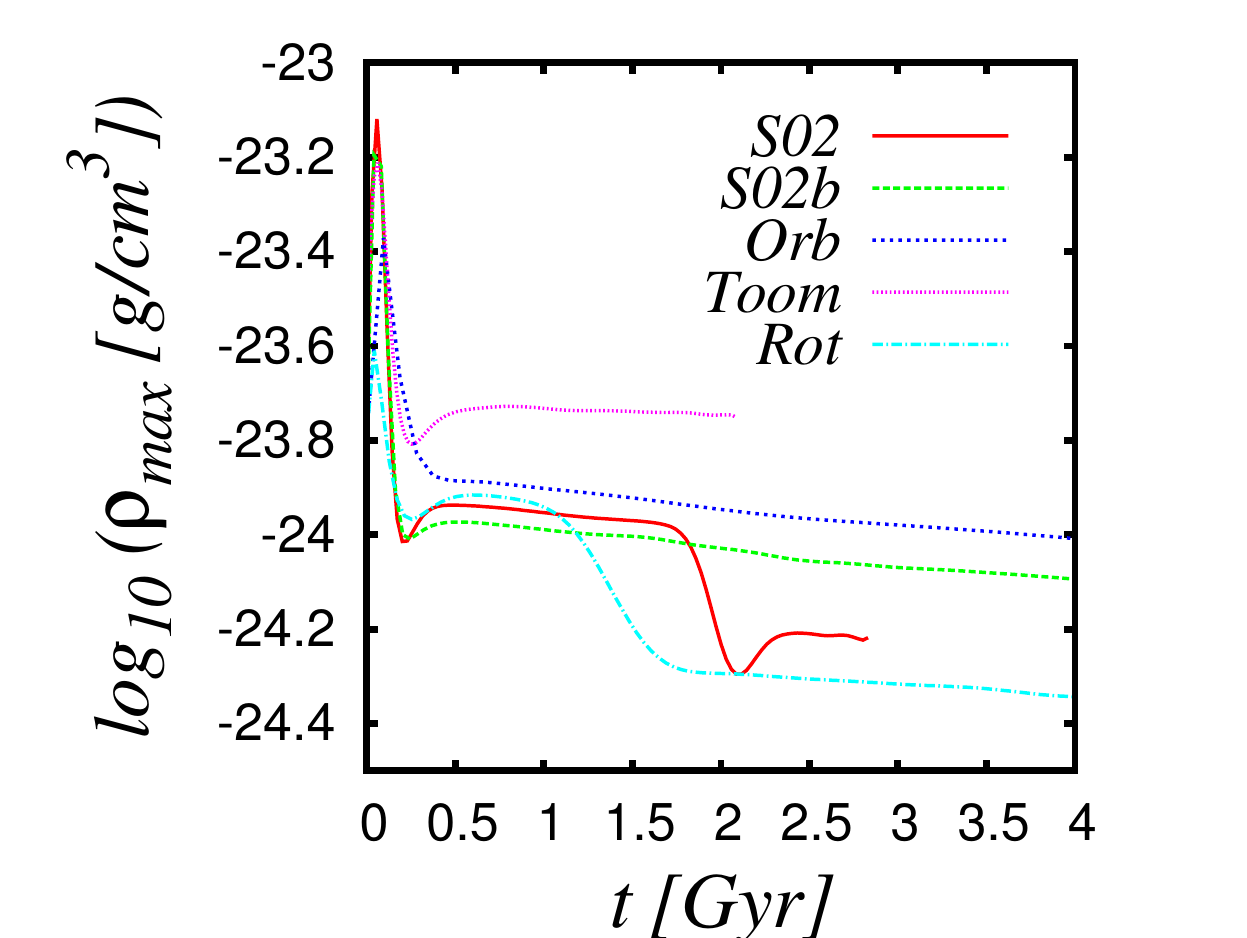}
\caption{\label{DenMaxColGalax2} Time evolution of the peak density for the gas in all the collision models.}
\end{center}
\end{figure}

One way to quantify the effects of this gas expansion of the galaxy model on the collision models is 
by calculating again the time evolution of the peak density of the gas for all the collision 
models, as has been done in Fig.~\ref{DenMaxColGalax2}. In this plot, one can see that all the peak density 
curves rise and fall very quickly at a very small radius and then a stabilization stage follows for large radius. 
This indicates that the most of the gas remains bounded gravitationally to the galaxy center, while a small 
fraction of gas manages to escape away.

\begin{figure}
\begin{center}
\begin{tabular}{cc}
\includegraphics[width=3.0 in]{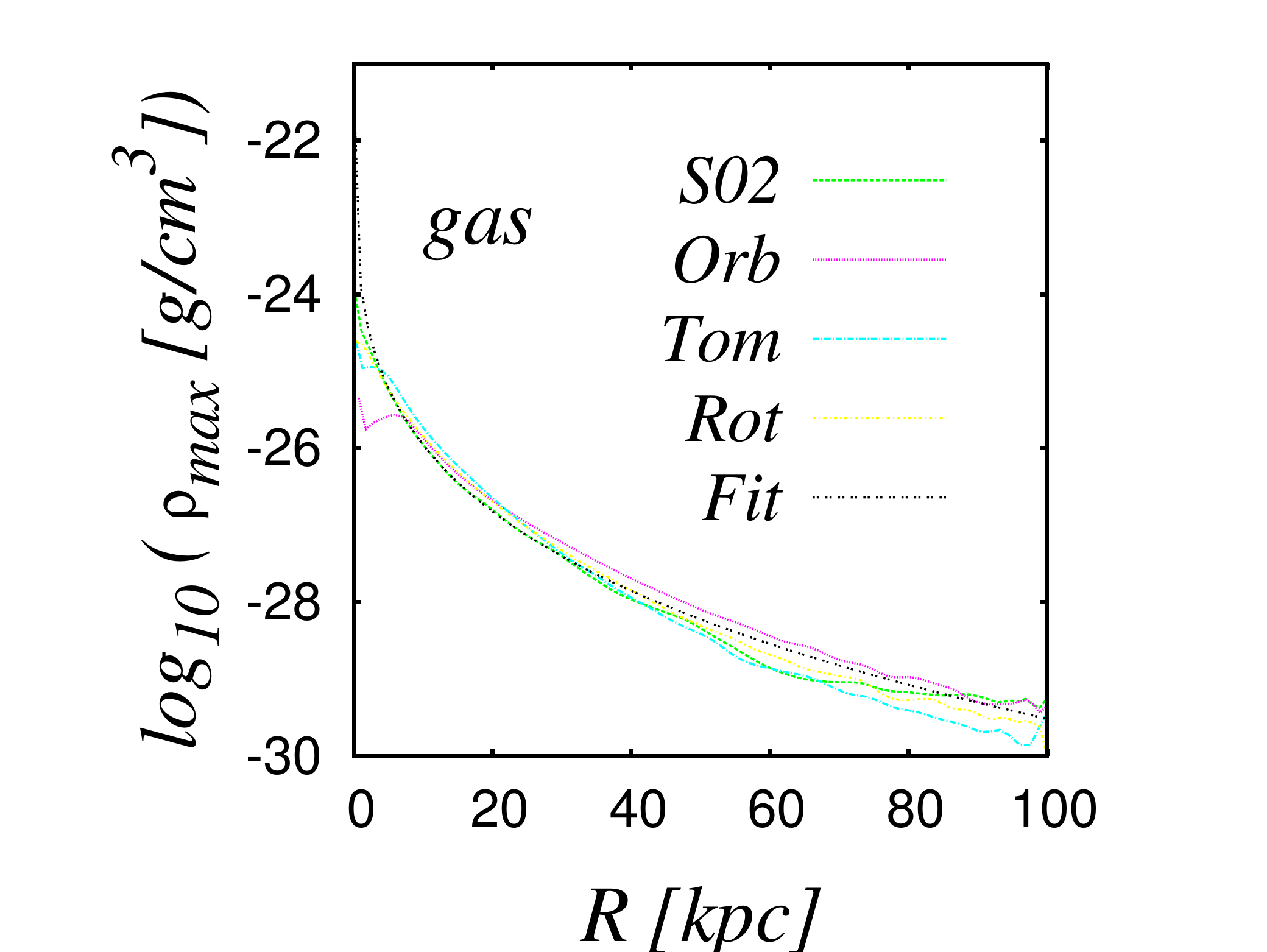} &\includegraphics[width=3.0 in]{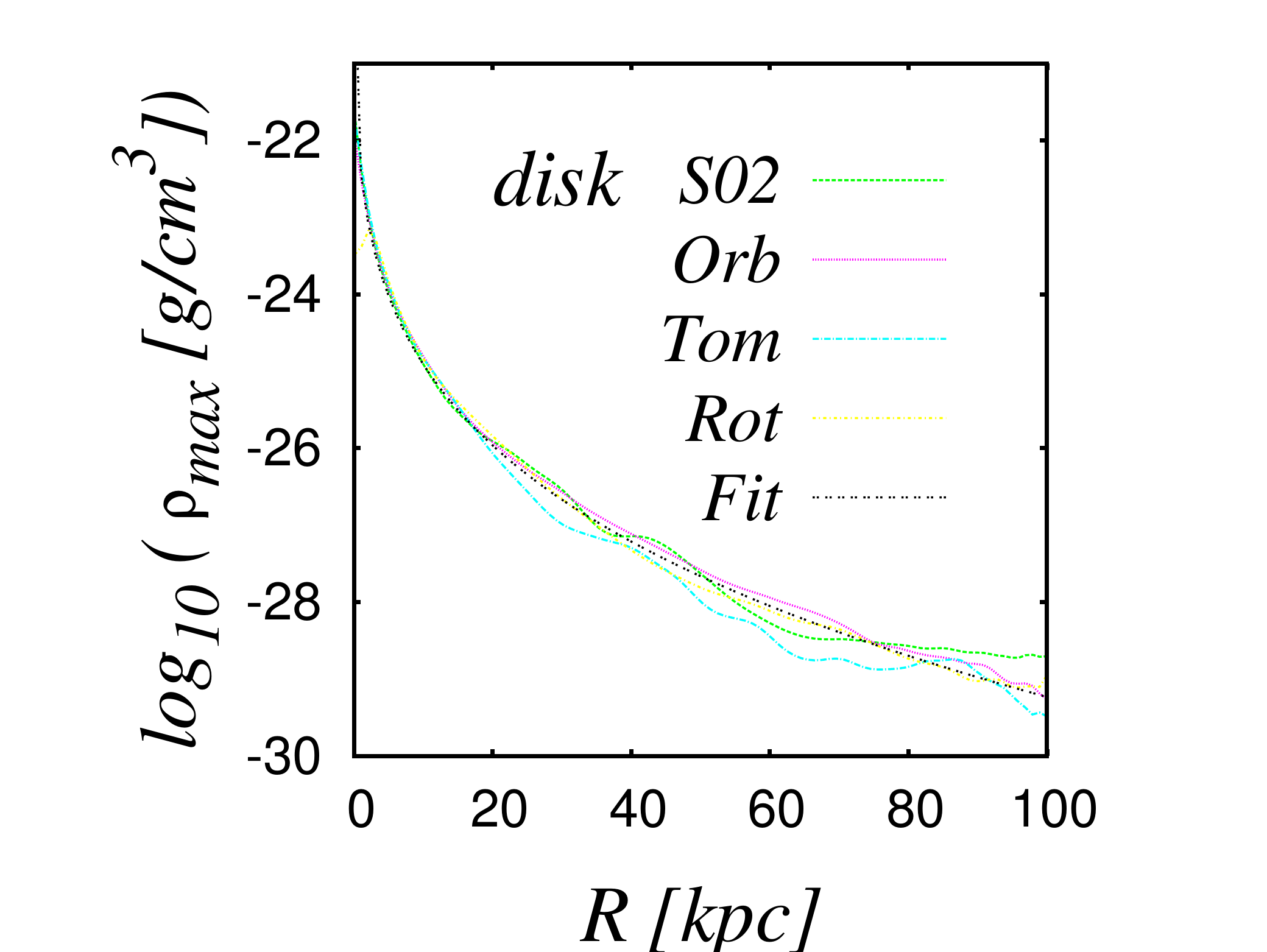}\\
\includegraphics[width=3.0 in]{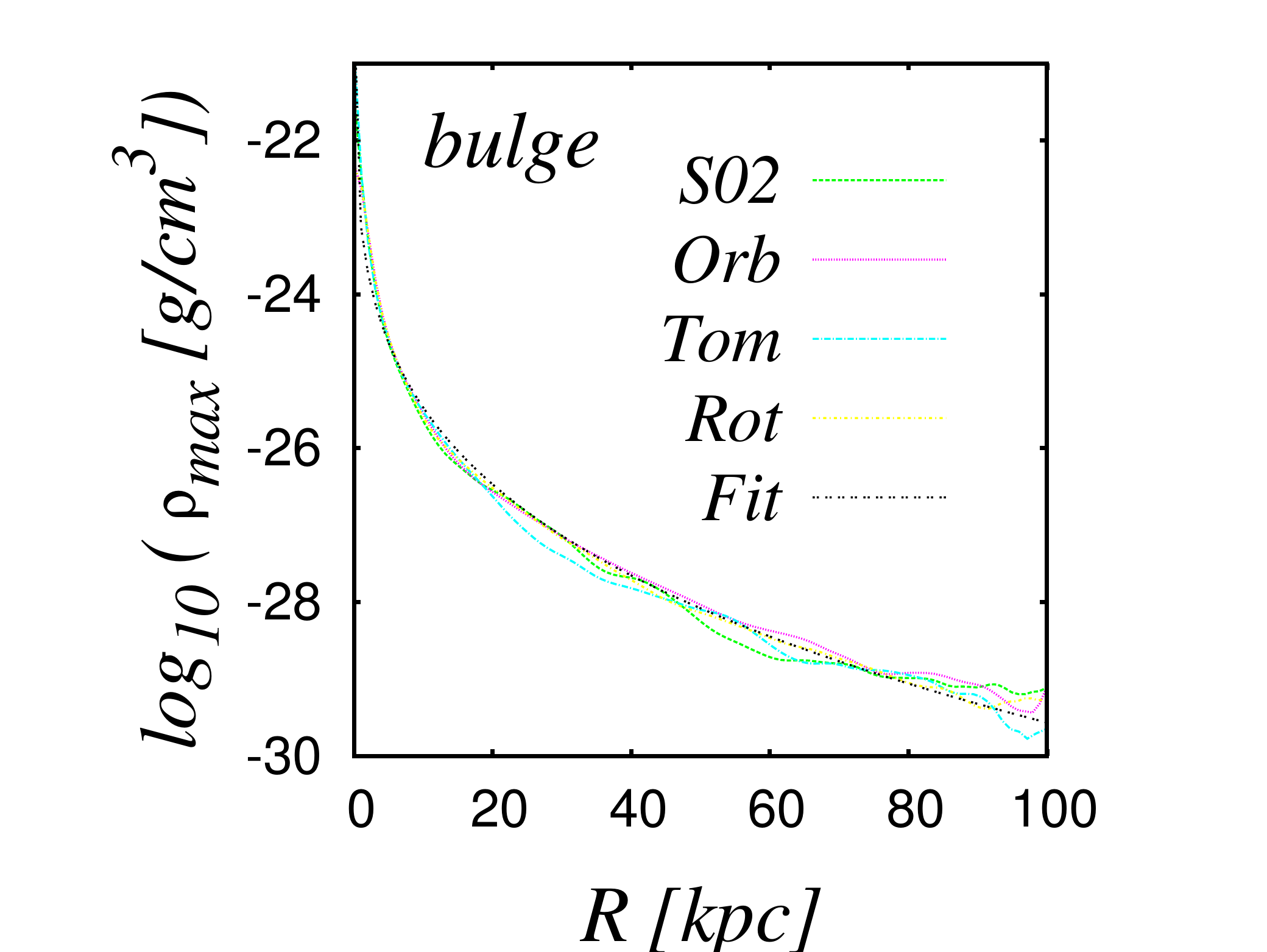} &\includegraphics[width=3.0 in]{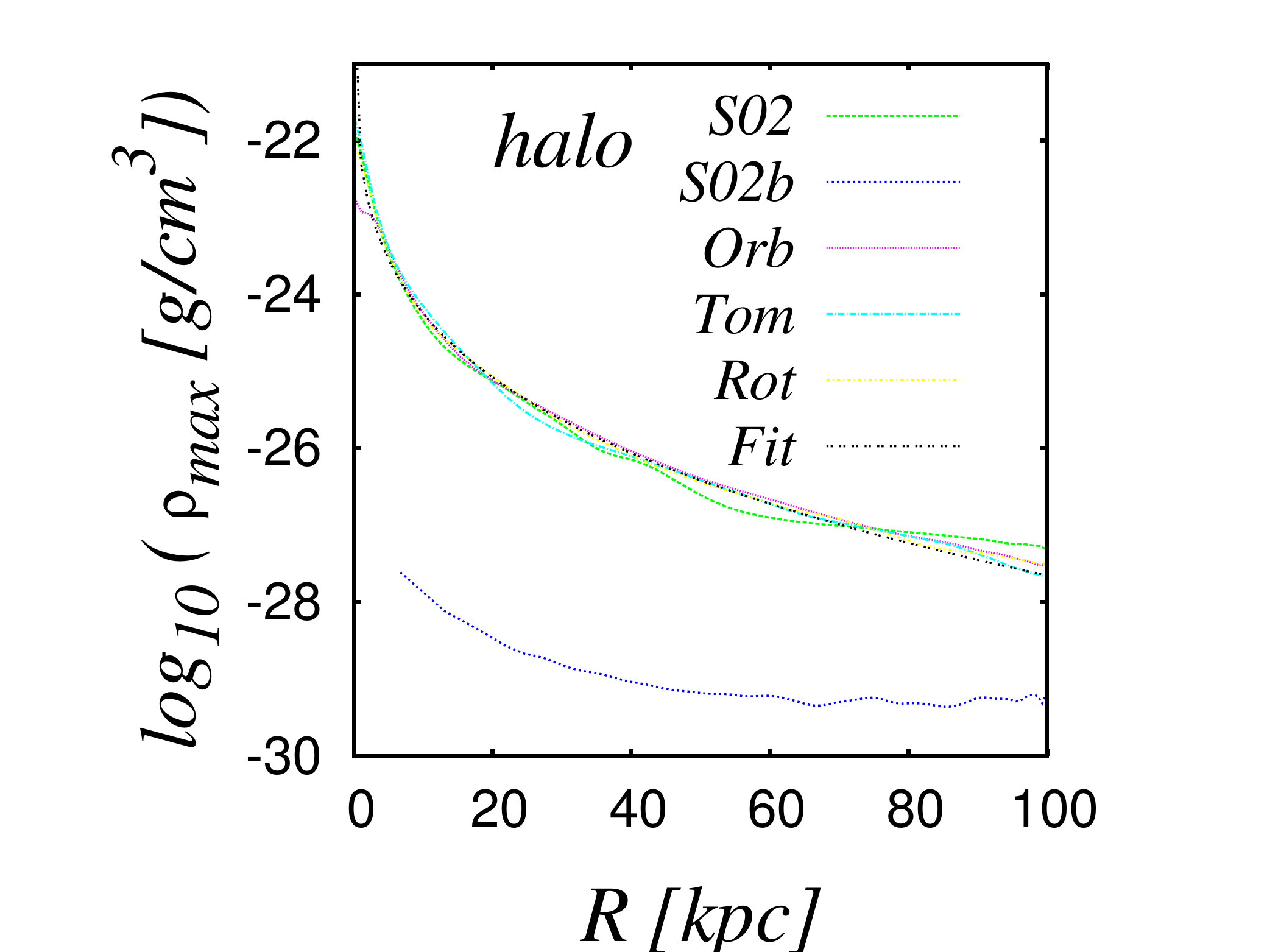}
\end{tabular}
\caption{\label{DenProColGalax2} The radial density profile of the merger remnants formed as a result of the collision models 
summarized in Table~\ref{tab:collision}. Fitting curves for all the models are also included.}
\end{center}
\end{figure}

It is in the central region of the galaxy model where the minimum of the gravitational potential well is 
generated by the most massive matter component, namely the dark-matter halo. Consequently, the densest 
gas is located in the central region of the galaxy model, and it remains rotating in the azimuthal direction 
while oscillating radially simultaneously.     

To achieve a further characterization, we next determine the radial 
profile of the peak density for all the collision models. To do this, we followed the same procedure 
outlined in Section~\ref{subs:evol} about a radial 
partition of $n_{\rm bin}$ bins started from the center 
of the mass of the merger remnants up to a maximum radius of 100 kpc. As was done previously, we accounted for all 
the particles contained in each radial bin taking into account their matter type. We then get the mass 
contained in each radial bin for each matter component and thus the 
density at the average radius of the bin. This density {\it versus} radius calculation is plotted in 
four panels in Fig.~\ref{DenProColGalax2}, so that each panel corresponds to a matter component 
and each curve to a collision model.   

It should be emphasized that due to this procedure, there are no radial density profile curves for the 
model SO2b in the first three panels, because there was no merging process in this model and therefore no new 
structure was formed. However, as the dark-matter component fills the entire volume in which the 
collision models take place, it is possible to determine the radial density 
profile for the collision model S02b in the case of the dark-matter component, which is 
shown in the four panels of Fig.~\ref{DenProColGalax2}.      

We observed no significant difference in the curves shown in Fig.~\ref{DenProColGalax2} with respect 
to the collision model, especially for large radii. In the interval 0--10 kpc, the curves for the collision 
model Orb show a small but noticeable difference with respect to the curves of the other collision models, as 
can be seen in the panel for the gas component. This behavior indicates that the process by which the mass is 
gathered does not make a significant difference because the mass is assembled in the new galactic structure 
driven by the gravitational force. Therefore, the mass is accumulated first at the central region, where the 
gravitational potential takes its deepest value, and later the mass is accumulated on the periphery. 

To take advantage of this result, in Fig.~\ref{DenProColGalax2} we also show the fitting curves for all of 
the matter components of all the collision models. This means that the free parameters of a de Vaucouleurs 
function have been calculated for each density profile curve shown in Fig.~\ref{DenProColGalax2} and 
averaged to have only an overall fitting curve per matter component describing the behavior of the 
radial density profile for each matter component, irrespective of the merging geometry. More 
details about this fitting process are given in Appendix~\ref{appeA}.   

\subsubsection{The circular velocity profile and the time evolution of the angular momentum for the collision models.}
\label{subs:angmomcolmod}

\begin{figure}
\begin{center}
\begin{tabular}{cc}
\includegraphics[width=3.0 in]{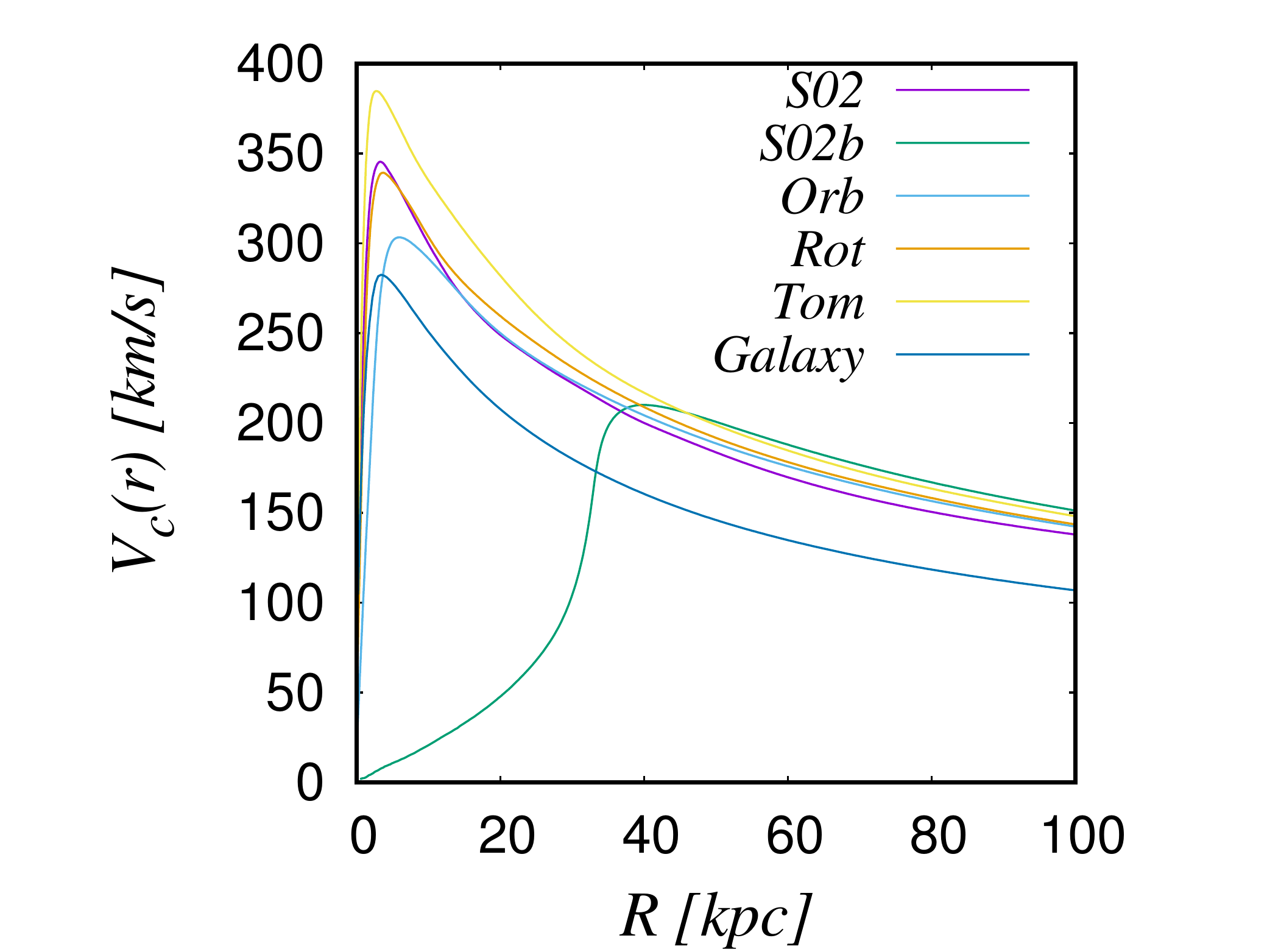}&\includegraphics[width=3.0 in]{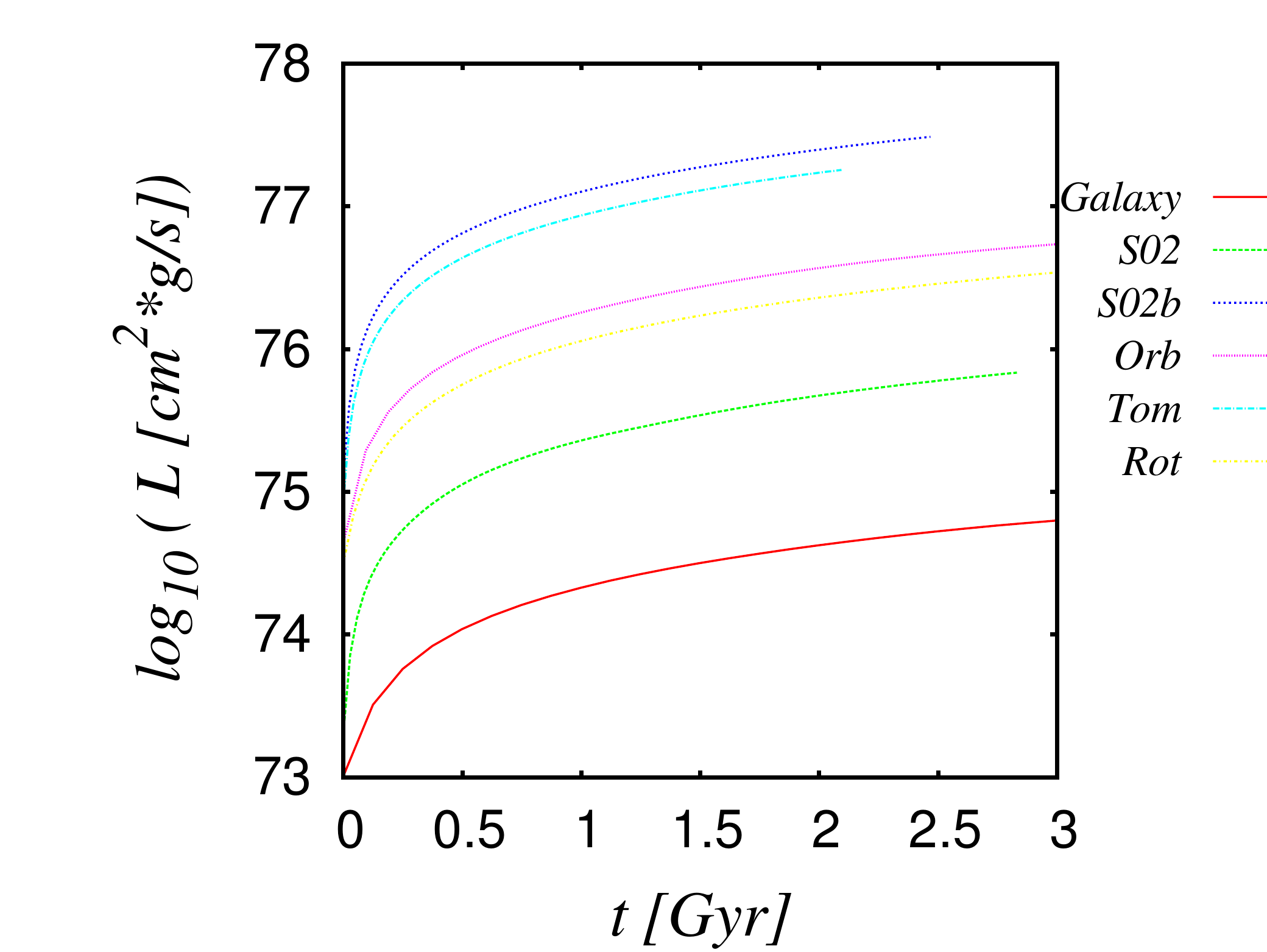}\\
\end{tabular}
\caption{\label{evolmomentoangular} (left) The circular velocity for the collision models. (right) Time evolution 
of the magnitude of the angular momentum L. In these plots all the curves include the 
contribution of all the particles irrespective of the matter component type. To allow comparison with the 
galaxy model, the curves here labeled ''Galaxy'' in these plots are taken from the curves 
labeled ''all'' in Fig. \ref{circularvel}.}
\end{center}
\end{figure}

Taking advantage of the radial partition described in Section~\ref{subs:evol}, in the left-hand panel  
of Fig.~\ref{evolmomentoangular} we now show the circular velocity curves of the collision models, so that 
the matter components are not considered separately. In addition, in the right-hand panel of 
Fig.~\ref{evolmomentoangular} we show the time evolution of the magnitude of the total angular momentum 
for the collision models. We emphasize that both panels of Fig.~\ref{evolmomentoangular} were 
calculated using all of the particles of the simulation, irrespective of the matter component 
type.

With regard to Fig.~\ref{evolmomentoangular}, two comments are in order. First, it should be noted that the 
curves labeled ''Galaxy'' in both panels of Fig.~\ref{evolmomentoangular} are 
those that were labeled ''all'' in Fig.~\ref{circularvel}, so that these curves have been repeated here 
for the sake of comparison between the results of the galaxy model with those shown 
here for the collision models. Second, as was mentioned in Section~\ref{subs:gasdyn}, for the collision 
model S02b, there is no a new structure in which a center of mass can be defined properly, so that 
there is no sense to the circular velocity curve for small radii. When the radius is large enough for 
the two galaxies to be included within this radius, then the circular velocity calculation does 
not realize that the galaxies are separated and the total galaxy mass generates the same behavior 
of the circular velocity curve, as the other collision models do and was observed for the galaxy model, which 
is that the circular velocity curves decrease as $1/\sqrt{R}$ for large $R$.                 

According to the right-hand panel of Fig.~\ref{evolmomentoangular}, the magnitude of the total angular momentum 
of the collision system increases systematically with the evolution of time. This could be due 
to the fact that the matter components expand radially, whence the lever arm length increases and 
although we expect a decrease in the magnitude of the circular velocity, the product of the two physical quantities 
increases. 

It should be noted that all the collision models 
substantially increase their angular momentum with respect to that determined for the galaxy model 
before the collision. The higher values of angular momentum are a consequence of the orbital 
motion developed in the collision models, so that for the models S02b and Tom the curves are at the top of 
the right-hand panel of Fig.~\ref{evolmomentoangular}. The models Orb 
and Rot, which have followed the same 2-body pre-collision path, have an angular momentum very similar 
in magnitude, so their curves are at the middle of the right-hand panel of Fig.~\ref{evolmomentoangular}. The collision 
model S02 shows the lowest angular momentum magnitude, so that its curve is located at the bottom.    
\section{Discussion}
\label{sec:disc}

The main purpose of this paper is to follow the evolution of a galaxy model, which was 
basically taken from the paper by~\citet{Gabassov2006}. However, the widths 
of the disk were very different, because these authors used 0.001 kpc while we used here 
1 kpc, that is, a much wider disk. We will consider a more slender galaxy model elsewhere. It must be 
emphasized that this difference in the galaxy models makes it difficult to compare the outputs, because a 
thin disk favors the growth of perturbations in the orbits of stars, which will result in the 
formation of a bar.    

As we mentioned earlier, an important improvement of our work with respect to 
that of~\citet{Gabassov2006} is that we have included gas in the galaxy model. While it is true that the mass 
fraction of the gas is very small with respect to the fractions of the other matter components, 
the gas dynamics observed in Section~\ref{subs:gasdyn} are interesting and very important to be followed
from the point of view of star formation. For example, ~\citet{Springel2005} found that when the gas fraction is small, the 
resulting merger remnant usually resembles an elliptical galaxy; while if the gas fraction is high enough, then 
other structures can be formed.

We observed that the gas component, initially located in a ring, is moved quickly to the center of the 
galaxy model. Consequently, the peak density of the gas increased significantly. Shortly after, the gas 
is expanded up to an equilibrium radius, which is indicated by the strong decrease of the peak 
density determined in the left-hand panel of Fig.~\ref{densityprofile}. From this moment, most of the gas 
evolves tied to the galaxy center while a small fraction of it managed to escape away, as is indicated by the 
smooth decrease of the curve shown in this panel for large evolution times. 

As we mentioned in Section~\ref{subs:galcol}, the galaxy model at the evolution time $t=0$, was used 
in all of the collision models. Therefore, the initial gas behavior of the galaxy model was observed to happen 
also in the collision models, as can be noticed by comparing the magnitude of the density peak observed 
for the galaxy model in the left-hand panel of Fig.~\ref{densityprofile} with those determined for the 
collision models, which are plotted in Fig.\ref{DenMaxColGalax2}.       

We next simulate some galaxy collision models in order to determine the effects on the distribution 
of the matter components in the new galaxy structures formed out of the merging process of the galaxy model.
With respect to the paper by~\citet{Luna2015}, we emphasize that in this work the number of particles 
used to build the galaxy model increased significantly: \citet{Luna2015} use 1024-29491-245760 particles to 
represent the bulge, disk and halo, respectively. In this paper we use the numbers 33205-99950-533082, which 
are little more than the double. \citet{Gabassov2006} presented a convergence study in 
the number of particles, so that the highest resolution simulation of these authors used 
65536-196608-1048576 particles to represent the components of the bulge, the disk and the halo, 
respectively. This means that our particle numbers in this paper represent the half. Consequently, we 
conclude that the simulations presented in this paper have a resolution comparable to the papers that 
have served us as motivation.

In this paper, we have not observed the formation of long tails in the collision models considered 
in Section~\ref{subs:galcol}. However, when we described the results obtained in some collision models, such as
Tom and Rot, we mentioned that some spiral arms have been formed. It must be emphasized that these structures can 
also be named tails, in the sense that they were formed in close encounters of the 
galaxy model, because the mutual tidal force made particles of the disk and bulge to be ejected from 
the central region. Thus, they can be named either tails or arms and these structures are small in 
length. The reason for the lack of long tails in our simulations was already 
explained by ~\citet{dmh}, so that the formation of long tails in interacting galaxies can be inhibited 
by the presence of a massive dark-matter halo. Later, ~\citet{Springel1999} demonstrated that a dark matter 
halo with a large enough spin parameter led to the formation of long tidal tails, otherwise, no tails are observed.          

Before the log scale is taken in the right-hand panel of Fig.~\ref{densityprofile}, this plot can be compared 
with the four panels of Fig.~\ref{DenProColGalax2}, we observed that the radial density profile in the 
galaxy model is very similar to that observed in new galactic structures formed after 
the collision process. It should be noted that the center of the merger remnants was 
defined as the center of mass, as was mentioned in Section \ref{subs:gasdyn}. Starting from this 
center, we made the radial partition to calculate the physical properties presented in 
Section~\ref{subs:dyncaracol}. The disadvantage of this procedure is that the centers of 
mass for the different matter components are slightly displaced. It should be mentioned that 
other choices for the center of the merger remnants are possible; for instance, the location 
of the particle with the minimum of the gravitational potential.

As was mentioned at the end of Section~\ref{subs:angmomcolmod}, we presented the de Vaucouleurs fitting curves 
for the radial density profiles shown in Fig.~\ref{DenProColGalax2}. The strategy followed was 
explained in detail in Appendix ~\ref{appeA}. The first point to be emphasized is that there is no need to 
fix $n$ at the value 4, as we did in this paper just for simplicity. Meanwhile, $n$ can be varied around 
4 so that the best least 
squares fit must be chosen. From this value of $n$, the value of $b_n$ can then be obtained from the 
approximated formula $b_n=2 n -0.327$, which was proposed by ~\citet{capa}. Next, from the values of 
$A$ and $B$ given by the least squares method, the parameters $\rho_e$ and $R_e$ can be obtained. In principle, 
with this strategy, no parameter is left undetermined. However, with the procedure outlined in 
Appendix ~\ref{appeA}, the parameter $R_e$ is left undetermined. To deal with this situation, we varied the 
parameter $R_e$ within the interval (0.5,20) kpc and obtained all the fitting curves parameters $\rho_e$ and $b_n$. 

We briefly mention the results of 
another case, ~\citet{Bournaud} determined the best fitting parameters of a S\'ersic function for the radial 
profile of the surface density for a set of compact spheroids, which were the outcome of a set of 
high-redshift galaxy merger simulations with high fractions of turbulent and clumpy gas: the average 
values found were $n=3.4$ and $R_{1/2}=4$ kpc. ~\citet{ferrarese} reported 
the isophotal parameters and the surface brightness profiles of 100 galaxies in the Virgo Cluster, and 
found that the surface brightness profiles are well described by a S\'ersic function. In addition, ~\citet{kormendy2009} 
also reported the values of the S\'ersic parameters for many elliptical and spheroidal galaxies also in the Virgo cluster.    

The second point that deserves attention is the use we made of the de Vaucouleurs function 
to describe directly a radial density profile. In fact, ~\citet{mellier} proposed a density function with the form 
$\rho (R)=\rho_e \left( \frac{R}{R_e} \right)^{\beta} \exp \left( \frac{R}{R_e} \right)^{\alpha}$, which was obtained 
as a deprojection of the de Vaucouleurs $\left( \frac{R}{R_e} \right)^{1/4}$ law. In this density 
function, $\alpha$ and $\beta$ are free parameters in addition to $\rho_e$ and $R_e$. 
It should be emphasized that the observables of a galaxy are line-of-sight projections 
of the corresponding three-dimensional physical quantities. In the general case, the projected quantity 
is related to the three-dimensional quantity by an integration along the line-of-sight 
spatial coordinate.  

In this paper we adopted a functional form for $\rho (R)$ as that of the de Vaucouleurs 
function and then constrained 
its free parameters by comparing it to the calculated radial density profile shown in 
Fig.~\ref{DenProColGalax2}, which is already a 
three-dimensional quantity. As was shown in Section~\ref{subs:dyncaracol}, the de Vaucouleurs 
function does an excellent fit with the radial density profiles shown in Fig.~\ref{DenProColGalax2}. The 
reason behind this success is that the de Vaucouleurs formula was designed to represent a central peak 
surrounded by a region where the variable of interest falls with the 1/4 power of 
the radius, just as the radial density profile does, irrespective of being a projected 
or three-dimensional quantity.             

It should be emphasized that ~\citet{aguilar} presented N-body simulations with an initial density 
profile of the de Vaucoulours form. At the final evolution time, they found that the density profile 
remains that of de Vaucoulours but with other parameters. This statement can be said in other terms, such as 
de Vaucouleurs surface brightness profile appears to be invariant under galaxy harassment. In this sense, it can 
then be considered that our paper confirms part of this result; as we mentioned earlier, the merger 
remnants manage to adopt a radial density profile of the de Vaucoulours form, irrespective of the 
pre-collision trajectory. It must be emphasized that of galaxy model used much more particles and matter components 
than the galaxy model of ~\citet{aguilar} because they used 3000 particles in their galaxy collision models. 

One last comment about some important physical elements of the gas that are 
missing in this paper, for instance ~\citet{Springel2000} presented 
simulations of interacting disk galaxies including star formation and feedback, where the gas 
is able to cool radiatively and to form collisionless stars. There is an extensive literature devoted to 
the study of galaxy formation and evolution by hydrodynamical 
simulations, some of which include this kind of complicated gas physics. 
  
Many efforts have been made to 
date to incorporate star formation and 
feedback in simulations of galaxy formation and evolution, see for 
instance, \citet{springelhernquist2003}. It has proven 
to be a difficult problem to be managed, as many recipes have been introduced and 
tested during many years, see 
for instance, \citet{stinson2006}. 
We have not even attempted to consider this complicated problem, which is beyond the scope
of this paper, since our interest at the moment is only to put the gas on a consistent basis 
in a general model of a galaxy. However, we would like to comment about the importance of 
the lack these gas physics on simulations of galaxy collisions, like the ones presented in this paper. 

As we have seen, the simulations show that the gas moves rapidly towards the central region of 
the dark-matter halo and therefore, the gas density increases. At some point, it will be very important to model 
the transformation of gas into stars. As expected, this transformation will modify the dynamics of all 
the matter components in the galaxy model. In addition, many simulations have demonstrated
that galaxy collisions in general augment the star formation rate, from low levels 
(a few times the star formation rate detected for the isolated galaxy model) to high values
(20-60 times the isolated galaxy case), see for instance \citet{matteo}. In this case, the 
amount of gas available will be reduced after the collision, so the curves for the radial 
density profile are expected to be different with respect to the isolated galaxy model when star 
formation be included somehow. 

\section{Consistency of our simulations with regard to other papers}
\label{sec:comp}

In this Section we will try to establish the consistency of the simulations 
presented in this paper by comparing their results with other simulations, with 
observations and with virtual observations.

\subsection{Comparison with other simulations}
\label{subsec:compsim}

\citet{barnes1996} determined the radial density profile of their collision models 
and found a set of curves falling systematically for the remnant's innermost region. 
It should be emphasized that in this paper we also found a similar behavior 
but we extend the radius up to 100 kpc from the remnant's center.            

For this reason, we claim that in the present paper we also observe this 
redistribution of gas to the central region and show that the gas is linked 
to the central region during almost all the evolution time even in cases in 
which a collision with another galaxy occurs. We did not observe the well defined 
spiral pattern of the gas in our galaxy model, as was observed by 
\citet{barnes1996}. It should be mentioned that this spiral pattern is
a transitional stage that ends quickly. We believe that this failure is due to 
the lack of radiative cooling in the gas component.   

In addition, \citet{HernquistMihos95} determined the time evolution of 
the total angular momentum of the galaxy components in their satellite 
merger model, and found that when the primary galaxy model includes 
a bulge, then the curves grow systematically. Meanwhile, the curves for the 
gas component fall systematically. Because the magnitude of the latter 
curves are quite smaller than that of the former curves, 0.03 {\it versus} 0.25 as can be 
appreciated in their Fig. 8, we conclude that the general behavior of the total 
angular momentum, when all the components are included together, should be 
similar to the curves shown in Fig.\ref{evolmomentoangular} of the present 
paper, that is, growing systematically with the evolution time.       

On the other hand, \citet{MihosHernquist96} showed that the geometry 
of the orbits of the approaching galaxies does not seem to be 
important in the determination of the resulting dynamics of the gas. It 
seems to be more important the internal structure of the galaxies, for 
instance, the presence or absence of a bulge in the galaxy model and 
its physical properties. Although using only a very limited 
collection of approaching orbits, in the present paper we confirm this 
result of \citet{MihosHernquist96}, as we mentioned at the end of the 
Section \ref{sec:disc}, where we connect our results with those of the paper 
by \citet{aguilar}.  

\subsection{Comparison with observations}
\label{subsec:compobs}

All of the simulations reported in 
the papers mentioned above in Section \ref{subsec:compsim}, contributed with different 
elements to support the idea of the formation of elliptical galaxies by 
means of collisions between spiral galaxies. As \citet{barnes1996} mentioned, it 
happens that centrally concentrated gas systems, like the ones observed in those 
simulations, have been detected by means of CO interferometer observations of 
galaxies, for instance Arp 220, see \citet{Scoville1986} and NGC 520, 
see \citet{Sanders1988b}, among others.

We believe that the present paper reaffirms this idea, as 
the collision remnants seen in Figs. \ref{colisionorb}, \ref{colisiontom} 
and \ref{colisionOrbRot} seem to be spheroidal systems supported by 
rotation, likely resembling the kind of systems usually classified as 
normal elliptical E, lenticular galaxies of the type SO, see \citet{kd89}, as the 
size and mass of our merger remnants are 
around 100 kpc of radius and the total mass contained up to this radius is 
around 3$\times 10^{11}\,M_{\odot}$, in which all the mass 
components have been included. 

On the other hand, by combining new surface photometry with
published data, \citet{kormendy2009} constructed composite brightness 
profiles over large radius ranges of all known elliptical galaxies in 
the Virgo cluster. It must be noted that \citet{kormendy2009} took a 
conclusion (see their Section 7.2), which seems to generalize and 
at the same time provides some observational support to the idea 
described at the end of Section\ref{sec:disc} of this paper, that is, the idea 
that the de Vaucouleurs profile curve fits well all of the 
resulting merger remnants of the galaxy collisions considered 
in this paper, irrespective of the collision geometry. 

Let us now quote that conclusion in \citet{kormendy2009} own words: ''One of 
the main conclusions of this paper is that S\'ersic functions fit the main
parts of the profiles of both elliptical and spheroidal galaxies
astonishingly well over large ranges in surface brightness. For
most galaxies, the S\'ersic fits accurately describe the major-axis
profiles over radius ranges that include ∼ 93–99 percent of the light
of the galaxies (see Figure 41). At small r, all profiles deviate 
suddenly and systematically from the best fits.'' Later, in their Section 9.2, 
\citet{kormendy2009} continued in this way: ''This result is 
remarkable because there is no astrophysical
basis for the S\'ersic function. We know no reason why violent
relaxation, dissipation, and star formation should 
conspire-surely in different ways in different galaxies-to produce so
simple and general a density profile.''

\subsection{Comparison with virtual observations}
\label{subsec:compvirobs}

The Illustris cosmological hydrodynamic simulation, which is described 
by \citet{illustris}, has successfully reproduced the distributions of galaxies 
in clusters, so that both spirals and elliptical galaxies can be 
distinguished morphologically for the first time, as far as we know.

By using a suite of simulations based on the Illustris simulation, \citet{taylor} 
determined the mass profile of a dark-matter halo in which 
there is embedded a galaxy with a mass comparable to the Milky Way's mass. In 
the left panel of their Fig.2 they show the circular velocity curves for the chosen 
mass systems (consider only the curves labeled as ''D12''). The curves grow rapidly 
for small radius, until a peak circular 
velocity is reached, from which the curve falls smoothly as the radius 
increases. It must be emphasized that the shape of these 
curves is very similar to those shown in the left panel 
of Fig.\ref{circularvel}, calculated in 
the present paper to characterize the galaxy model. 

When they include all their matter components (dark-matter, gas and stars) in the 
calculation of the circular velocity, their peak velocity 
is a little below 200 km/s; in our case, when we included all the four matter 
components, we obtained a peak value around 270 km/s, see the curve 
labeled as ''all'' in the left panel 
of Fig.\ref{circularvel}. The peak velocity of the curve for the dark-matter halo is
a little up 200 km/s, see the curve labeled as ''halo'' again in the left panel 
of Fig.\ref{circularvel}.

When they separate their mass components in their calculation of the circular 
velocity, see the right panel of their Fig.2, they obtained for the gas a curve 
around 50 km/s. In our case, our corresponding curve for the gas component 
is around 30 km/s. 

We can compare their circular velocity curves with the 
ones we obtained for the merger remnants, which are shown in the left panel 
of Fig.\ref{evolmomentoangular}. Because the mass assembled in the 
remnants is a bit more massive than the galaxy model, all the curves 
are higher in magnitude than those of \citet{taylor}. The peak circular 
velocity of the curves shown in Fig.\ref{evolmomentoangular} occurs for 
a radius around 10 kpc, while that radius of \citet{taylor} for their curves is around 20 kpc. 

As the radial profile of the circular velocity is a good indicator of 
the mass distribution of a system, then we can conclude, on the basis of the 
previous comparison, that we have roughly modeled a system of a 
similar mass and size of those chosen for \citet{taylor} to model the Milky Way galaxy.            
\section{Concluding Remarks}
\label{sec:conclu}

In this paper we implemented a galaxy model that proved to be stable over a long evolution time. 
As an improvement over the papers of ~\citet{Gabassov2006} and ~\citet{Luna2015}, on which this work is based, 
here we included a gas component. 

Using this galaxy model, we then explored several collision models of equal-mass galaxies to study the 
effects of different interaction scenarios on the dynamic of the matter components of the new 
structures formed after such merging process. As expected, the interaction between galaxies produce notable 
changes in the galaxies participating and in their physical properties. We here focused specifically on the 
density profile. Some of the conclusions to be emphasized from the models calculated here follow: 

\begin{enumerate}

\item The galaxy model has proved to be stable (in the sense that it has reached a state of dynamic 
equilibrium) up to an evolution time of 14 Gyr.

\item Most of the gas of the galaxy model evolved strongly tied to the galaxy center, while a small fraction of 
it managed to escape away with the evolution of time.

\item The collision models considered in this paper are not intense enough to 
significantly eject the gas from the central mass distribution of the galaxy model.

\item The gas is gravitationally bounded to the center of each of the galaxies even during 
the process of collision.

\item The gas shows an interesting dynamics, despite the fact that it is always bounded gravitationally to
the central region of the newly formed system.       

\item It should be noted that all the collision models substantially increase their angular momentum with 
respect to that of the individual galaxy before the collision. 

\item The dynamic variables studied in Section~\ref{subs:dyncaracol} (e.g. peak density, density profile 
and angular momentum) do not capture the moment at which the merging process of the two galaxies takes place.

\item It seems to be that the radial density profile does not make any difference with respect to the 
collision process that gather the mass together. Therefore it is possible to obtain an overall general 
de Vaucouleurs curve to describe the general behavior of the radial density profile of the merger 
remnants. The fitting parameters of the de Vaucouleurs curve are describe in Table~\ref{tab:sersic} and in 
Fig.~\ref{berhoeProm}. 

\item The four-parameters formula considered in Appendix B also fits well in general the radial density profile 
for radii within 10-20 kpc. But, for some collision models, like the Orb y Tom, there are considerable deviations 
for very small radii.     

\end{enumerate}

\appendix
\section{The de Vaucouleurs fitting curves} 
\label{appeA}

Let us thus adapt the de Vaucouleurs function to describe the radial profile of the 
peak density calculated in Section~\ref{subs:gasdyn}, which in this paper will be considered to be 
given as

\begin{equation}
\rho(R)=\rho_e \, \exp \left\{ -b_e  \left[ \left( \frac{R}{R_e} \right)^{1/4} -1 \right] \right\}
\label{sersic}
\end{equation}
\noindent In the case of the surface brightness $I$, $R_e$ is known as the effective radius or 
half-light radius, because this indicates the radius within which the brightness of the elliptical galaxy 
includes half the light of the image. For this paper, $\rho_e$, $b_e$ and $R_e$ are free parameters of 
Eq.\ref{sersic}, which must be determined.       

By taking the natural logarithm 
on both sides of Eq.~\ref{sersic} and defining $x=R^{1/4}$, $x_e=R_e^{1/4}$, 
$y=\log{\rho}$ and $y_e=\log{\rho_e}$ we get 

\begin{equation}
y= A + B\, x
\label{fitmodel}
\end{equation}
\noindent where $A$ and $B$ are parameters to be determined by the least squares method applied 
to the data shown in the plots of Fig.~\ref{DenProColGalax2}, where $y$ and $x$ take the values 
$y_i$ and $x_i$ with $i=1..n_{\rm bin}$, as was done in Section~\ref{subs:evol}, where the 
radial partition was described. These parameters are related to the de Vaucouleurs parameters 
by means of

\begin{equation}
\begin{array}{l}
A=b_e + y_e \; \vspace{0.1 cm}\\
B=-\frac{b_e}{x_e}
\end{array}
\label{param}
\end{equation}
\noindent 
so that there are three free parameters in the right-hand side of Eq.~\ref{param}, while the least 
squares method gives us only two parameters in the left-hand side of Eq.~\ref{param}. To solve 
this issue, we consider the following strategy, which is obviously not unique, see the end of 
Section~\ref{sec:disc}. We make a partition in the 
radial parameter $R_e$ so that we scan a relevant interval, for example from 0.5 to 20 kpc. Then having 
the values of $A$ and $B$ by using the procedures of \citet{numrecip} and fixing the value for $R_e$, we 
then obtain the corresponding values of de Vaucouleurs parameters $b_e$ and $\rho_e$ by means of

\begin{equation}
\begin{array}{l}
b_e = - x_e \, B  \; \vspace{0.1 cm}\\
y_e= A-b
\end{array}
\label{solparam}
\end{equation} 
          
It is interesting to mention that this strategy led us to a unique curve from 
all the curves for each $R_e$. This means that the increment 
in the value of the parameter $R_e$ produces a change in the values of the parameters $b_e$ 
and $\rho_e$, so that the three parameters produce the same curve by means of Eq.\ref{sersic}, for every 
value of $R_e$ within the scanned interval.

\begin{table}[ph]
\caption{The averaged parameters of the de Vaucouleurs fitting curves for the radial density profile. The parameter 
$R_e=0.5$ kpc.}
{\begin{tabular}{|c|c|c|} 
\hline
matter component   & $b_e$      & $\rho_e$\\
\hline
\hline
gas      &  4.97  &  2.767 $\, \times 10^{-24}$ \\
\hline
disk     &  6.03  &  1.006 $\, \times 10^{-22}$ \\
\hline
bulge    &  5.85  &  1.925 $\, \times 10^{-23}$ \\
\hline
halo     &  4.75  &  1.108 $\, \times 10^{-22}$ \\
\hline
\hline
\end{tabular} }
\label{tab:sersic}
\end{table}

\begin{figure}
\begin{center}
\begin{tabular}{cc}
\includegraphics[width=3.0 in]{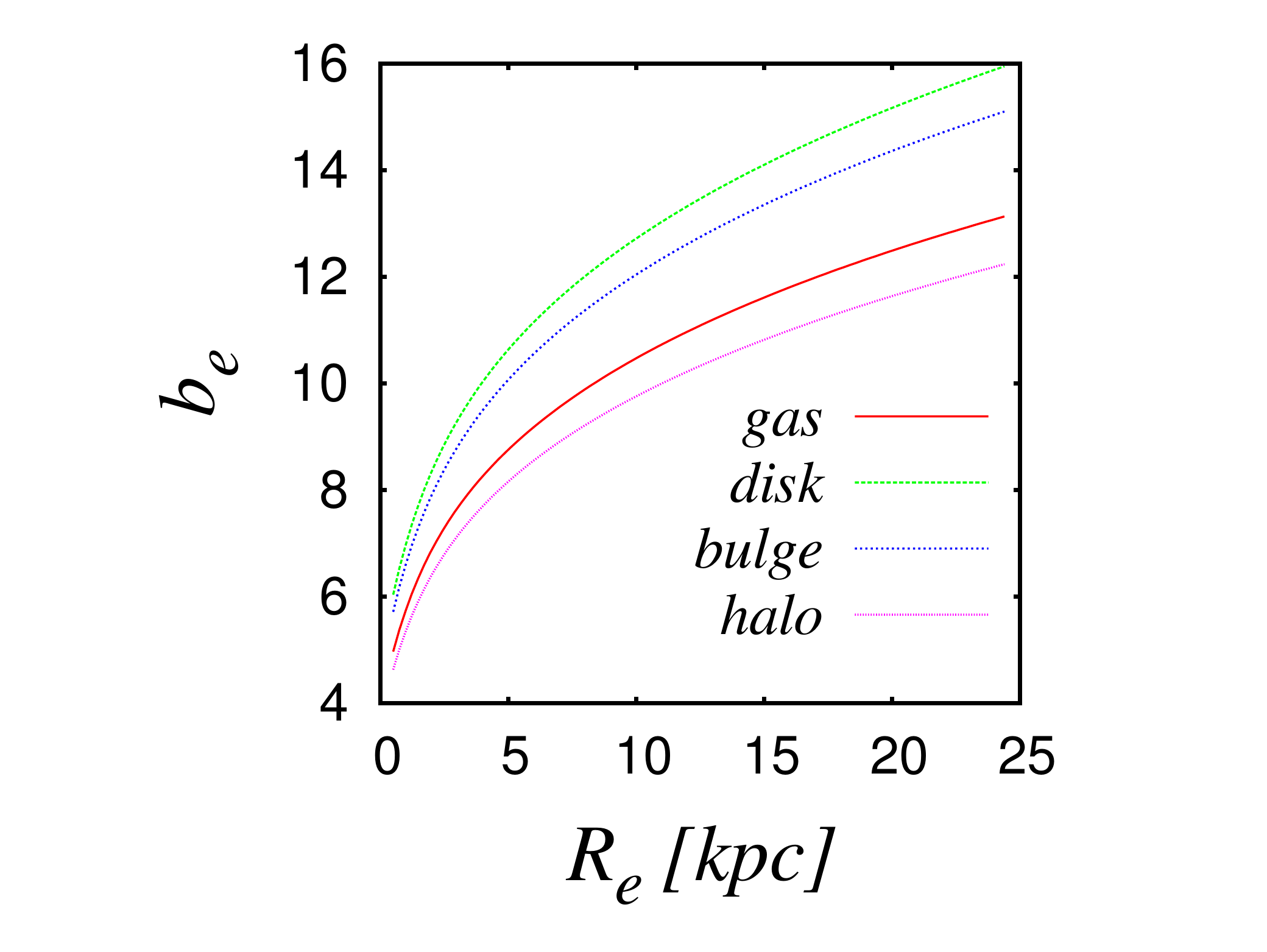} & \includegraphics[width=3.0 in]{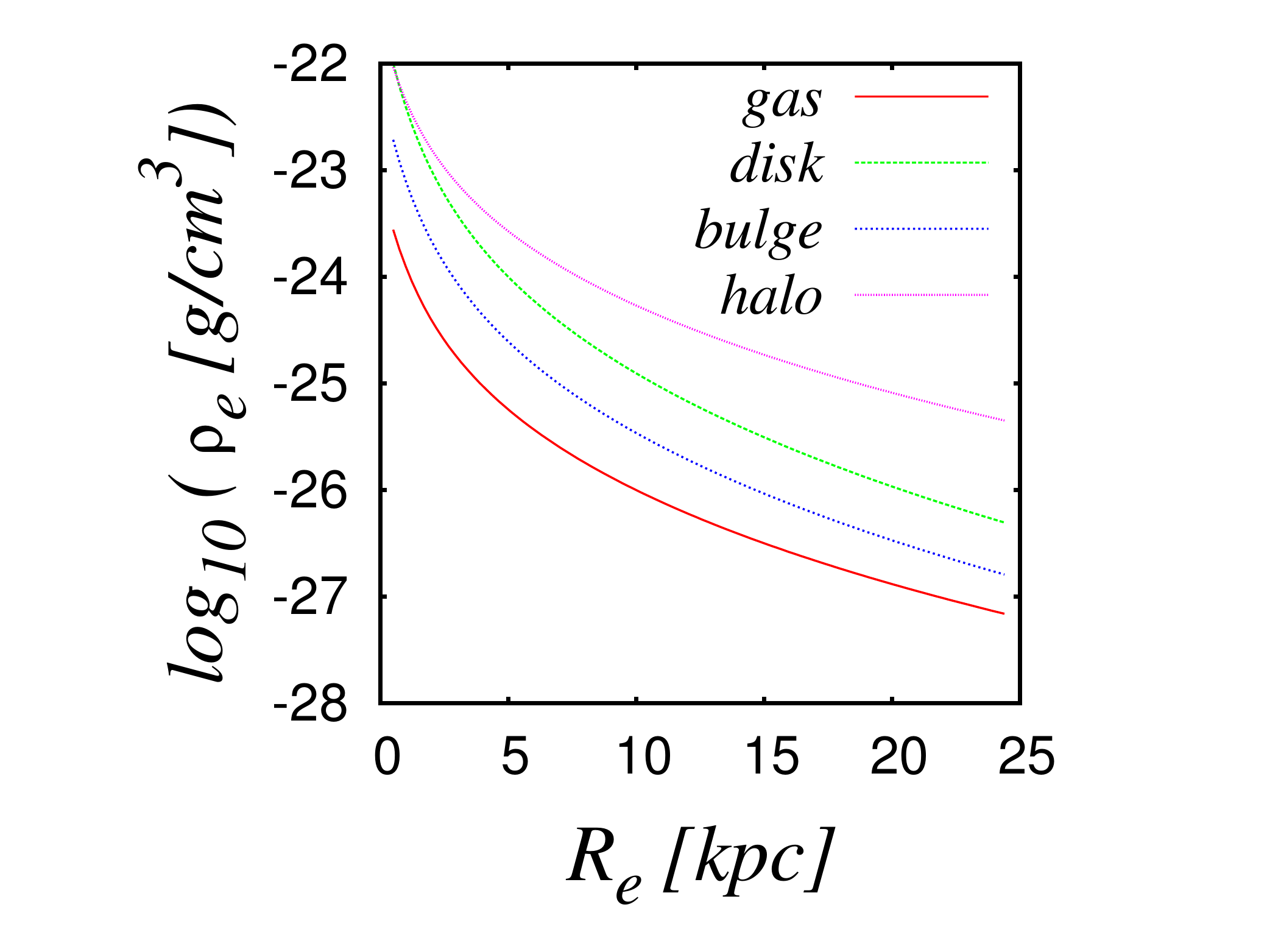}\\
\end{tabular}
\caption{\label{berhoeProm} Averaged values of the de Vaucouleurs parameters as a function of the value of the parameter $R_e$, 
(left) $b_e$ and (right) $\rho_e$.}
\end{center}
\end{figure}
 
Finally, given that there are four matter components and four collision models, once we have the 
fitting curve for every model and matter component, we then take the 
averaged values of the parameters $b_e$ and $\rho_e$ for a fixed value 
of $R_e$, so that the resulting averaged fitting curves have been plotted 
in each panel of Fig.~\ref{DenProColGalax2}. It should be noticed that these averaged values for the 
de Vaucouleurs parameters $b_e$ and $\rho_e$ are reported in Table~\ref{tab:sersic}.

With the procedure followed in this Appendix \ref{appeA}, the parameter $R_e$ is left 
undetermined. To alleviate this issue, in Fig.~\ref{berhoeProm} we show the values of $b_e$ and $\rho_e$ 
obtained as a function of $R_e$.

\section{A four-parameters fitting curves} 
\label{appeB}

The objective of this Appendix \ref{appeB} is twofold. First, to complement the curves of the 
radial density profile shown in Section \ref{subs:gasdyn}, which were constructed by using a radial 
partition up to a maximum radius of 100 kpc, so that now we shorten this radial range up to 
40 kpc, to show the gas distribution of the innermost region of the merger remnants with more 
detail. Second, to complement the results of 
Appendix \ref{appeA}, where a de Vaucouleurs function was proposed to describe 
the radial density profile, so that now we test another radial formula 
which has given good results as a fitting model, as we explain below.    

\begin{figure}
\begin{center}
\begin{tabular}{cc}
\includegraphics[width=3.0 in]{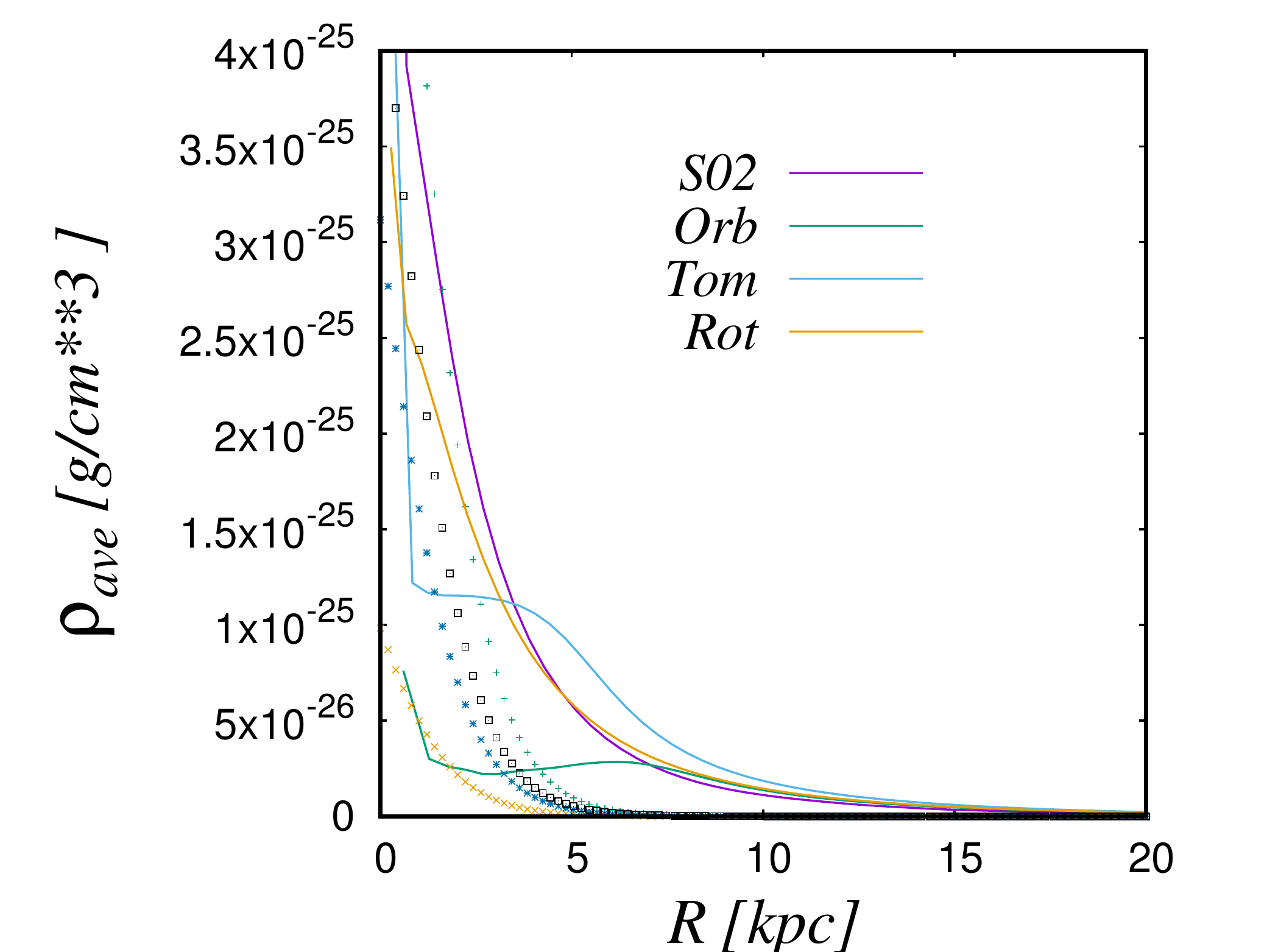} & \includegraphics[width=3.0 in]{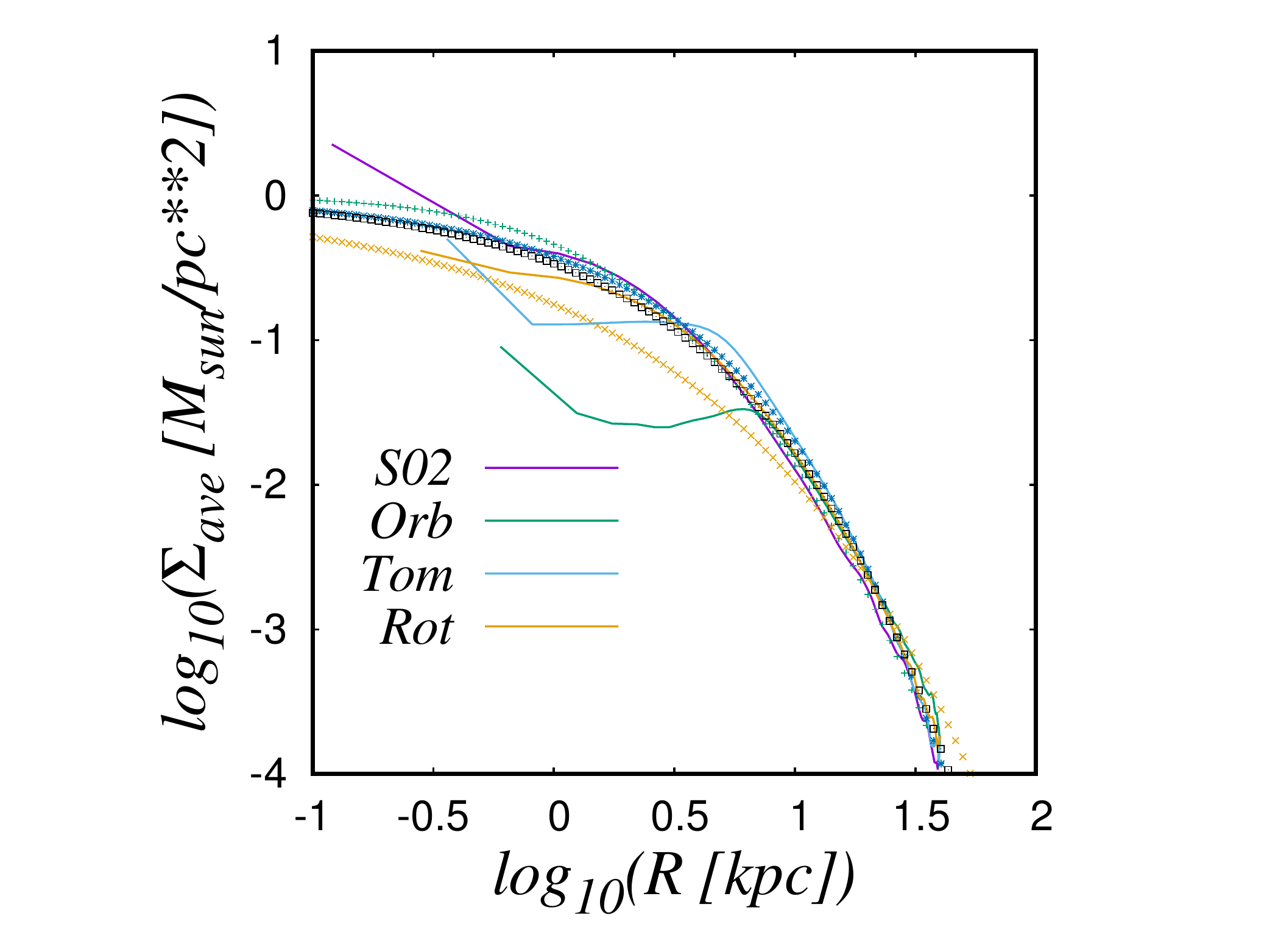}\\
\end{tabular}
\caption{\label{fitting4par} The radial density profile of the merger remnants (shown with lines) and the fitting 
curves (shown with points) for the collision models.}
\end{center}
\end{figure}

Recently, \citet{wang} focused on understanding the radial distribution of the gas in a sample of 
spiral galaxies, showed that there is a mathematical function that works well as a fitting model
for the radial profiles of the HI surface density for the 42 galaxies, which are part 
of a sample of galaxies of the Bluedisk project, see \citet{bluedisk}. The formula is 
shown in Eq.1 of \citet{wang} and we repeat it here for the reader's convenience:

\begin{equation}
\Sigma(x)=\frac{ I_1 \, \exp \left( -x/r_s \right) } {1 + I_2 \, \exp \left( -x/r_c \right)}
\label{fourparamformula}
\end{equation}
\noindent where $\Sigma$ is the surface density and $I_1$, $r_s$, $I_2$, $r_c$ are free parameters to be 
determined by adjusting the curve to the data available. By using an ANSI C 
translation of the MPFIT program, see \citet{webmpfit}, we calculate the best fitting parameters to solve the
least-squares problem applied to the data $\rho_{ave}(R)$ {\it versus} $R$, where $\rho_{ave}(R)$ is the average density 
for a thin radial shell centered around the 
radius $R$ with a width given by $\delta R$. The resulting curve is shown in the left panel of 
Fig.\ref{fitting4par}. The set of parameters per each collision model is 
shown in Table\ref{tab:fourparam}. 

In this case, the function $\Sigma(x)$ defined in Eq. \ref{fourparamformula} has been identified directly with the mass 
density $\rho(R)$ and the independent variable $x$ with the 
radius $R$. Let us compare our results with those of \citet{aumer}, who presented simulations of gas disk 
formation and evolution, so that they located gas at redshift 1.3 in some dark-matter halos chosen from a 
dark-matter-only simulation. The gas evolved up to a redshift zero in a zoom-in cosmological re-simulation. In the right 
panel of the second line of their Fig.12, they reported the surface 
gas density (in units of $10^{10}\,\times \,\frac{M_{\odot}}{kpc^3}$) {\it versus} the disk radius (in kpc), so we include 
in the sixth column of Table\ref{tab:fourparam} the values obtained for the following combination of fitting parameters: 
$\rho_{ave}^{\Sigma}(0)=\frac{I_1}{1+I_2}$, which corresponds to the value of the fitting curve at $R=0$, see 
Eq.\ref{fourparamformula}. \citet{aumer} reported a value of $10^{-1}$ at $R=0$, which is very high compared to our values, 
which are in the range from $1.2 \, \times 10^{-3}$ to $6.95 \, \times10^{-4}$.            

In order to compare with the paper of \citet{wang}, we calculate an approximate surface density profile based on the procedure  
already explained above to obtain the radial density profile (see also Sections \ref{subs:evol} 
and \ref{subs:gasdyn}). As we mentioned, 
we made a radial partition of the spherical galaxy in terms of spherical shells centered on a radius $R_i$, so that 
the number and type of the particles contained in each radial shell was accounted for and we simply divide it 
by the surface area of the shell at the radius, which is  $4 \pi R_i^2$ on average. We present our results in a log-log plot 
and changed the units of the surface density to make comparison easier.

It must be noted that we have applied the fitting process on 
the log-log data directly to determine the parameters of the best fitting curve. In this case, we 
have identified the $\Sigma(x)$ defined in Eq. \ref{fourparamformula} with $\log (\Sigma_{ave})$ and 
the $x$ with $\log (R)$. The fitting curve is shown in the right panel of Fig.\ref{fitting4par} and 
the parameters are reported in Table \ref{tab:fourparamSigLL}.       

The average values of the parameters $r_s$ and $r_c$ for the 43 galaxies reported by \citet{wang} are 
$<r_s>=5.77$ kpc and $r_c=4.23$ kpc, while our average values obtained for the plot shown in the left panel of 
Fig.\ref{fitting4par} are $<r_s>=0.9$ kpc and $r_c=1.17$ kpc. While it is true that the curves of the right panel of 
Fig.\ref{fitting4par} show a similar shape to the curves reported by \citet{wang} in their Figs. 1, 4 and 5, our 
curves are quite below their curves, as can be seen in the sixth column of Table \ref{tab:fourparamSigLL}, in which we 
show the expected value of our fitting curve $\Sigma_{ave}(0)$ ( now without the log scale ) at r=0. \citet{wang} showed a 
value of the surface density at r=0 within 1 to 10, while our values are always below 1.         

\begin{table}[ph]
\caption{Parameters of the four-parameters fitting curves shown in Eq.\ref{fourparamformula} for the radial density profile.}
{\begin{tabular}{|c|c|c|c|c|c|} 
\hline
model   & $I_1$ [$g/cm^3$] & $r_s$ [kpc] & $I_2$ [] & $r_c$ [kpc] & $\rho_{ave}^{\Sigma}(0)$[ $10^{10}\,\times \,\frac{M_{\odot}}{kpc^3}$] \\
\hline
\hline
S02     & $2.01\, \times 10^{-24}$ & 0.948 & 1.34  & 1.13 & $1.27 \, \times 10^{-3}$ \\
\hline
Orb     & $2.28\, \times 10^{-25}$ & 0.9489 & 1.32  & 1.21  &  $1.45\, \times 10^{-4}$ \\
\hline
Tom     & $7.29\, \times 10^{-25}$ & 0.94 & 1.34  & 1.17  & $4.6\, \times 10^{-4}$ \\
\hline
Rot     & $1.1\, \times 10^{-24}$ & 0.94 & 1.34 & 1.15   & $6.95\, \times 10^{-4}$ \\
\hline
\hline
\end{tabular} }
\label{tab:fourparam}
\end{table}


\begin{table}[ph]
\caption{Parameters of the four-parameters fitting curves shown in Eq.\ref{fourparamformula} applied 
directly to the log-log data of the radial density profile.}
{\begin{tabular}{|c|c|c|c|c|c|} 
\hline
model   & $I_1$ [$\log_{10}( \frac{M_{\odot}}{pc^2})$] & $r_s$ [$\log_{10}(kpc)$] & $I_2$ [] & $r_c$ [$\log_{10}(kpc)$] & $\Sigma_{ave}(0)$[ 
$\frac{M_{\odot}}{pc^2}$]\\
\hline
\hline
S02     & -0.94  & -1.05 & 1.8   & 0.56 & 0.46 \\
\hline
Orb     & -30.89 & -1     & 40  & 760   & 0.17 \\
\hline
Tom     & -5.5 & -0.9 & 12.12   & 3.13  & 0.38 \\
\hline
Rot     & -7.19 & -1.7 & 14.22   & 1.17  & 0.33 \\
\hline
\hline
\end{tabular} }
\label{tab:fourparamSigLL}
\end{table}
\section*{Acknowledgements}
The author gratefully acknowledges the computer resources, technical expertise, and
support provided by the Laboratorio Nacional de Superc\'omputo del Sureste de M\'exico
through grant number O-2016/047. The author would like to thank to the referee for his/her report on this 
manuscript, which has helped a lot in improving its content.


\begin{thebibliography}{99}

\bibitem[Aguilar and White (1986)]{aguilar} Aguilar, L. and White, S.D.M., 1986, The Astrophysical Journal, 307, pp.97-109.

\bibitem[Aumer and White (2013)]{aumer} Aumer, L. and White, S.D.M., 2013, MNRAS 428, pp.1055–1076.

\bibitem[Athanassoula and Bosma (2019)]{atha} Athanassoula, E. and Bosma, A., 2019, Nature Astronomy,  3, pp.588-589.

\bibitem[Balsara (1995)]{balsara1995} Balsara, D., 1995, J. Comput. Phys. 121, 357.

\bibitem[Barnes and Hernquist (1991)]{barnes1991} Barnes, J.E. and Hernquist, L.E., 1991, ApJ, 370, L65.

\bibitem[Barnes and Hernquist (1996)]{barnes1996} Barnes, J.E. and Hernquist, L.E., 1996, ApJ, 471, 115.

\bibitem[Binney and Tremaine (1994)]{BinneyTremaine} Binney, J. and Tremaine, S., \textit{Galactic Dynamics}, Princeton 
University Press, New Jersey, 1994.

\bibitem[Bournaud et al. (2011)]{Bournaud} Bournaud, F., Chapon, D., Teyssier, R., Powell, L.C.,  Elmegreen, B.G.,
Elmegreen, D.M., Duc, P.A., Contini, Epinat, T.B. and Shapiro, K.L., 2011, The Astrophysical Journal, 730, 4.  

\bibitem[Burkert et al. (2008)]{burkert} Burkert, A., Naab, T., Johansson, P.H. and Jesseit, R,2008, The 
Astrophysical Journal, 685, pp.897-903. 

\bibitem[Caon et al. (1993)]{capa} Caon, N., Capaccioli, M. and D'Onofrio, M., 1993, MNRAS, 265, 1013.

\bibitem[de Vaucouleurs (1948)]{vau} de Vaucouleurs, (1948), Ann. Astrophysics, Vol. 11, p.247.
 
\bibitem[Dehnen (1993)]{Dehnen} Dehnen, W., (1993), MNRAS, 265-250.

\bibitem[Dubinski at al. (1996)]{dmh} Dubinski, J., Mihos, J.C. and Hernquist, L., (1996), ApJ, 462, 576.

\bibitem[Ferrarese (2006)]{ferrarese} Ferrarese, L., Cot\'e, P., Jord\'an, A., Peng, E.W., Blakeslee, J.P., Platek, S., Mei, S., Merrit, D., Milosavljevi\'c, M., Tonry, J.L. and West, M.J., (2006), The Astrophysical Journal Supplement Series, 164, pp.334-434.

\bibitem[Freeman (1970)]{Freeman} Freeman, K.C., (1970), The Astrophysical Journal, 160-811.

\bibitem[Gabbasov at al. (2006)]{Gabassov2006} Gabbasov, R.F., Rodriguez-Meza, M.A., Cervantes-Cota, J.L. and 
Klapp, J. (2006), Astron.Astrophys, 449-1043.

\bibitem[Garbow et al. (2013)]{webmpfit} Garbow, B., Hillstrom, K. and More, J., \url{https://pages.physics.wisc.edu/~craigm/idl/cmpfit.html}. 

\bibitem[Hernquist (1990)]{Hernquist1990} Hernquist, L., (1990), The Astrophysical Journal, 356:359-364.

\bibitem[Hernquist and Mihos (1995)]{HernquistMihos95} Hernquist,L.E. and Mihos,J.C., 1995, ApJ, 448, 41.

\bibitem[Hohl (1971)]{Hohl} Hohl, F. (1971), The Astrophysical Journal, 168, 343-359.

\bibitem[Kormendy and Djorgovski (1989)]{kd89} Kormendy, J. and Djorgovski, S. Annu. Rev.Astron.Astrphys. 1989, 27, pp. 235-277.

\bibitem[Kormendy at al. (2009)]{kormendy2009} Kormendy, J., Fisher, D.B., Cornell M.E. and Bender, R., (2009), The 
Astrophysical Journal Supplement Series, 182, 216-309.

\bibitem[Kuijken at al. (1995)]{Kuijken} Kuijken, K. y Dubinski, J., (1995),  Mon. Not. R. Astron. Soc. 277, 1341-1353.

\bibitem[Liu (2003)]{liu} Liu, G.R. y Liu, M.L., \textit{Smoothed Particle Hydrodynamics: A Meshfree Particle Method}, World 
Scientific Publishing Company, 2003.

\bibitem[Luna at al. (2015)]{Luna2015} Luna Sanchez, J.C., Rodriguez Meza, M.A., Arrieta, A. and Gabbasov, R.,(2015), 
\textit{Numerical Simulations of Interacting Galaxies: Bar Morphology}, Springer International
Publishing Switzerland 2015J. Klapp et al. (eds.), Selected Topics of Computational and Experimental Fluid Mechanics, Environmental Science and Engineering, DOI 10.1007-978-3319-11487-3-42.

\bibitem[Mateo at al. (2007)]{matteo} Matteo, P. Di, Combes, F.,  Melchior, A.L. and B. Semelin, B., 2007, Astron.Astrophy, 468, pp.61–81. 

\bibitem[Mayya at al. (2009)]{Mayya} Mayya, Y.D. and Carrasco, L., (2009), RevMexAA, 37, 44-55.

\bibitem[Mellier and Mathez (1987)]{mellier} Mellier, Y., and Mathez, G., (1987), Astronomy and  
Astrophysics, 175, pp.1-3.
 
\bibitem[Meza at al. (2003)]{Meza} Meza, A., Navarro, J.F., Steinmetz, M. and Eke, V., (2003), The 
Astrophysical Journal, 590, pp.619-635.

\bibitem[Mihos and Hernquist (1996)]{MihosHernquist96} Mihos, J. C., and Hernquist, L.E., 1996, ApJ, 464, 641.

\bibitem[Moster at al. (2011)]{moster} Moster, B. P., Macci\'o, A.V., Somerville, R.S., Naab, T. and Cox, T.J., 
(2011), MNRAS, 415, pp. 3750-3770.

\bibitem[Naab at al. (2006)]{naab} Naab, T., Jesseit, R. and Burkert, A., (2006), MNRAS. 372, pp.839-852.

\bibitem[Negroponte and White (1983)]{negroponte83} Negroponte, J. and White, S.D.M., 1983, MNRAS, 205, 1009.

\bibitem[Noguchi (1988)]{noguchi} Noguchi, M., 1988, Astron.Astrophy, 203, 259.

\bibitem[Press et al. (1992)]{numrecip} Press, W.H., Teukolsky, S.A., Vetterling, W.T. and Flannery, B.P., 1992,
\textit{Numerical Recipes in Fortran 77}, Second Edition, Cambridge University Press.

\bibitem[Paraview (2013)]{paraview} Paraview, an open-source, multi-platform data analysis and visualization
application, http://www.paraview.org/.

\bibitem[Quinn et al. (1993)]{quinn} Quinn, P. J., Hernquist, L. and Fullagar, D. P., 1993, ApJ, 403, 74.

\bibitem[Toomre at al. (1972)]{Toomre1972} Toomre, A. and Toomre, J., (1972), J. Astrophys. J., 178, 623-666.

\bibitem[Sanders et al. (1988)]{Sanders1988b} Sanders, D. B., Soifer, B. T., Elias, J. H., Madore, B. F., Mathews, K.,
Neugebauer, G. and Scoville, N. Z., 1988, ApJ, 325, 74.

\bibitem[Schneider (2006)]{extra} Schneider, P., 2006, \textit{Extragalactic Astronomy and Cosmology}, Springer-Verlag, Berlin Heidelberg 2006.

\bibitem[Scolville et al. (1986)]{Scoville1986} Scoville, N. Z., Sanders, D. B., Sargent, A. I., Soifer, B. T., Scott, S. L., and  Lo, K. Y., 1986, ApJ, 311, L42.

\bibitem[S\'ersic (1968)]{sersic} S\'ersic, J.L., (1968), Atlas de Galaxias Australes, Boletin de 
la Asociaci\'on Argentina de Astronom\'{\i}a, Vol. 6, p.41.

\bibitem[Springel and White (1999)]{Springel1999} Springel, V. and White, S.D.M., (1999), MNRAS, 307,162-178.

\bibitem[Springel (2000)]{Springel2000} Springel, V., (2000), MNRAS, 312,859-879.

\bibitem[Springel (2005)]{Springel} Springel, V., (2005), MNRAS, 364,4,1105-1134.

\bibitem[Springel and Hernquist (2003)]{springelhernquist2003} Springel, V. and Hernquist, L., (2003), MNRAS, 339, pp.289-311.

\bibitem[Springel and Hernquist (2005)]{Springel2005} Springel, V. and Hernquist, L., (2005), ApJ, 622,L9-L12.

\bibitem[Stinson et al. (2006)]{stinson2006} Stinson,G., Anil Seth, A.,  Katz,N., Wadsley, J., Governato, F. and Quinn, T., 2006, MNRAS, 373, Issue 3, pp.1074-1090.

\bibitem[Taylor et al. (2016)]{taylor} Taylor, C., Boylan-Kolchin, M., Paul Torrey, P. and Vogelsberger, M., 2016, MNRAS 461, pp.3483–3493.

\bibitem[Villalobos and Helmi (2008)]{villa} Villalobos, A. and Helmi, A., 2008, Mon. Not. R. Astron. Soc. 391, 1806–1827.

\bibitem[Wang et al. (2014)]{wang} Wang, Fu, J., Aumer, M., Kauffmann, G., J'ozsa, G.I.G, Serra, P., 
Huang, M.L., Brinchmann, J., Hulst, T. and Bigiel, F., 2014, MNRAS 441, pp.2159–2172.

\bibitem[Wang et al. (2013)]{bluedisk} Wang,J., Kauffmann,G., J'ozsa, G.I.G, Serra, P.,
Hulst, T., Bigiel, F., Brinchmann, J., Verheijen, M.A.W., Oosterloo, T., Wang, E., Li, C., 
Heijer, M. and Kerp, J., 2013, MNRAS 433, 270–294.

\bibitem[Vogelsberger et al. (2014)]{illustris} Vogelsberger, M., Genel, S., Springel, V., Torrey, P., 
Sijacki, D., Xu, D., Snyder, G., Bird, S., Nelson, D. and Hernquist, L., 2014, Nature, 509, Issue 7499, pp. 177-182.

\end{thebibliography}
\end{document}